\documentclass[structabstract]{aa}

\usepackage{natbib}

\usepackage{hyperref}
\usepackage{amsmath}
\hypersetup{colorlinks=true,citecolor=blue}
\usepackage{draftcopy}

\usepackage{url}
\usepackage{amssymb}
\usepackage{amsfonts}
\usepackage[ruled]{algorithm2e}
\usepackage{graphicx,epsfig,graphics}
\usepackage{aas_macros}
\usepackage{tikz}
\usepackage{breqn}

\newcommand{\bt}{\mbox{\boldmath{$\theta$}}}

\newcommand{\bx}{{\mathbf x}}
\newcommand{\by}{{\mathbf y}}

\newcommand{\norm}[1]{\left\lVert #1 \right\rVert}
\newcommand{\abs}[1]{\left\lvert #1 \right\rvert}
\newcommand{\scalp}[2]{\left\langle #1,#2 \right\rangle}

\DeclareMathOperator*{\argmin}{argmin}
\DeclareMathOperator{\prox}{prox}
\DeclareMathOperator{\dom}{dom}

\begin{document}

\label{firstpage}
\title{A Compressed Sensing Approach to 3D Weak Lensing} 

\author{A. Leonard \thanks{adrienne.leonard@cea.fr} \and F.-X. Dup\'e \thanks{francois-xavier.dupe@lif.univ-mrs.fr} \and J.-L. Starck }
\institute{Laboratoire AIM, UMR CEA-CNRS-Paris 7, Irfu, SAp/SEDI, Service d'Astrophysique, CEA Saclay, F-91191 GIF-SUR-YVETTE CEDEX, France.}


\abstract{Weak gravitational lensing is an ideal probe of the dark universe. In recent years, several linear methods have been developed to reconstruct the density distribution in the Universe in three dimensions, making use of photometric redshift information to determine the radial distribution of lensed sources.}{ In this paper, we aim to address three key issues seen in these methods; namely, the bias in the redshifts of detected objects, the line of sight smearing seen in reconstructions, and the damping of the amplitude of the reconstruction relative to the underlying density. We also aim to detect structures at higher redshifts than have previously been achieved, and to improve the line of sight resolution of our reconstructions.}{We consider the problem under the framework of compressed sensing (CS). Under the assumption that the data are sparse or compressible in an appropriate dictionary, we construct a robust estimator and employ state-of-the-art convex optimisation methods to reconstruct the density contrast. For simplicity in implementation, and as a proof of concept of our method, we reduce the problem to one-dimension, considering the reconstruction along each line of sight independently. We also assume an idealised survey in which the redshifts of sources are known.}{Despite the loss of information inherent in our one-dimensional implementation, we demonstrate that our method is able to accurately reproduce cluster haloes up to a redshift of $z_{\rm cl} = 1.0$, deeper than state-of-the-art linear methods. We directly compare our method with these linear methods, and demonstrate minimal radial smearing and redshift bias in our reconstructions, as well as a reduced damping of the reconstruction amplitude as compared to the linear methods. In addition, the CS framework allows us to consider an underdetermined inverse problem, thereby allowing us to reconstruct the density contrast at finer resolution than the input data.}{The CS approach allows us to recover the density distribution more accurately than current state-of-the-art linear methods. Specifically, it addresses three key problem areas inherent in linear methods. Moreover, we are able to achieve super-resolution and increased high-redshift sensitivity in our reconstructions.}


\maketitle

\section{Introduction}

Weak gravitational lensing has become a powerful tool to study the dark universe, allowing us to place constraints on key cosmological parameters, and offering the possibility to place independent constraints on the dark energy equation of state parameter, $w$ \citep{lb09,hj08,munshietal08,albrechtetal06,peacocketal06,schneider06,vwm03}.

Until recently, weak lensing studies considered the shear signal, and recovered the mass distribution, in two dimensional projection \citep[see][for a review of weak lensing]{schneider06}. However, with improved data quality and wide-band photometry, it is now possible to recover the mass distribution in three dimensions by using photometric redshift information to deproject the lensing signal along the line of sight \citep{sth09, masseyetal07b, masseyetal07a, tayloretal04}. 

Under the assumption of Gaussian noise, various linear methods have been developed to recover the 3D matter distribution, which rely on the construction of a pseudo-inverse operator to act on the data, and which include a penalty function encoding the prior that is to be placed on the signal \citep{vanderplasetal11, sth09, chk05, tayloretal04, bt03, hk02}.

These methods produce promising results; however, they show a number of problematic artefacts. Notably, structures detected using these methods are strongly smeared along the line of sight, the detected amplitude of the density contrast is damped (in some cases, very strongly), and the detected objects are shifted along the line of sight relative to their true positions. Such effects result from the choice of method used; \cite{sth09} note that their choice of filter naturally gives rise to a biased solution, and \cite{vanderplasetal11} suggest that linear methods might be fundamentally limited in the resolution attainable along the line of sight as a result of the smearing effect seen in these methods.

Furthermore, these methods are restricted to deal solely with the overdetermined inverse problem. In other words, the resolution obtainable on the reconstruction of the density is limited to be, at best, equal to that of the input data. Thus, the resolution of the reconstruction is entirely limited by the quality of the data and its associated noise levels, with no scope for improvement by judicious choice of inversion or denoising method.

In this paper, we consider the deprojection of the lensing signal along the line of sight as an instance of compressed sensing, where the sensing operator models the line-of-sight integration of the matter density giving rise to the lensing signal. For simplicity in implementation, and as a proof of concept of our method, we consider only the line of sight transformation, reducing the 3D weak lensing problem to a one-dimensional inversion, with each line of sight in an image treated as independent. We note, however, that the algorithm presented here can be cheaply generalised to three dimensions.

We adopt a sparse prior on our reconstruction, and use a state-of-the art iterative reconstruction algorithm drawn from convex analysis, optimisation methods and harmonic analysis set within a compressed sensing framework. This enables us to find a robust estimator of the solution without requiring any direct prior knowledge of the statistical distribution of the signal. We note that the compressive sensing framework allows us to consider an underdetermined inverse problem, thereby allowing us to obtain higher resolution on our reconstructions than that provided by the input data. 

This method produces reconstructions with minimal bias and smearing in redshift space, and with reconstruction amplitudes $\sim75\%$ of the true amplitude (or better, in some cases). This is a significant improvement over current linear methods, despite the adoption of a simplified, one-dimensional algorithm. In addition, our method exhibits an apparent increased sensitivity to high redshift structures, as compared with linear methods. Our reconstructions do exhibit some noise, with false detections appearing along a number of lines of sight. However, these tend to be localised to one or two pixels, rather than coherent structures, and we expect improved noise control with a full three-dimensional implementation. 

We note that we do not include photometric redshift errors in our simulations and, consequently, the simulations shown here should be considered to be idealised. \cite{maetal06} have presented a method to account for such errors in lensing measurements. This method has been used by various authors in studies on real data \citep[see, e.g.][]{simonetal11}, and is straightforward to implement in the algorithm presented here. A full treatment of photometric redshift errors, which is essential in order for the method to be useful on real data, will be presented in an upcoming work.

This paper is structured as follows. In \S~\ref{sec:theory}, we outline the weak lensing formalism in three dimensions, and outline several linear inversion methods to solve the 3D weak lensing problem. We introduce our compressed sensing framework and describe our proposed algorithm in \S~\ref{sec:nonlin}. In \S~\ref{sec:impl}, we discuss practical considerations in implementing the method, and describe our simulated dataset. In \S~\ref{sec:results} we demonstrate the performance of our algorithm in reconstructing simulated cluster haloes at various redshifts. We conclude with a discussion of our results and future applications in \S~\ref{sec:discuss}.

Throughout the text, we assume $\Lambda$CDM cosmology, with $\Omega_\Lambda = 0.736,\ \Omega_{\rm M} = 0.264,\ h=0.71,\ \sigma_8 = 0.801$, consistent with the WMAP-7 results \citep{larsonetal11}.

\section{3D Weak Lensing}
\label{sec:theory}

The distortion of galaxy images due to the weak lensing effect is described, on a given source plane, by the Jacobian matrix of the coordinate mapping between source and image planes:

\begin{equation}
  A = \left(\begin{array}{cccc}  1-\kappa-\gamma_1 & -\gamma_2\\ -\gamma_2 & 1-\kappa+\gamma_1\end{array}\right)\ ,
\end{equation}
where $\kappa$ is the projected dimensionless surface density, and $\gamma = \gamma_1+\rm{i}\gamma_2$ is the complex shear. The shear is related to the convergence via a convolution in two dimensions: 
\begin{equation}
  \label{eq:gamkap}
  \gamma(\bt) = \frac{1}{\pi}\int\ d^2\bt^\prime {\cal D}(\bt - \bt^\prime)\kappa(\bt^\prime)\ ,
\end{equation}
where
\begin{equation}
  {\cal D}(\bt) = \frac{1}{(\bt^\ast)^2}\ ,
\end{equation}
$\bt = \theta_1+{\rm i}\theta_2$, and an asterisk $^\ast$ represents complex conjugation.

The convergence, in turn, can be related to the three-dimensional density contrast $\delta(\boldsymbol{r}) \equiv \rho(\boldsymbol{r})/\overline{\rho} - 1$ by\small
\begin{eqnarray}
  \kappa(\bt,w) = \frac{3H_0^2\Omega_M}{2c^2} \int_0^w dw^\prime \frac{f_K(w^\prime) f_K(w-w^\prime)}{f_K(w)}\frac{\delta[f_K(w^\prime)\bt,w^\prime]}{a(w^\prime)}\ ,\nonumber\\ 
  \label{eq:kapconv}
\end{eqnarray}\normalsize
where $H_0$ is the hubble parameter, $\Omega_M$ is the matter density parameter, $c$ is the speed of light, $a(w)$ is the scale parameter evaluated at comoving distance $w$, and 
\begin{equation}
  f_K(w) = \begin{cases}  K^{-1/2}\sin(K^{1/2}w), & K>0 \\ w, & K=0 \\ (-K)^{-1/2}{\rm sinh}([-K]^{1/2}w) & K<0 \end{cases}\ ,
\end{equation}
gives the comoving angular diameter distance as a function of the comoving distance and the curvature, $K$, of the Universe.

If the shear (or convergence) data is divided into $N_{\rm sp}$ redshift bins, and the density contrast reconstruction is divided into $N_{\rm lp}$ redshift bins (where $N_{\rm sp}$ is not necessarily equal to $N_{\rm lp}$), we can write the convergence $\kappa^{(i)}$ on each source plane as 
\begin{equation}
  \label{eq:kapQ}
  \kappa^{(i)}(\bt) \simeq \sum_{\ell = 1}^{N_{\rm lp}} Q_{i\ell}\delta^{(\ell)}(\bt)\ ,
\end{equation}
where
\begin{equation}
  \label{eq:Qil}
  Q_{i\ell} = \frac{3H_0^2\Omega_M}{2c^2}\int_{w_\ell}^{w_{\ell+1}}dw \frac{\overline{W}^{(i)}(w)f_K(w)}{a(w)}\ ,
\end{equation}
and
\begin{equation}
  \label{eq:wbar}
  \overline{W}^{(i)}(w) = \int_0^{w^{(i)}} dw^\prime \frac{f_K(w-w^\prime)}{f_K(w^\prime)}\left(p(z)\frac{dz}{dw}\right)_{z = z(w^\prime)}\ .
\end{equation}

Thus, for each line of sight, equation \eqref{eq:kapQ} describes a matrix multiplication, encoding a convolution along the line of sight. It is the inversion of this transformation:
\begin{equation}
  \kappa(z) = \mathbf{Q}\delta(z)\ ,
\label{eq:kapdelta}
\end{equation}
that is the focus of this paper. We note that the inversion of equation \eqref{eq:gamkap} can be straightforwardly performed on each source plane in Fourier space.

\subsection{Linear Inversion Methods}
\label{sec:linmethods}

We focus here on the methods presented in \cite{sth09} and \cite{vanderplasetal11}. For a review of other linear methods, the reader is referred to \cite{hk02}.

The three dimensional lensing problem is effectively one of observing the density contrast convolved with the linear operator $\mathbf{R}$, and contaminated by noise, which is assumed to be Gaussian. Formally, we can write
\begin{equation}
  \label{eq:1}
  \boldsymbol{d} = \mathbf{R} \boldsymbol{s} + \varepsilon,\quad \varepsilon\sim\mathcal{N}(0,\sigma^2)~,
\end{equation}
where $\boldsymbol{d}$ is the observation, $\boldsymbol{s}$ the real density and $\varepsilon$ the Gaussian noise. 

The general idea behind linear inversion methods is to find a linear operator $\mathbf{H}$ which acts on the data vector to yield a solution which minimises some functional, such as the variance of the residual between the estimated signal and the true signal, subject to some regularisation or prior-based constraints. 

The simplest instance of such a linear operation is an inverse variance filter \citep{aitken34}, which weights the data only by the noise covariance, and places no priors on the signal itself:
\begin{equation}
\label{eq:invar}
\boldsymbol{\hat{s}}_{IV} = [\mathbf{R^\dagger} \boldsymbol{\Sigma}^{-1}\mathbf{R}]^{-1}\mathbf{R^\dagger} \boldsymbol{\Sigma}^{-1}\boldsymbol{d}\ ,
\end{equation}
where $\Sigma\equiv\left\langle\boldsymbol{n}\boldsymbol{n}^\dagger\right\rangle$ gives the covariance matrix of the noise. 

This method proves problematic when the matrices involved are non-invertible, such as when there are degeneracies inherent in the allowed solution. In order to make the problem invertible, some regularisation must be introduced. \cite{sth09} opt to use a Saskatoon filter \citep{tegmark97, tegmarketal97}, which combines a Wiener filter and an inverse variance filter, with a tuning parameter $\alpha$ introduced that allows switching between the two. This gives rise to a minimum variance filter, expressed as:
\begin{equation}
\boldsymbol{\hat{s}}_{MV} = [\alpha\mathbf{1} + \mathbf{SR^\dagger} \boldsymbol{\Sigma}^{-1}\mathbf{R}]^{-1}\mathbf{SR^\dagger} \boldsymbol{\Sigma}^{-1}\boldsymbol{d}\ ,
\end{equation}
where $\mathbf{S}\equiv\left\langle\boldsymbol{s}\boldsymbol{s}^\dagger\right\rangle$ encodes prior information about the signal covariance, and $\mathbf{1}$ is the identity matrix.

\cite{vanderplasetal11} have recently proposed a filter based on the singular value decomposition (SVD) of the inverse variance filter of Equation \eqref{eq:invar}. Under this formalism, we can write 
\begin{equation}
\boldsymbol{\hat{s}}_{IV} = \mathbf{V}\boldsymbol{\Lambda^{-1}}\mathbf{U^\dagger} \boldsymbol{\Sigma}^{-1/2}\boldsymbol{d}\ ,
\end{equation}
where we have decomposed the matrix $\widetilde{\mathbf{R}} \equiv \boldsymbol{\Sigma^{-1/2}}\mathbf{R} = \mathbf{U}\boldsymbol{\Lambda}\mathbf{V^\dagger}$, $\mathbf{U^\dagger U} = \mathbf{V^\dagger V} = \mathbf{1}$ and $\boldsymbol{\Lambda}$ is the square diagonal matrix of singular values $\lambda_i = \Lambda_{ii}$. 

Their filtering consists in defining a cutoff value $\sigma_{\rm cut}$, and truncating the matrices to remove all singular values with $\sigma_i<\sigma_{\rm cut}$. This effectively reduces the noise by removing the small singular values, which translate into large values in the inversion. 

VanderPlas et al. note that the SVD decomposition is computationally intensive, and while they do describe a method to speed up the process, it may not be practical to use this method on large images. 

Similar considerations must be made when using any of the linear methods described above, as these all involve matrix inversion, which is an $\mathcal{O}(N^3)$ process. While optimisations can be found, these methods become excessively time- and computer-intensive when large datasets are considered. This, in effect, limits the resolution attainable using these methods. 

Moreover, as discussed extensively in \cite{sth09} and \citet{vanderplasetal11}, these linear methods give rise to a significant bias in the 
location of detected peaks, damping of the peak signal and a substantial smearing of the density along the line of sight. The Compressed Sensing (CS) theory, described below, allows us to address the lensing inversion problem under a new perspective, and we will show that these three aspects are significantly improved using a non-linear CS approach.

\section{Compressed Sensing Approach}
\label{sec:nonlin}

Linear methods are easy to use, and the variance of each estimator is rather direct to compute. Furthermore, these methods generally rely on very common tools with efficient implementation. However, they are not the most powerful, and including non-Gaussian priors is difficult, especially when such priors imply non-linear terms. Obviously, using better-adapted priors is required for building a more robust estimator.  In this paper, we adopt a compressed sensing approach in
order to construct an estimator that exploits the sparsity of the signal that we aim to reconstruct. The estimator is modelled as a optimisation problem that is solved using recent developments from convex analysis and splitting methods.

\subsection{Compressed sensing theory}
\label{sec:compress-sensing}

We consider some data $Y_i$ ( $i \in \left[1, .., m\right]$) acquired through the linear system \begin{equation}\label{eq:pb2} Y = \boldsymbol{\Theta} X~,\end{equation} where $\boldsymbol{\Theta}$ is an $m\times n$ matrix.
Compressed Sensing \citep{CandesTao04,donoho:cs} is a sampling/compression theory based on the sparsity of the observed signal, which shows that, under certain conditions, one can exactly recover a $k$-sparse signal (a signal for which only $k$ pixels have values different from zero, out of $n$ total pixels, where $k < n$) from $m<n$ measurements. 

Such a recovery is possible from undersampled data only if the sensing matrix $\boldsymbol{\Theta}$ verifies the \textit{Restricted Isometry Property} (RIP) \citep[see][for more details]{CandesTao04}.  This property has the effect that each measurement $Y_i$ contains some information about all of the pixels of $X$; in other words, the sensing operator $\boldsymbol{\Theta}$ acts to spread the information contained in $X$ across many measurements $Y_i$. 

Under these two constraints -- sparsity and a transformation meeting the RIP criterion -- a signal can be recovered exactly even if the number of measurements $m$ is much smaller than the 
number of unknown $n$. This means that, using CS methods, we will be able to far outperform the well-known Shannon sampling criterion. 

The solution $X$ of \eqref{eq:pb2} is obtained by minimizing 
\begin{equation}
\label{eq:mincs}
   \min_{X}  \norm{X}_1\ ~~ s.t. ~~ Y= \boldsymbol{\Theta} X
\end{equation}
where the $\ell_1$ norm is defined by  $\norm{X}_1 = \sum_i \mid X_i \mid$. The $\ell_1$ norm is well-known to be a sparsity-promoting function; i.e. minimisation of the $\ell_1$ norm yields the most sparse solution to the inverse problem. Many optimisation methods have been proposed in recent years to minimise this equation. Further details about CS and $\ell_1$ minimisation algorithms can be found in \citet{smf10}.

In real life, signals are generally not ``strictly" sparse, but are \textit{compressible}; i.e. we can represent the signal in a basis or frame (Fourier, Wavelets, Curvelets, etc.) in which the curve obtained by plotting the obtained coefficients, sorted by their decreasing absolute values, exhibits a polynomial decay. Note that most natural signals and images are compressible in an appropriate basis. 

We can therefore reformulate the CS equation above (Equation \eqref{eq:mincs}) to include the data transformation matrix $\boldsymbol{\Phi}$:
\begin{equation}
  \label{eq:min_dico}
  \min_{\alpha}  \norm{\alpha}_1\ ~~ s.t. ~~ Y=  \boldsymbol{\Theta \Phi} \alpha
\end{equation}
where $X = \boldsymbol{\Phi^*} \alpha$, and $\alpha$ are the coefficients of the transformed solution $X$ in $\boldsymbol{\Phi}$, which is generally referred to as the {\em dictionary}. Each column represents a vector (also called an \textit{atom}), which ideally should be chosen to match the features contained in $X$. If $\boldsymbol{\Phi}$ admits a fast implicit transform (e.g. Fourier transform, Wavelet transform), fast algorithms exist to minimise Equation \eqref{eq:min_dico}. 

One problem we face when considering CS in a given application is that very few matrices meet the RIP criterion. However, it has been shown that accurate recovery can be obtained as long the mutual coherence between $\boldsymbol{\Theta}$ and $\boldsymbol{\Phi}$, $\mu_{\boldsymbol{\Theta}, \boldsymbol{\Phi}} = \max_{i,k} \big| \Big< \boldsymbol{\Theta}_i, \boldsymbol{\Phi}_k,  \Big>  \big|$, is low \citep{candes2010}.
The mutual coherence measures the degree of similarity between the sparsifying basis and the sensing operator. Hence, in its relaxed definition, we consider a linear inverse problem $Y = \boldsymbol{\Theta\Phi} X$ as being an instance of CS when:
\begin{enumerate}
\item the problem is underdetermined,
\item the signal is compressible in a given dictionary $\boldsymbol{\Phi}$,
\item The mutual coherence $\mu_{\boldsymbol{\Theta}, \boldsymbol{\Phi}}$ is low. This will happen every time the matrix $\mathbf{A} = \boldsymbol{\Theta \Phi}$ has the effect of spreading out the coefficients $\alpha_j$ of the sparse signal on all measurements $Y_i$.
\end{enumerate}
Most CS applications described  in the literature are based on such a soft CS definition.
CS was introduced for the first time in astronomy for data compression \citep{bobin2008,Barbey11}, and a direct link between CS and radio-interferometric image reconstruction 
was recently established in \citet{wiaux2009}, leading to dramatic improvement thanks to the sparse $\ell_1$ recovery \citep{Cornwell2011}.
 
\subsubsection{CS and Weak Lensing}
\label{sec:lensing-operator}

The 3D weak lensing reconstruction problem can be seen to completely meet the soft-CS criteria above. Indeed,  
\begin{enumerate}
\item the problem is undetermined as we seek a higher resolution than initially provided by the observations, which are noise-limited,
\item the matter density is seen to be sparsely distributed, showing clusters connected by filaments surrounding large voids,
\item the lensing operator encoded by the matrix $\mathbf{Q}$ in Equation \eqref{eq:kapdelta} (or equivalently, the combination $\mathbf{P}_{\gamma\kappa}\mathbf{Q}$, where $\mathbf{P_{\gamma\kappa}}$ encodes the convolution in Equation \eqref{eq:gamkap}) spreads out the information about the underlying density in a compressed sensing way.
\end{enumerate}

\begin{figure}
\includegraphics[width=0.45\textwidth]{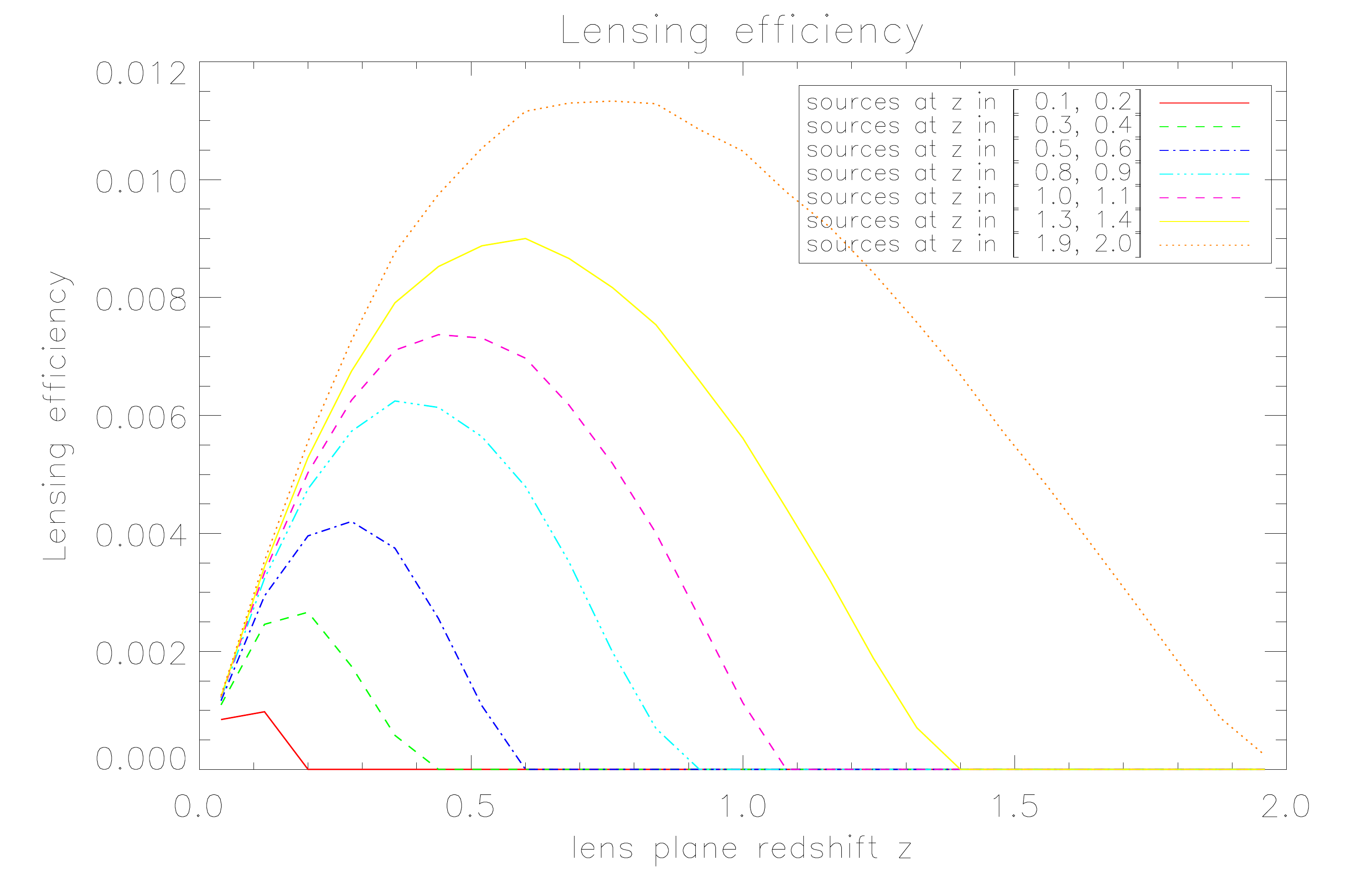}
\includegraphics[width=0.45\textwidth]{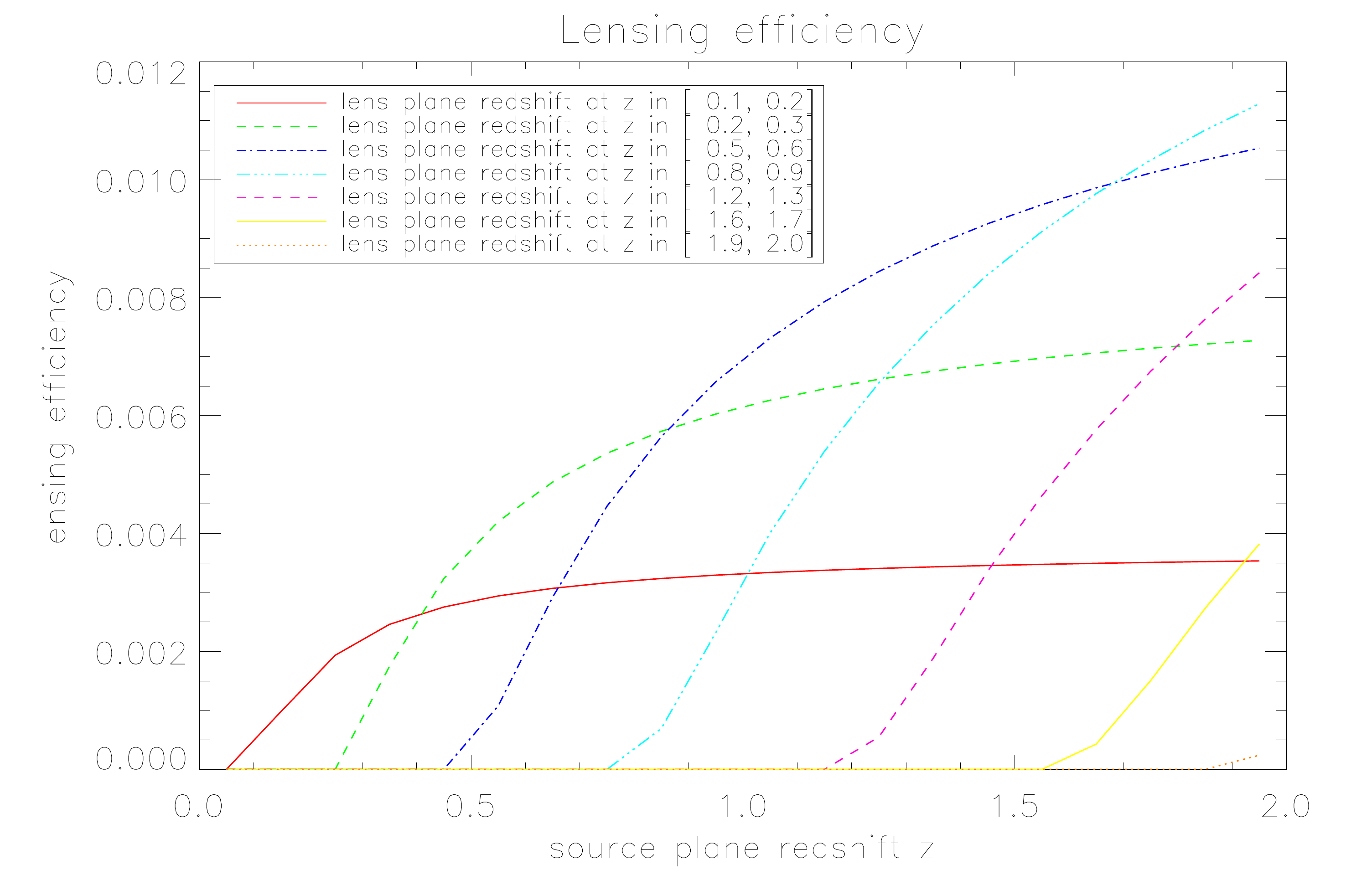}
\caption{
\textit{Top Panel}: Lensing efficiency kernel for given source planes as a function of lens
  redshift. 
  \textit{Bottom Panel}: Lensing efficiency of a given lens plane as a function of source
  redshift. 
  }
 \label{fg:qmatrix}
\end{figure}

To highlight point (3) above, Figure \ref{fg:qmatrix} shows the rows (top panel) and columns (bottom panel) of the transformation matrix, $\mathbf{Q}$, encoding weak lensing along the line of sight. The top panel shows the lensing efficiency kernel, which reflects the sensitivity of a given source plane to the shearing effects of lenses at various redshifts. This is the broad convolution kernel of Equation \eqref{eq:kapconv}. The bottom panel shows the effect of a discrete lens at a given redshift on source planes, and demonstrates that localised lenses give rise to effects that are non-local and affect all sources at redshifts greater than the lens redshift. 

\subsubsection{Sparsity prior}
\label{subsec:sparsity}

Sparse priors have been shown to be very useful in regularising ill-posed inverse problems \citep[see][and references therein]{fs09}. In addition, a sparse prior using a wavelet basis has been used in many areas of signal processing in astronomy, such as denoising, deconvolution and inpainting to recover missing data in images \citep{smf10} . The idea underlying such priors is that there exists a dictionary in which a given dataset is sparsely represented. The dictionary used should therefore match as closely as possible the shapes of the structures that we aim to detect.

Many experiments, such as N-body simulations, have shown that the matter density in the universe is largely distributed in localised clusters, which are connected by thin filaments. Because structures in the Universe appear to be physically sparse, we may therefore assume that the matter density is sparse in a domain adapted for cluster- and curve-like structures. Such domains exist and can be constructed by gathering several well chosen transforms inside a dictionary (e.g. a combination of wavelets and curvelets). The dictionary should be chosen in order to match as closely as possible the type of structure we aim to recover (e.g. isotropic structures for wavelets, filaments for curvelets), and may be adapted to include structures that are not sparse in the direct domain, but which may be sparsely represented in an appropriate dictionary.

Regularisation using a sparsity constraint can be understood through a Bayesian framework. We assume that the distribution of the solution in the sparsifying dictionary has a Laplacian distribution. This is equivalent to constraining the $\ell_1$ norm, which promotes sparsity. A Gaussian assumption, in contrast, constrains the $\ell_2$ norm, and leads to the standard Wiener filtering, a Tikhonov solution or an SVD solution, depending on the way we constrain the solution.

For specific classes of inverse problem, it has even been shown that the sparse recovery leads to the exact solution of the problem (compressed sensing). Such a behaviour does not exist with any other prior than sparsity. Obviously, the sparse recovery will be optimal when the signal is sparse, in the same way that the Wiener filter is optimal when the signal (and noise) is Gaussian. Because the Laplacian assumption (in a appropriate space such as wavelets) is more applicable than the Gaussian distribution, restoration of astronomical data is generally much more efficient using sparsity.

This explains why wavelets have been so successful for astronomical image restoration/detection. For the reconstruction of clusters along the line of sight, we are in a perfect situation for sparse recovery since clusters are not resolved due to the bin size, and they can therefore be modelled as Dirac $\delta-$functions. We therefore take $\Phi$ to be a $\delta-$function dictionary. Clearly, in this case, the pixel domain is especially appropriate for sparse recovery. We believe that the sparse prior is a much better model for this kind of data compared to previous methods with implicit Gaussian assumptions. Our results in this paper appear to support this claim.

We note that the method presented here is somewhat similar to the point-source reconstruction method described in \cite{hk02}, though they use $\ell_2$ minimisation combined with a strict prior on the number of haloes present along a line of sight. Our method, in contrast, places no priors on the number of structures along the line of sight.

\subsection{Problem statement}
\label{subsec:problem-statement}

Under the CS framework, the reconstruction of the matter density amounts to finding the most sparse solution that is consistent with the data. There are many different ways to formulate such an optimisation problem, and we opt for the following:
\begin{equation}
  \label{eq:minalt}
  \min_{\boldsymbol{s}\in \mathbb{R}^n} \norm{\boldsymbol{\Phi}^*\boldsymbol{s}}_1\ s.t.\ \tfrac{1}{2}\norm{\boldsymbol{d} - \mathbf{R}\boldsymbol{s}}^2_{\boldsymbol{\Sigma}^{-1}}\le \epsilon\quad \boldsymbol{s} \in \mathcal{C}\ ,
\end{equation}
where the term to minimise is a sparsity-penalty function over the dictionary coefficients. The second term in Equation \eqref{eq:minalt} above is a data fidelity constraint, with $\boldsymbol{\Sigma}$ being the covariance matrix of the noise and $\epsilon$ the allowed distance between the estimation and the observation, while the final term forces the solution to have values inside a given interval, usually $\mathcal{C} = [-1,+\infty[^n$ for matter overdensity. 

Note that this latter constraint, encoding a hard minimum on the signal to be recovered, is not possible with linear methods, and is therefore an additional strength of our method. Enforcing such physical constraints on our solution helps to ensure the recovery of the most physically compelling solution given the data.

In the compressed sensing literature, we can find equivalent writing of the problem \eqref{eq:minalt}, but with a focus on
the data fidelity. We prefer to directly seek for the sparsest solution for a given freedom (i.e. $\varepsilon$) on the
distance to the observation, as such distance is usually easier to tune than an equilibrium between a regularization and
the data fidelity terms.

Note this optimisation problem can be equivalently expressed in a Bayesian framework, assuming Gaussian noise and a Laplacian distribution of the dictionary coefficients. 


In order to solve \eqref{eq:minalt}, we use the primal-dual splitting method of \cite{cp10}, and is described in full in Appendix~\ref{sec:solv-optim-probl}. The algorithm is iterative, and effectively splits the problem into two parts, applying the two constraints in equation \eqref{eq:minalt} separately.

On each iteration, the estimate of the reconstruction is compared with the data, and the data fidelity constraint (the second constraint in equation \eqref{eq:minalt}) is applied by projection of the residual onto an $\ell_2$ ball of radius $\epsilon$. Independently, the estimate of the solution is projected onto the sparsifying basis, and sparsity is imposed through soft-thresholding, which minimises the $\ell_1$ norm of the basis coefficients. The threshold level is set by the parameter $\lambda$, which aids in controlling the noise. The noise modelling is discussed in section \ref{subsec:noise}, whilst the algorithm, and all practical considerations related to its implementation, are discussed in detail in Appendix \ref{sec:implementation}. Figure \ref{fg:alg_schem} shows a simplified schematic of the algorithm used.

\begin{figure}[h]
\includegraphics[width = 0.5\textwidth]{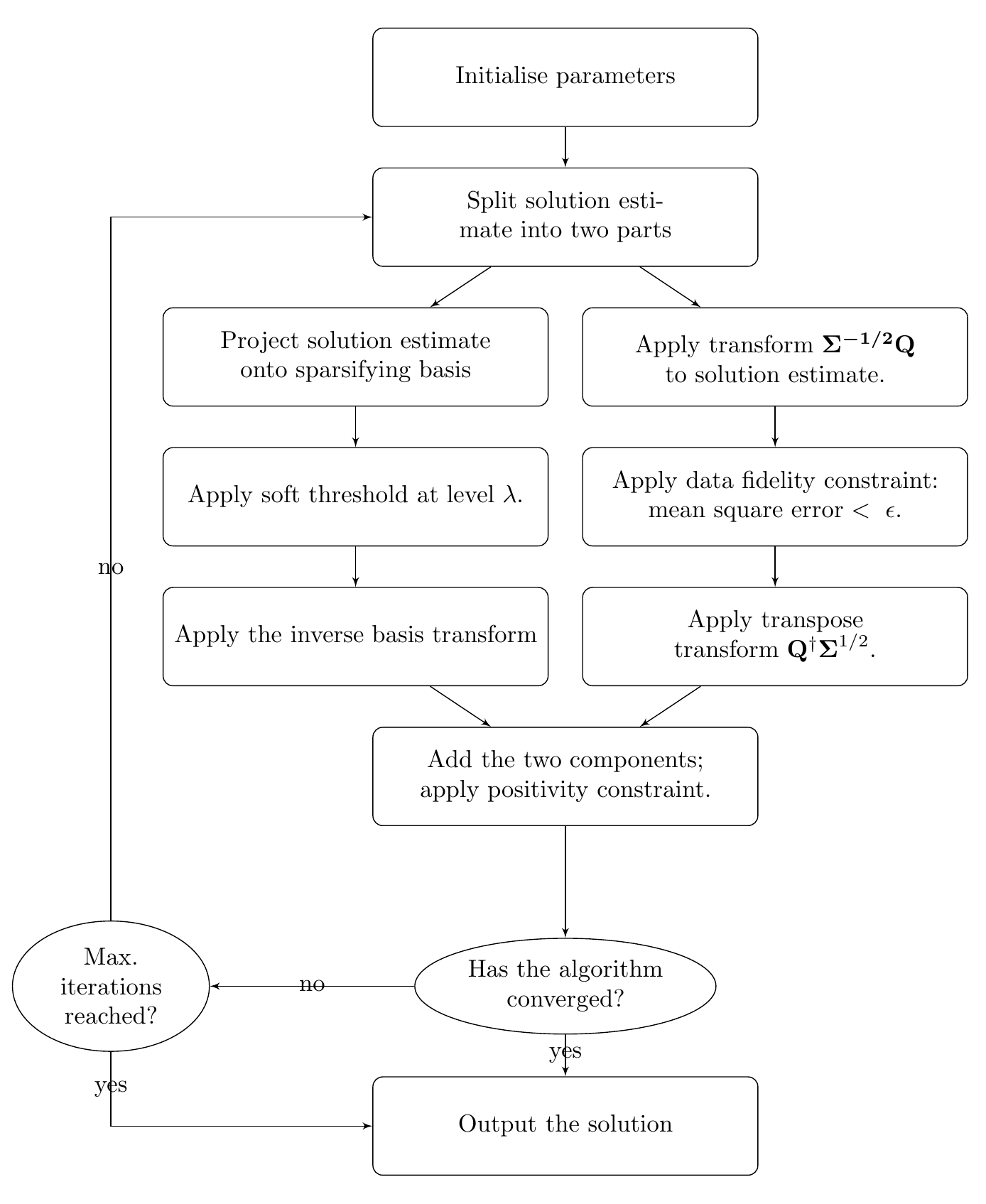}
\caption{Simplified schematic of the reconstruction algorithm described in the text.\label{fg:alg_schem}}
\end{figure}

\section{Implementation of the Algorithm}
\label{sec:impl}

We first consider the simulated data to be used in the remainder of this paper, before discussing some practical considerations important when implementing the algorithm.

\subsection{Cluster Simulations}
\label{subsec:sims}

\begin{figure}
\includegraphics[width=0.45\textwidth]{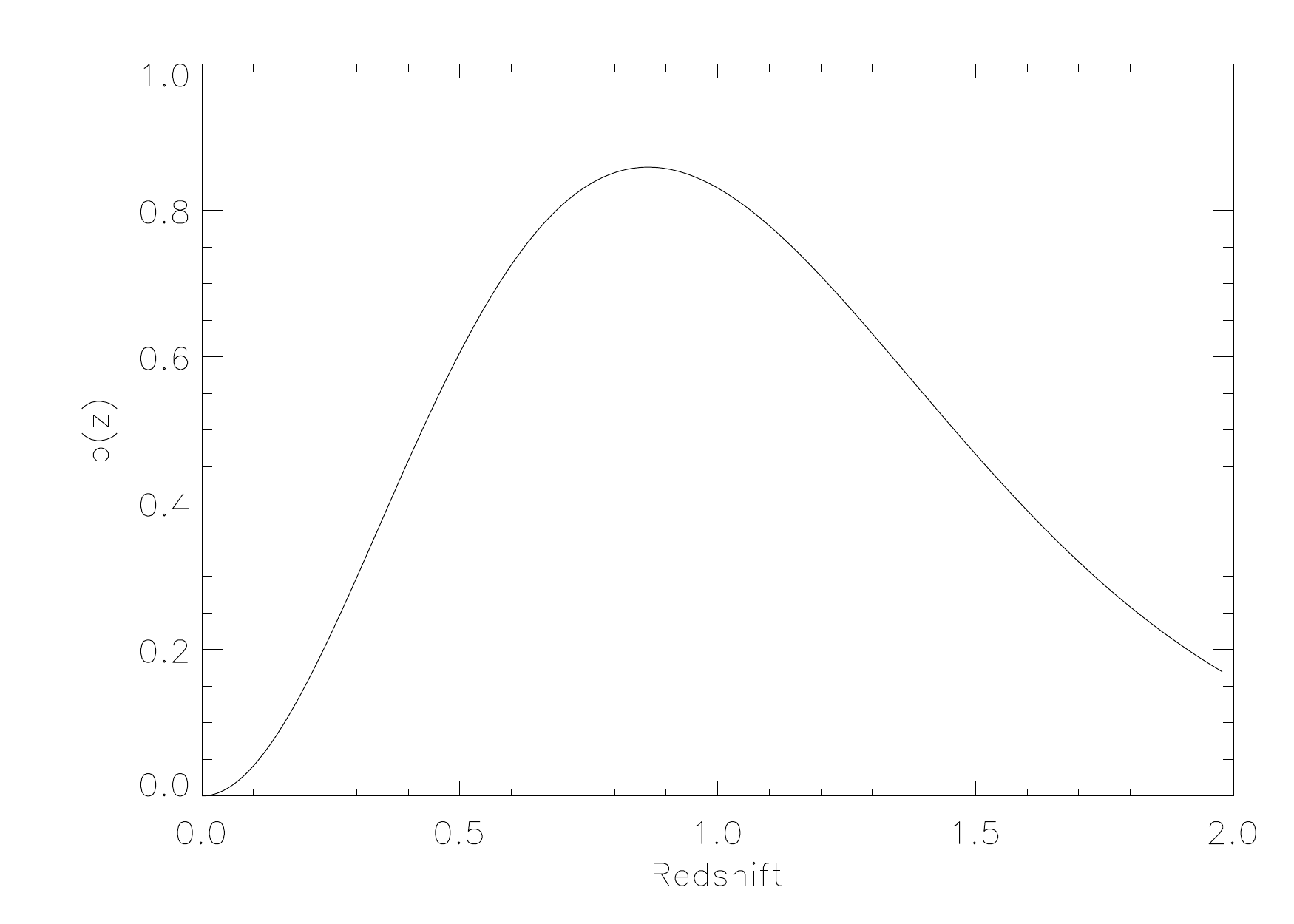}
\caption{Redshift probability distribution $p(z)$ of the sources as a function of source redshift $z$ for the simulations described in the text. \label{fg:zdist}}
\end{figure}

In order to test our method, we need to simulate a realistic data set. To this end, we consider here a fiducial survey with a background galaxy number density $n_g = 100\,{\rm arcmin}^{-2}$ distributed in redshift according to 
\begin{equation}
p(z) \propto z^2 {\rm e}^{-(1.4z/z_{0})^{1.5}} \ ,
\end{equation}
with $z_0 = 1.0$ \citep{tayloretal07, kitchingetal11}, and the distribution truncated at a maximum redshift of $z_{max} = 2$. Figure \ref{fg:zdist} shows this probability distribution, normalised arbitrarily.  

We take the intrinsic dispersion in shear measurements to be $\sigma_\gamma = 0.2$, and consider a field of $1^\circ \times 1^\circ$ divided into a grid of $60 \times 60$ pixels. These parameters are chosen to mimic the data quality expected from next-generation surveys such as Euclid \citep{euclid} and LSST \citep{lsst1, lsst2}.

In each simulated image, one or more clusters are generated following an NFW density profile with $M_{200} = 10^{15}\,M_\odot,\ c=3$ binned into $N_{\rm sp}$ redshift bins. The effective convergence and shear are computed by integrating the lensing signal within each source redshift bin, and Gaussian noise is added, scaled appropriately by the number density of galaxies within that bin. 

\subsection{Reconstructions in 1D}

As noted in \S~\ref{sec:theory}, the 3D weak lensing problem can be reduced to a one-dimensional problem, by taking as our data vector the (noisy) lensing convergence along each line of sight, which is related to the density contrast through Equation \eqref{eq:kapdelta}. Therefore, we take $\boldsymbol{d} = \boldsymbol{\kappa}_{ij}(z)$ and $\mathbf{R} = \mathbf{Q}$, and consider each line of sight in our images independently. Further, as discussed previously, we take $\boldsymbol{\Phi}$ to be a $\delta$-function dictionary.

In our simulations, clusters are placed into a region where the mean density in the absence of the cluster is equal to the mean density of the Universe at that redshift. In other words, $\delta$ is constrained to be greater than zero in all our simulations. Therefore, the projection onto the convex set $\mathcal{C}$ in algorithm \ref{alg:inversion} applies a positivity constraint at each iteration. 

Clearly, a one-dimensional implementation throws away information, as we do not account at all for the correlation between neighbouring lines of sight that will arise in the presence of a large structure in the image; however, reducing the problem to a single dimension is fast and easy to implement, and allows us to test the efficacy of the algorithm using a particularly simple basis function through which we impose sparsity. A fully three-dimensional treatment of the problem, with more accurate noise modelling (see below) will be the subject of a future work.

However, the algorithm used is entirely general; therefore, with appropriate choice of a three-dimensional basis set and taking $\boldsymbol{d} = \boldsymbol{\gamma}(\bt,z)$ and $\mathbf{R} = \mathbf{P_{\gamma\kappa}Q}$, one can implement this algorithm as a fully three-dimensional treatment of the data with no modification to the algorithm itself.

\subsection{Noise Modelling and Control}
\label{subsec:noise}

\subsubsection*{Noise Model for the Data}
We assume that the redshifts of the sources are known exactly, so there is no correlation between the noise in each source bin. Therefore, the covariance matrix of the noise along the line of sight is diagonal, with 
\begin{equation}
\boldsymbol{\Sigma}_{ii} = \sigma_g^2(z_i) = \frac{\sigma_\gamma^2}{n_g(z_i)A_{pix}}\ ,
\end{equation}
where $A_{pix}$ is the pixel area, $n_g(z_i)$ is the number density of sources in the bin at redshift $z_i$, and $\sigma_\gamma$ is the intrinsic dispersion in galaxy ellipticity, taken throughout to be $0.2$. 

This covariance matrix is used in the evaluation of the data fidelity constraint in our algorithm above. Note that the covariance matrix is only diagonal if the galaxy redshifts are known exactly. In practice, photometric redshift errors mean that each redshift slice in the data is likely to be contaminated with a few galaxies whose redshift error bars overlap with neighbouring redshift bins. In this case, the covariance matrix will have additional, non-diagonal elements that are non-zero. This is straightforward to model, however, for the chosen method of photometric redshift estimations, and our algorithm is entirely general regarding the form of the covariance matrix. Therefore, the problem of photometric redshift errors is readily tractable in our method, and will be presented in a future work.

\subsubsection*{Noise Control in the Algorithm}
The noise in the reconstruction is controlled and suppressed by two parameters in the algorithm described in Figure \ref{fg:alg_schem} and Appendix \ref{sec:implementation}. The first, and most important of these parameters is the data fidelity control parameter, $\epsilon$.

\begin{figure}[h]
\includegraphics[width = 0.25\textwidth]{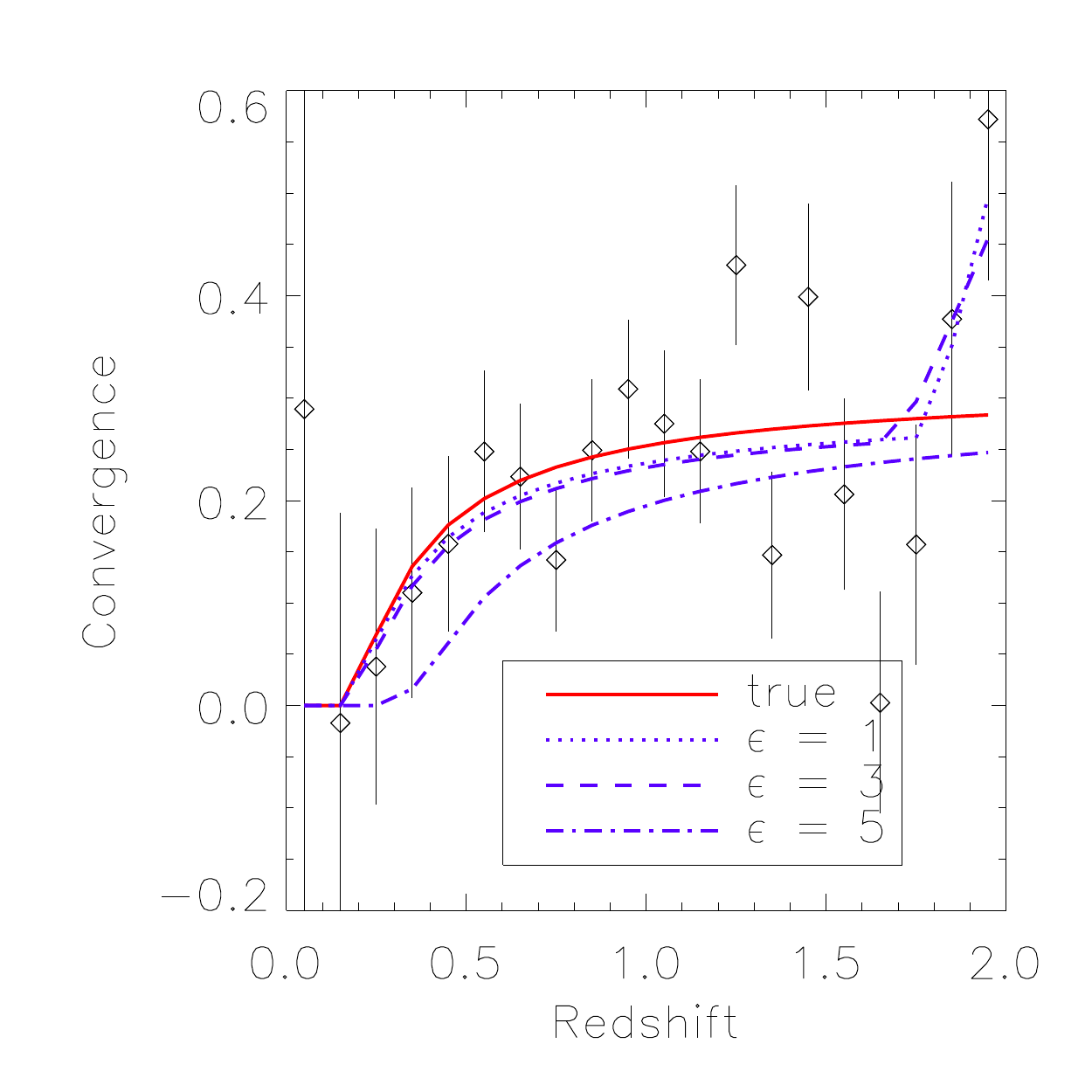}\includegraphics[width=0.25\textwidth]{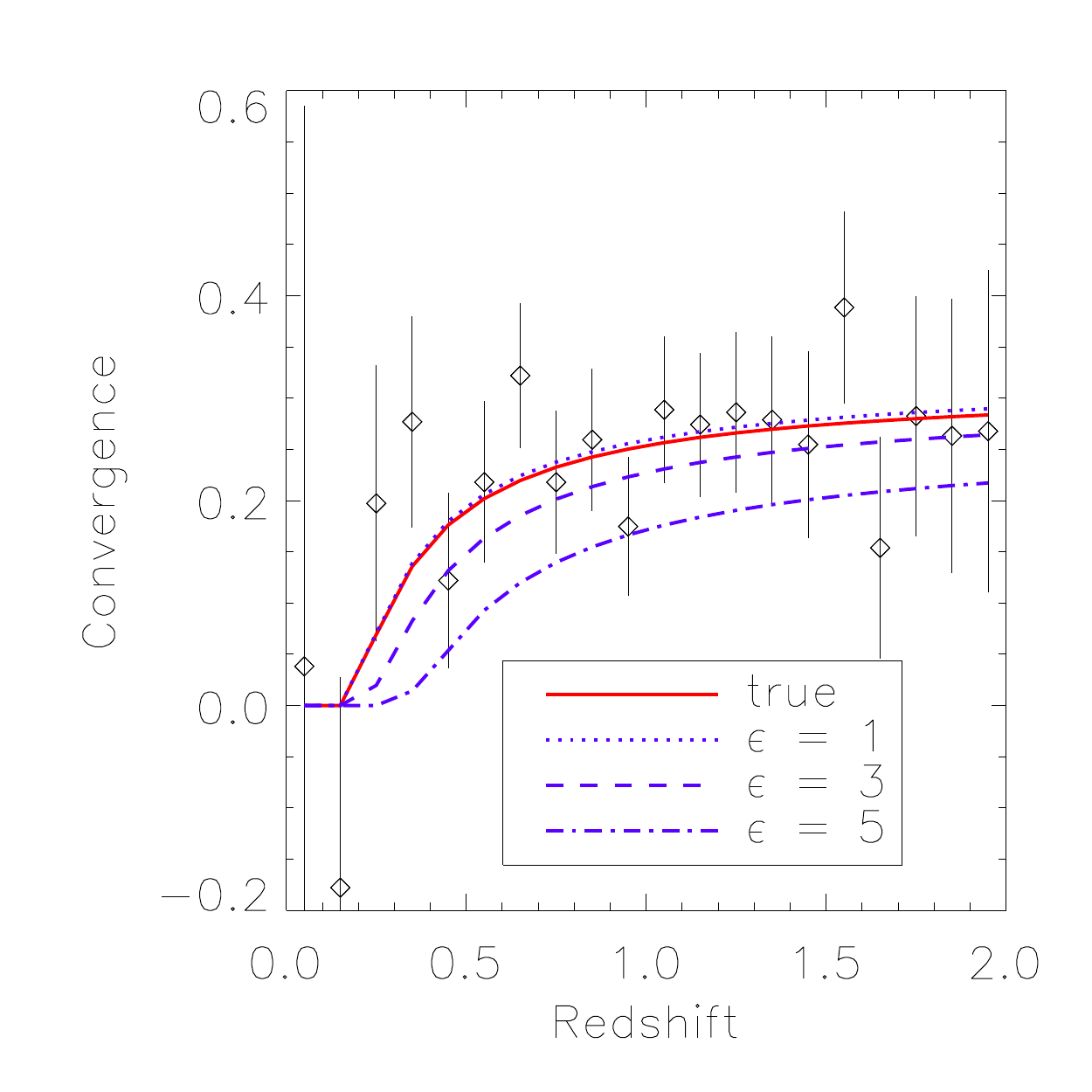}
\includegraphics[width = 0.25\textwidth]{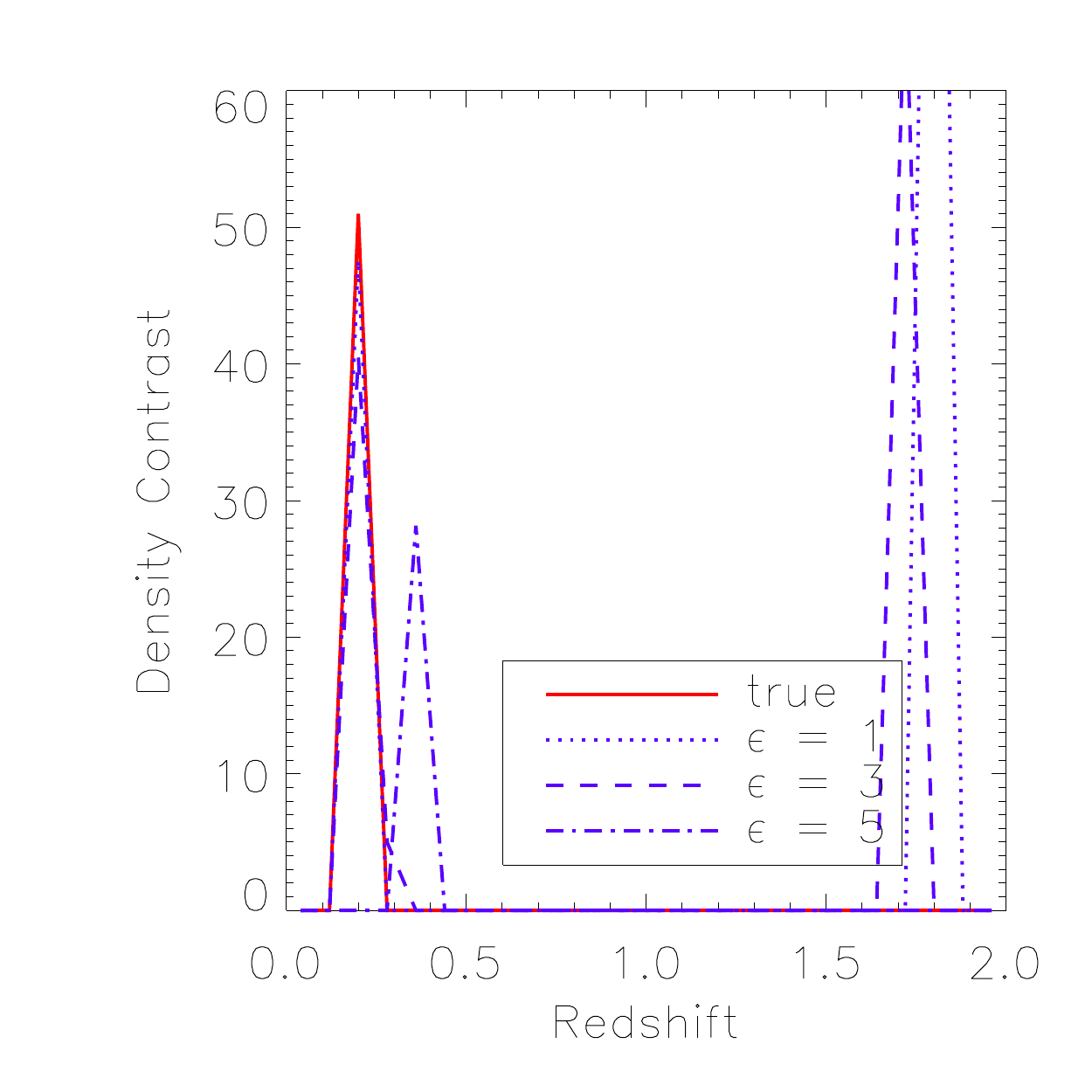}\includegraphics[width=0.25\textwidth]{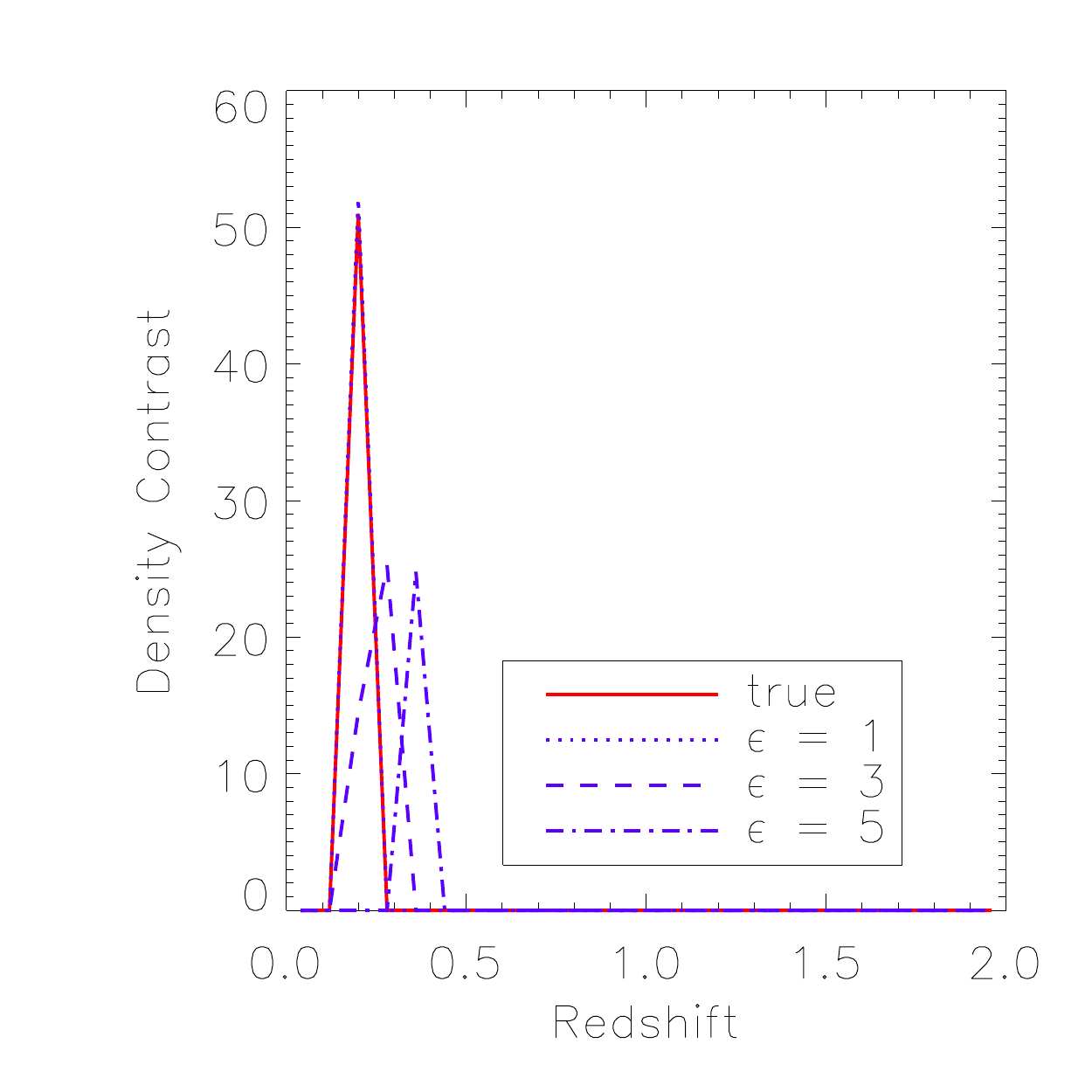}
\caption{Reconstructions for two lines of sight obtained with varying $\epsilon$. The top row shows the input data and reconstructed convergence vector, 
while the bottom row shows the reconstructed density contrast.}
\label{fg:epsilon}
\end{figure}

This parameter controls how well the data are fit by the reconstruction, with $\epsilon = 0$ implying a perfect fit to the data, which is not possible in the presence of noise. Figure \ref{fg:epsilon} demonstrates the effect of varying $\epsilon$ in the reconstruction of two lines of sight from our simulations. 

Clearly, when $\epsilon$ is small, the algorithm attempts to fit each data point more closely which, in the presence of noise, can result in overfitting of the data (as seen in line of sight 1) and hence false detections along the line of sight. On the other hand, a large $\epsilon$ may result in a solution that is not a good fit to the data (as seen in line of sight 2). 

The second parameter used to control the noise is the soft threshold parameter $\lambda$, which is used in the algorithm to impose the sparsity prior. A threshold set excessively high will result in a null solution, whilst a threshold set fairly low will allow for more false detections of noise peaks along a given line of sight. The appropriate value for this threshold should be related to the expected fluctuations in the density contrast resulting from noise variations, and should scale with the signal to noise in the image.

Note that while $\epsilon$ strongly affects the accuracy of the estimation in reproducing the underlying density contrast, $\lambda$ simply affects the sparsity of the solution. In other words, changing $\lambda$ will not greatly affect the reconstructions of true density peaks, but may affect the number of false detections and noise peaks seen. Also note that a thresholding $\lambda$ does not imply that density peaks with $\delta < \lambda$ will not be detected, as soft thresholding is only applied to one part of the estimate of the solution.

\section{Results}
\label{sec:results}

\subsection{Comparison with Linear Methods}

Firstly, to demonstrate the effectiveness of our method, we compare our method directly with the linear methods of \cite{sth09} and \cite{vanderplasetal11}. As these linear methods are only defined for $N_{\rm lp} \le N_{\rm sp}$, we consider the case of $N_{\rm lp} = N_{\rm sp} = 20$, and generate a single cluster halo at a redshift of $z_{\rm cl} = 0.25$ following an NFW halo profile with $M_{200} = 10^{15}M_\odot$ and $c=3$. 

In the SVD method of \cite{vanderplasetal11}, we take $v_{cut}  = 1 - \sum_{i=1}^n\sigma_i^2/\sum_i^{n_{max}} \sigma_i^2 = 0.01$, and in the transverse and radial Wiener filtering methods of \cite{sth09}, we take the tuning parameter to be $\alpha = 0.05$ in both cases. For our CS approach, we take the soft threshold parameter $\lambda=8$, and the data fidelity control parameter $\epsilon = 3$.

Note that while the linear methods take $\boldsymbol{d} = \boldsymbol{\gamma}(\bt, z),\ \mathbf{R} = \mathbf{P_{\gamma\kappa}Q}$, our method takes $\boldsymbol{d} = \boldsymbol{\kappa}(\bt,z),\ \mathbf{R} = \mathbf{Q}$ as before. The noise levels in each case are identical.

\begin{figure}
  \includegraphics[width=0.24\textwidth]{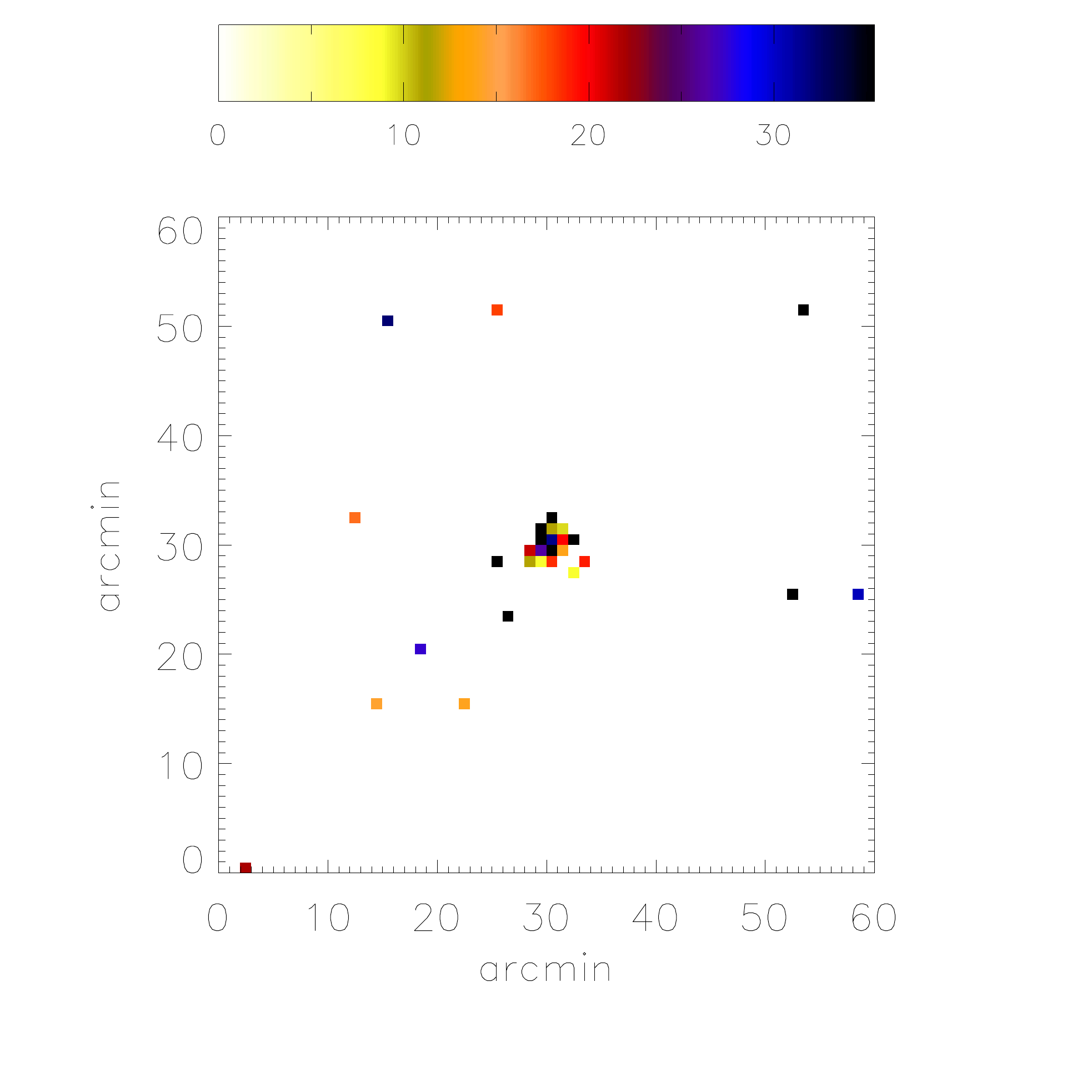}\includegraphics[width=0.24\textwidth]{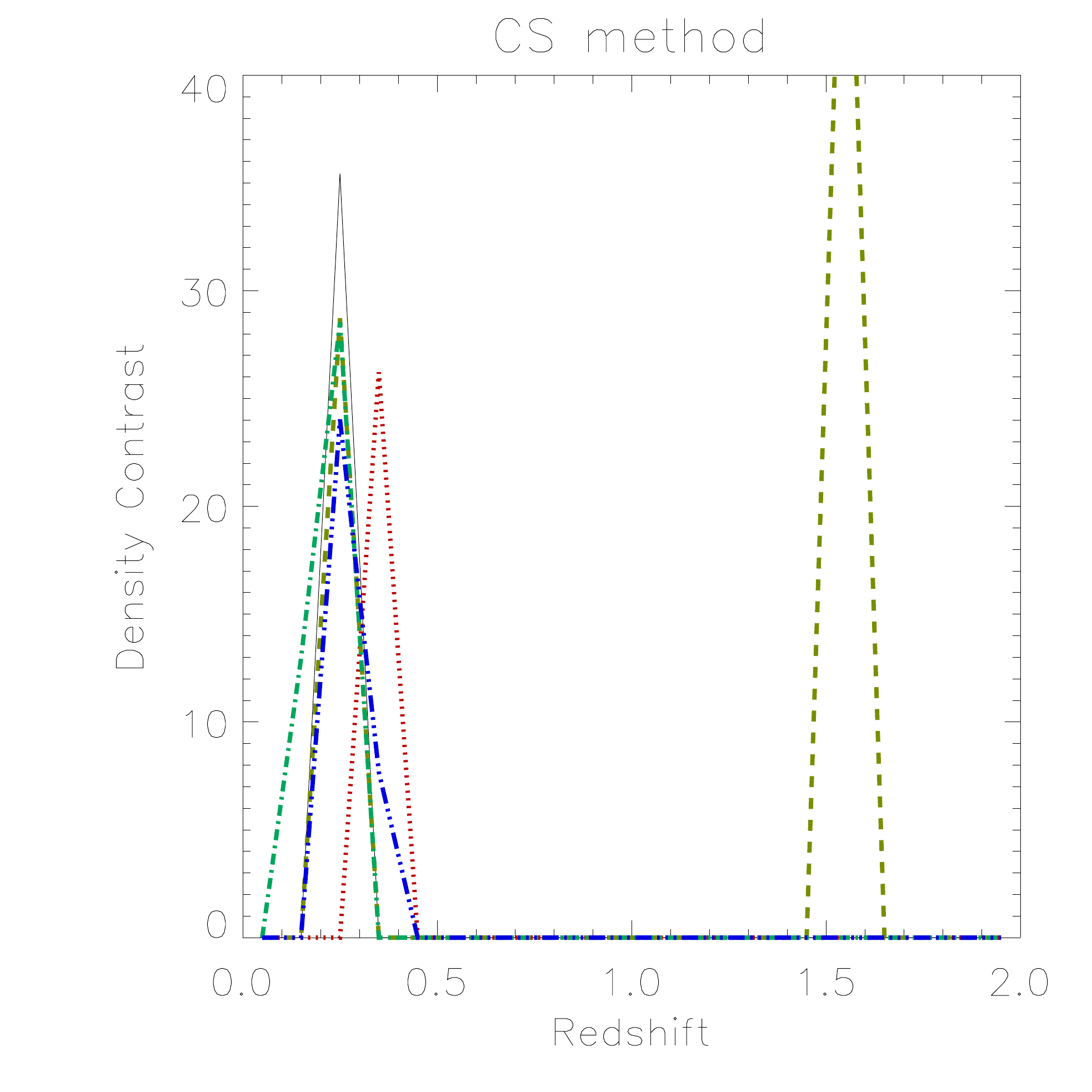}
  \includegraphics[width=0.24\textwidth]{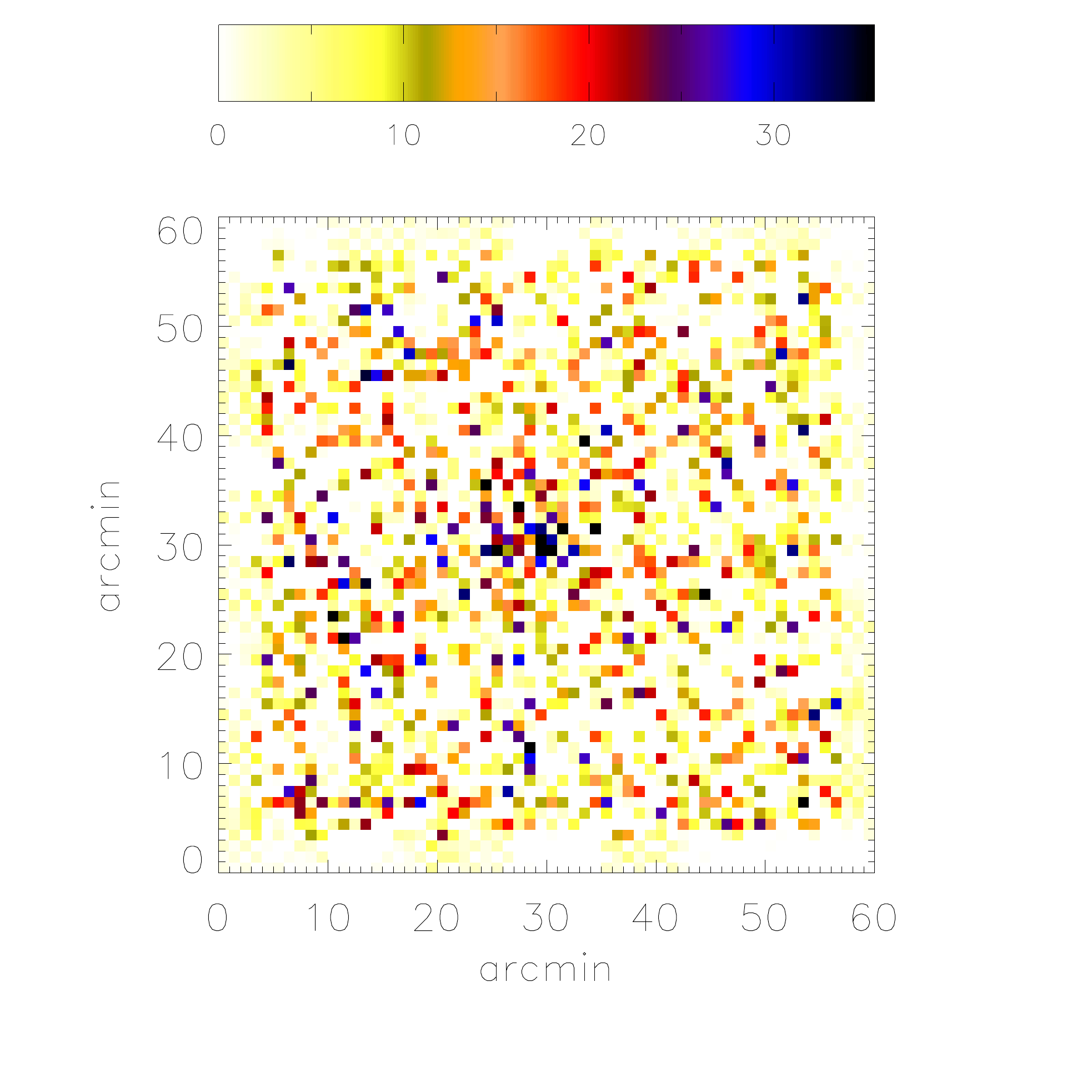}\includegraphics[width=0.24\textwidth]{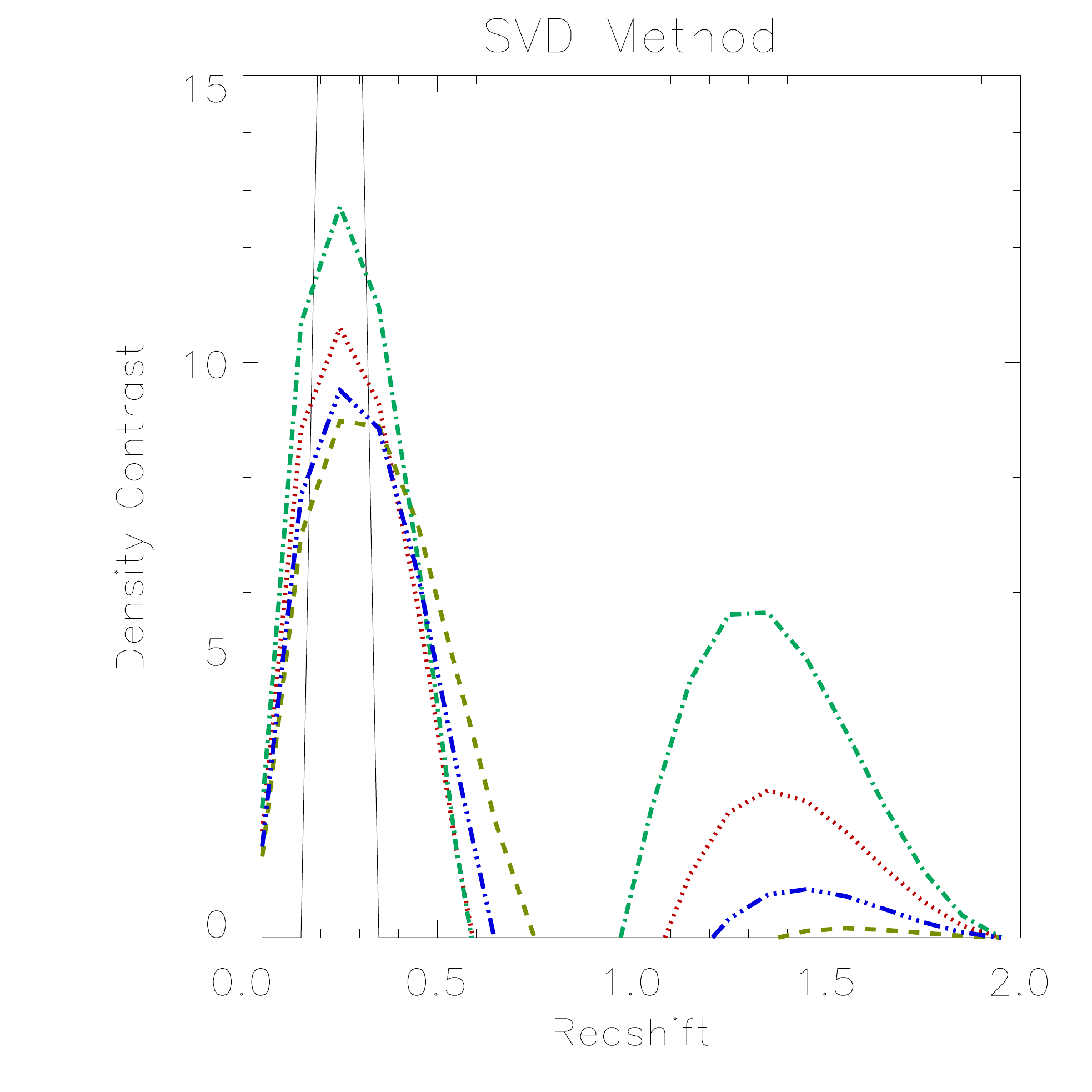}
  \includegraphics[width=0.24\textwidth]{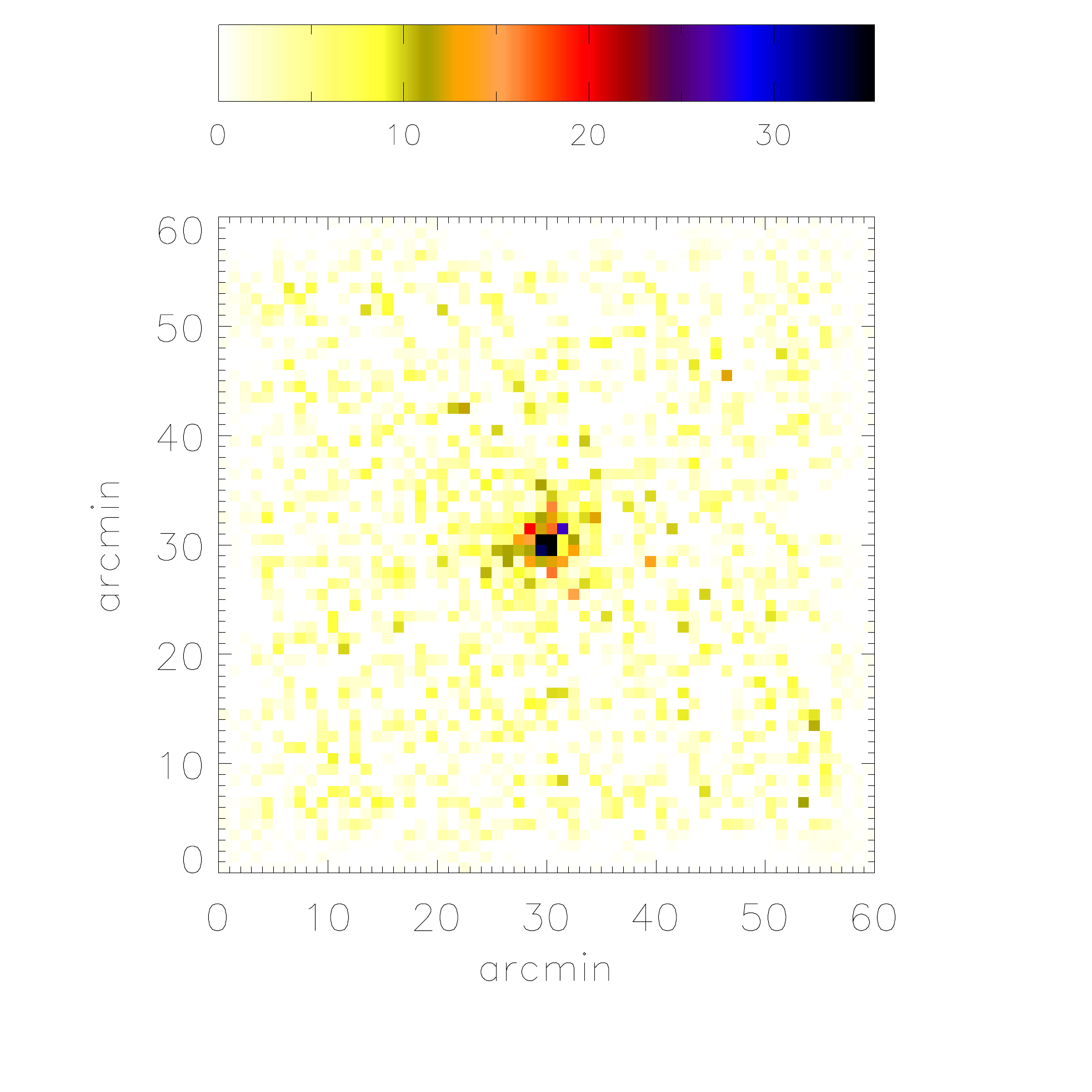}\includegraphics[width=0.24\textwidth]{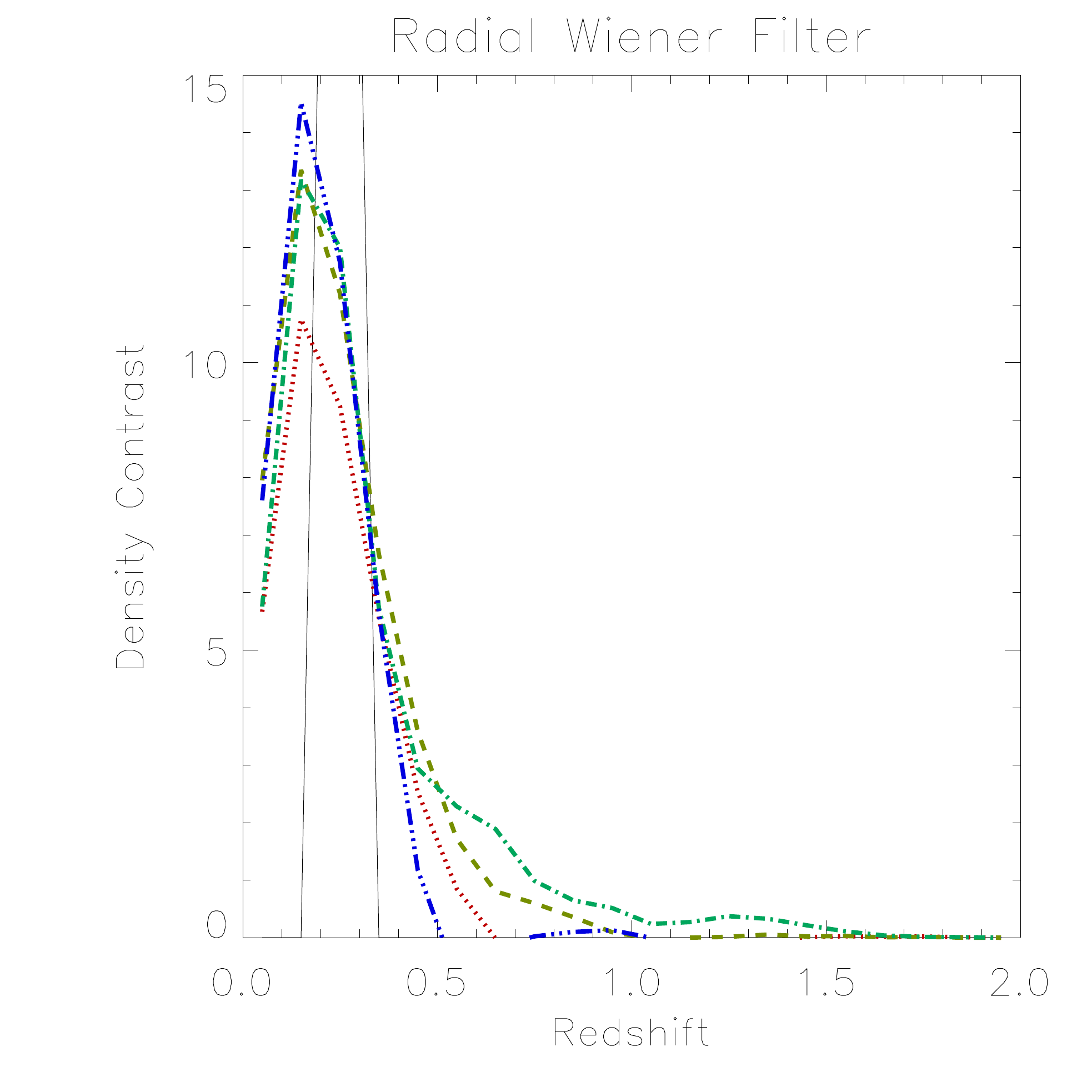}
  \includegraphics[width=0.24\textwidth]{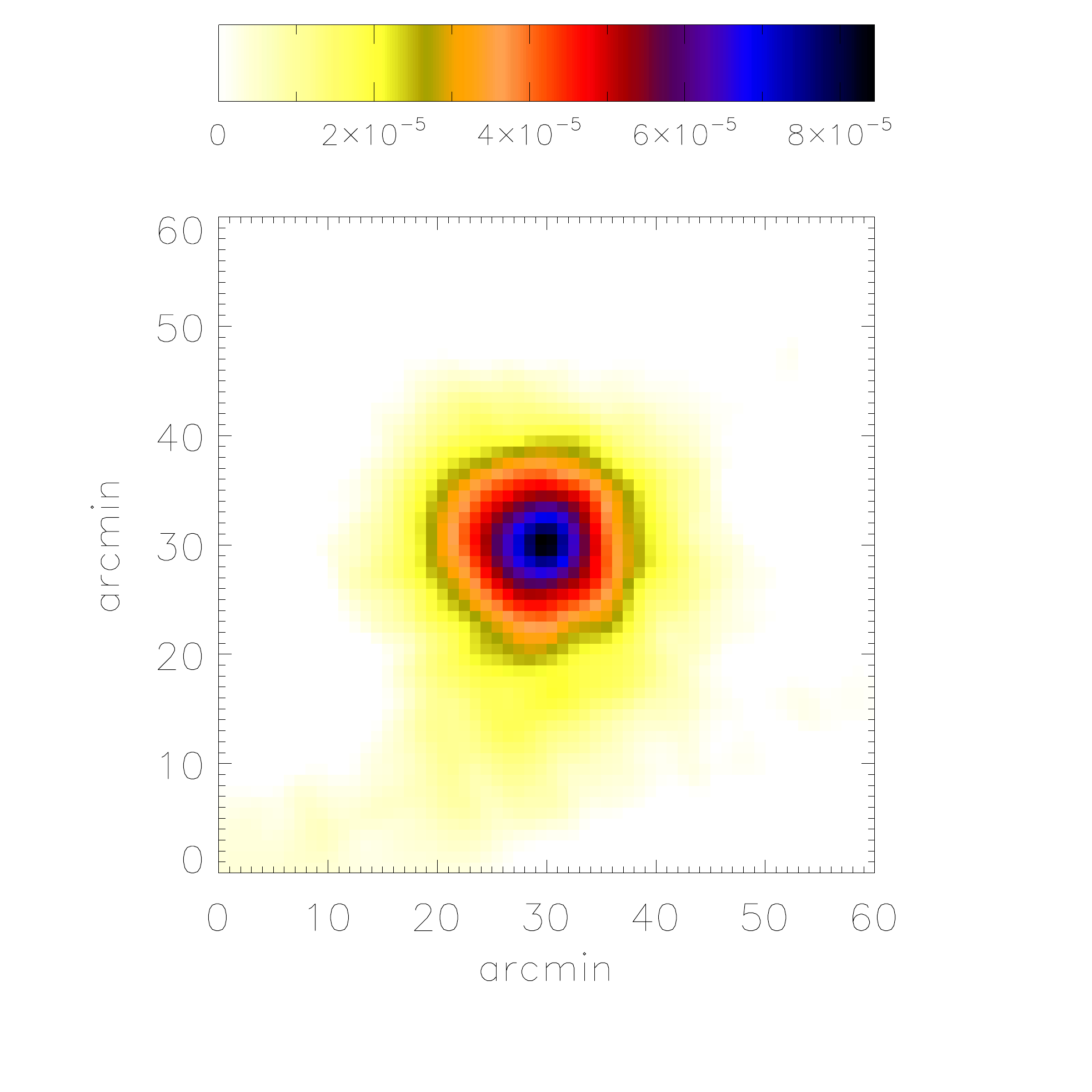}\includegraphics[width=0.24\textwidth]{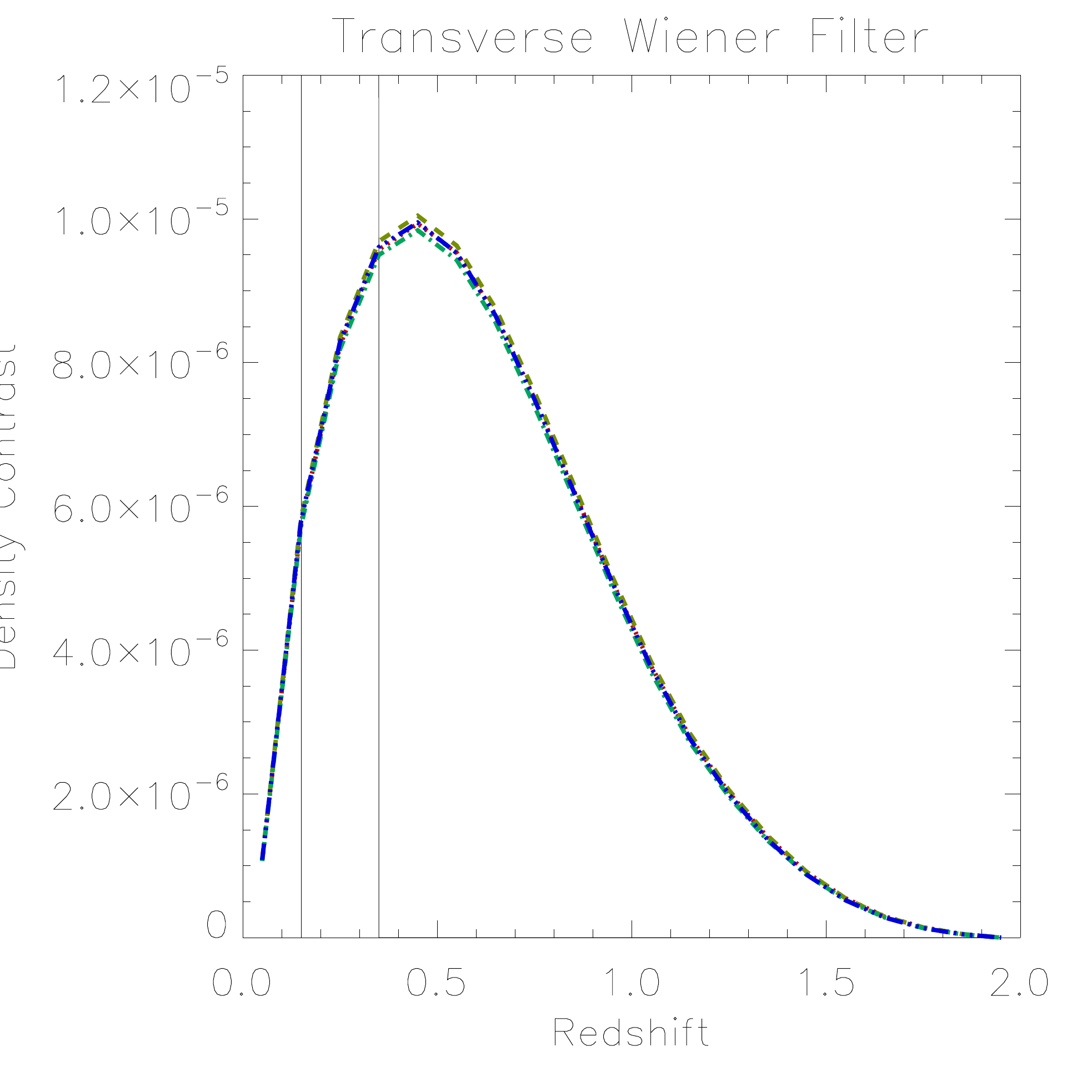}
  \caption{Comparison of our method with the linear methods as labelled. The left column shows the 2D projection of the reconstruction, while the right column shows the 1D reconstructions along the four central lines of sight (dashed lines). Note that, due to the amplitude damping effect in SVD and Wiener reconstructions, the y-axis scaling is different in each of the line of sight plots.
  \label{fg:complin1}}
\end{figure}

\begin{figure*}[hp]
  \centering
  \includegraphics[width=0.6\textwidth]{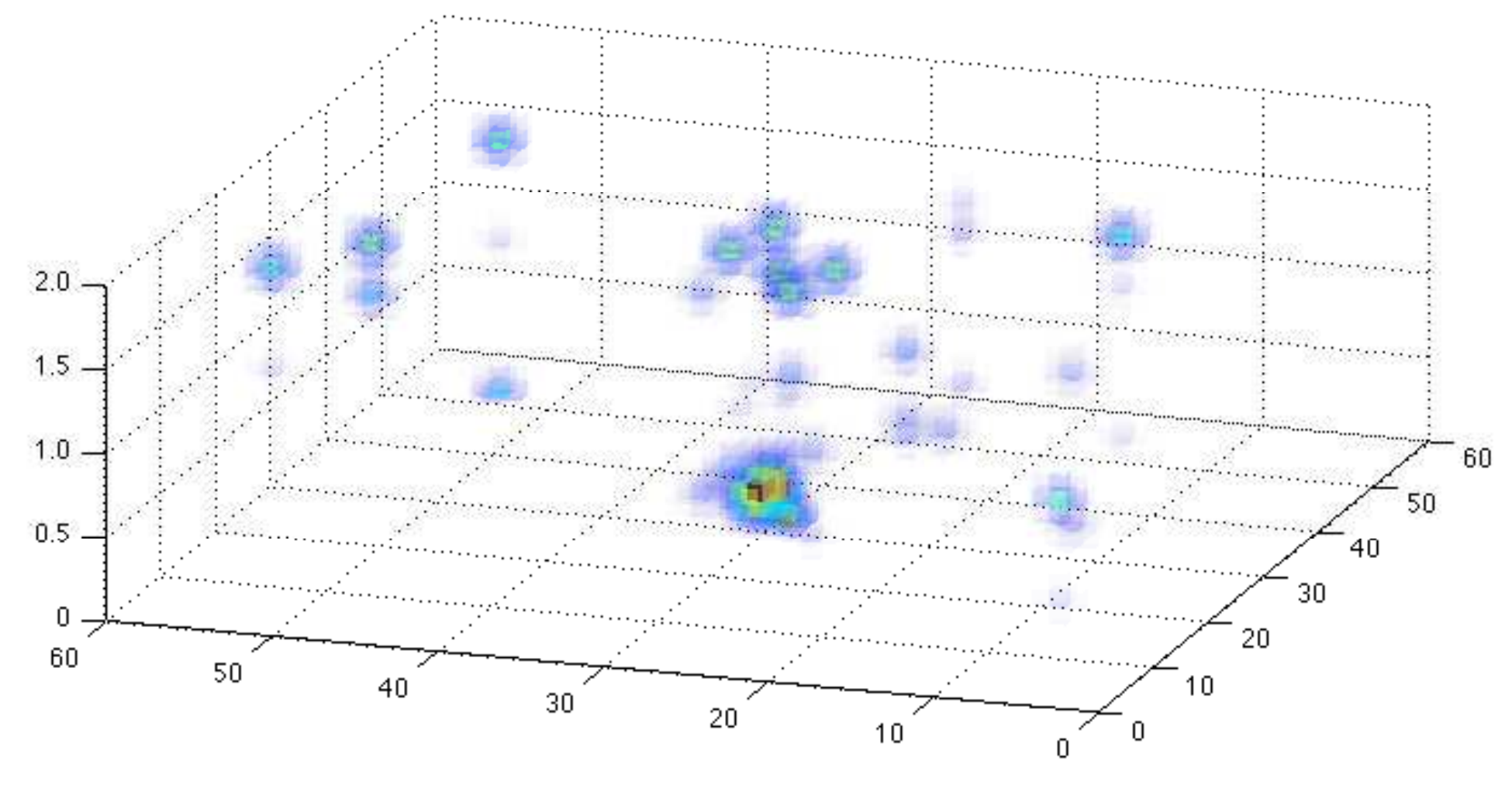} \\
  \includegraphics[width=0.6\textwidth]{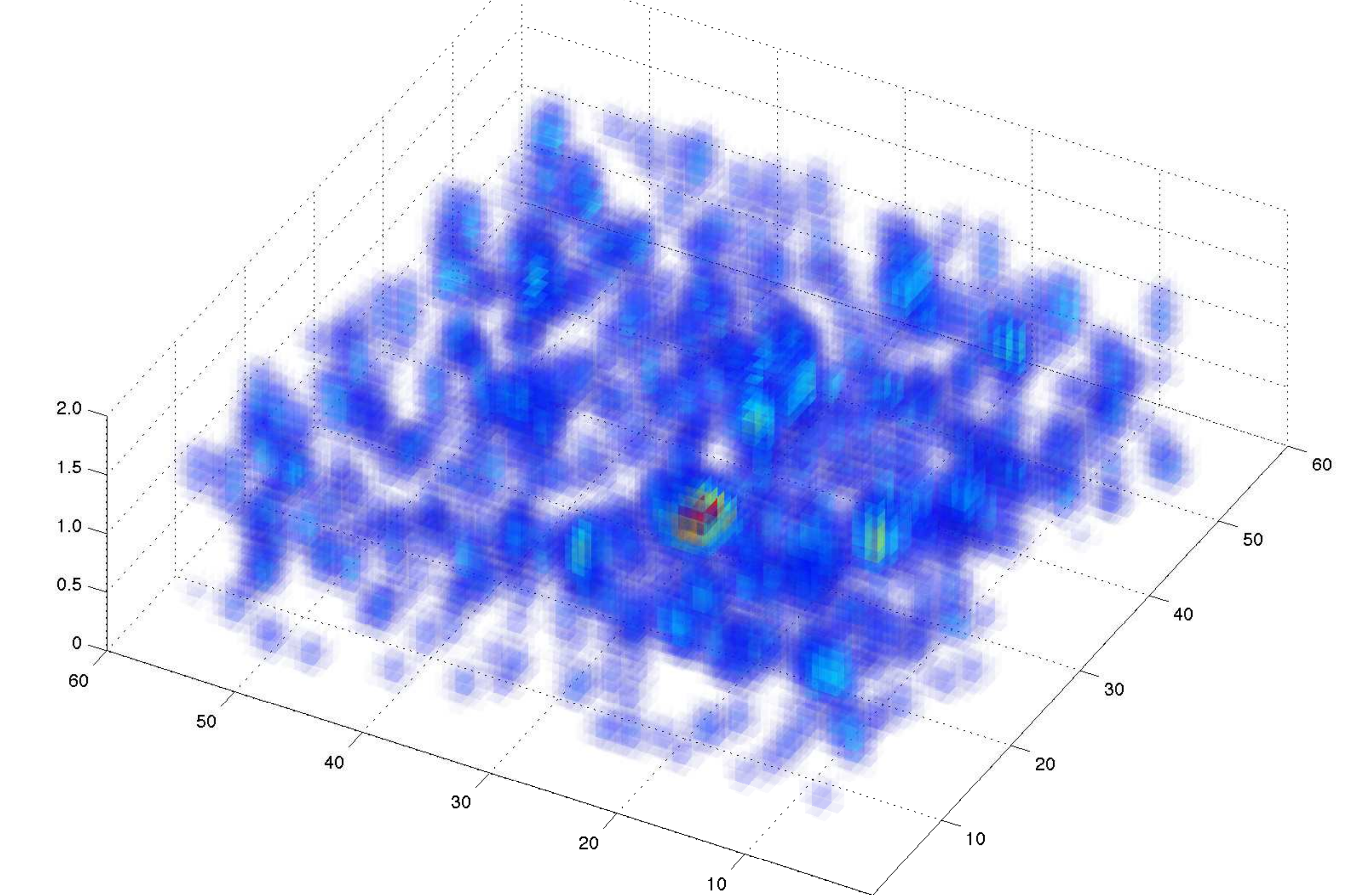} \\
  \includegraphics[width=0.6\textwidth]{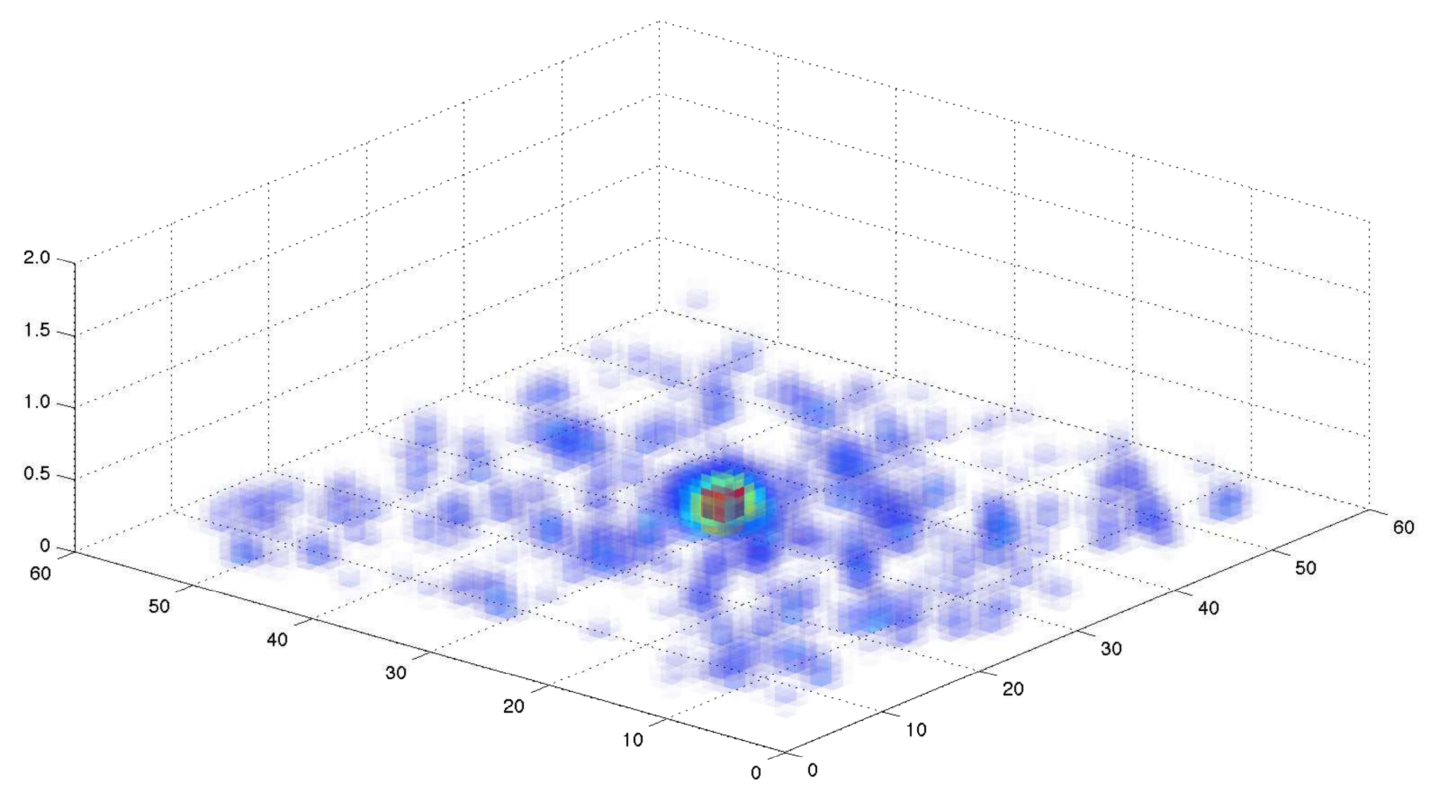} \\
  \includegraphics[width=0.6\textwidth]{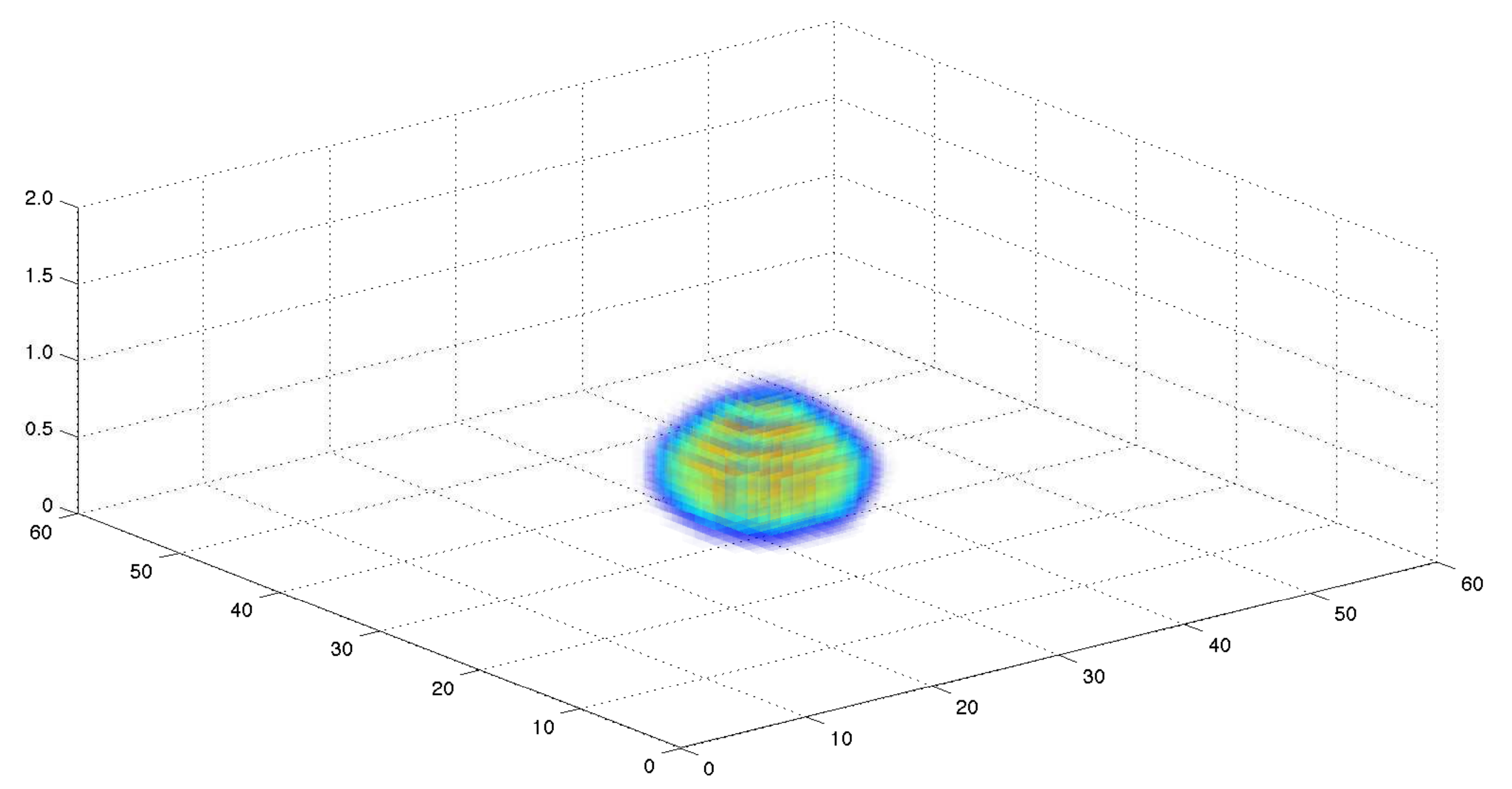} 
  \caption{3D rendering of the reconstruction of a $z_{\rm cl} = 0.25$ cluster using our method (top), the SVD method (2nd row,  $v_{cut} = 0.01$), radial Wiener filter method (3rd row, $\alpha = 0.05$) and the transverse Wiener filter method (bottom, $\alpha = 0.05$) as described in the text. \label{fg:complin2}}
\end{figure*}

The results are presented in figures \ref{fg:complin1} and \ref{fg:complin2}. Figure \ref{fg:complin1} presents the 2D projections of the reconstructions, computed by integrating the reconstruction along each line of sight, and the 1D reconstructions along the four central lines of sight. In the 3D renderings of Figure \ref{fg:complin2}, the reconstructions from our method, the SVD method and the radial Wiener filter method are thresholded at $\delta = 3$ (i.e. the plot only shows $\delta_{\rm rec} \ge 3$), and each is smoothed with a Gaussian of width $\sigma = 0.7\,$pix in all three directions. The reconstruction from the transverse Wiener filter method is heavily damped with respect to the amplitude of the density contrast; a threshold of $\delta = 5 \times 10^{-6}$ is chosen in this case, and no smoothing is applied as the reconstruction already shows a very smooth distribution. 

The SVD method appears, in the 1D plots, to identify the correct redshift of the cluster, with a small amount of line of sight smearing, but the plots show a prominent high-redshift peak along the line of sight. We note that it may be possible to remove this false detection by raising $v_{cut}$, but at the cost of increased line of sight smearing \citep[see][]{vanderplasetal11}. The two-dimensional projection consisting of the integrated signal along each line of sight is seen to be more noisy than our results. The three-dimensional rendering shows that the SVD method does well at identifying and localising the cluster, but the resulting reconstruction does suffer from widespread noise at a moderate level. 

The radial and tangential Wiener filter methods show very little noise in the 2D projections, and the radial Wiener filter shows very little smearing of the reconstruction along the line of sight. However, neither Wiener method recovers correctly the redshift of the cluster. While the transverse Wiener filter shows very little noise, it exhibits a broad smearing in both the radial and transverse directions, and a heavy damping of the amplitude of the reconstruction. The results obtained using the radial Wiener filter are considerably better, with very little noise seen in the reconstruction, and with the cluster seen to be well-localised in the radial direction. However, the amplitude of the cluster reconstruction is a factor of $\sim 2.5$ smaller than the input density.

Our results are seen to suffer from several prominent, pixel-scale false detections along noisy lines of sight not associated with the cluster. However, along the four central lines of sight an excellent correlation between the input density contrast and our reconstruction is seen. One line of sight exhibits a prominent high-redshift false detection; however this does not appear in the remaining three lines of sight and the overall amplitude of the reconstruction is around $75\%$ of the true value. The three-dimensional rendering demonstrates that the noise in our reconstruction shows very little coherent structure (i.e. tends to be restricted to isolated pixels), and is largely low-level. Moreover, the cluster is incredibly well-localised in redshift space, with the smearing seen in the figure primarily arising from the applied smoothing.

\subsubsection*{Reconstruction with Noisier Data}

The simulations described in \S~\ref{subsec:sims} represent a rather optimistic set of data. Realistically, one might expect a much smaller number density of galaxies, so it is important to consider how our method fares with noisier data. To this end, weak lensing data were simulated for the same cluster as described above, but with $n_g = 30$ galaxies per square arcminute. This represents a factor of $\sim1.8$ reduction in the signal to noise of our data. In order to account for this reduction in signal to noise, we increased our soft threshold parameter $\lambda$ by a similar factor, taking $\lambda = 15$, whist we again used $\epsilon = 3$. 

\begin{figure}[t]
  \includegraphics[width=0.24\textwidth]{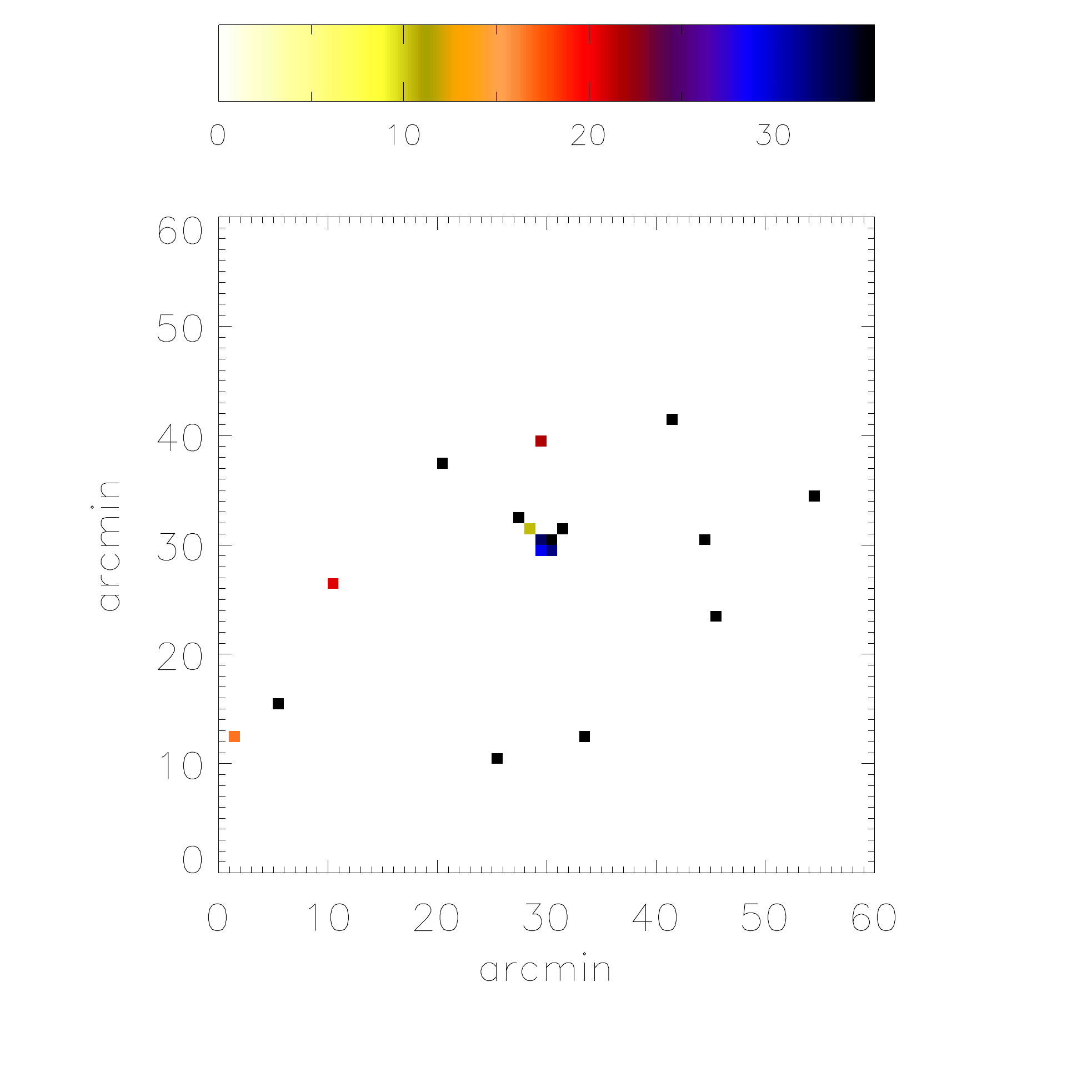}\includegraphics[width=0.24\textwidth]{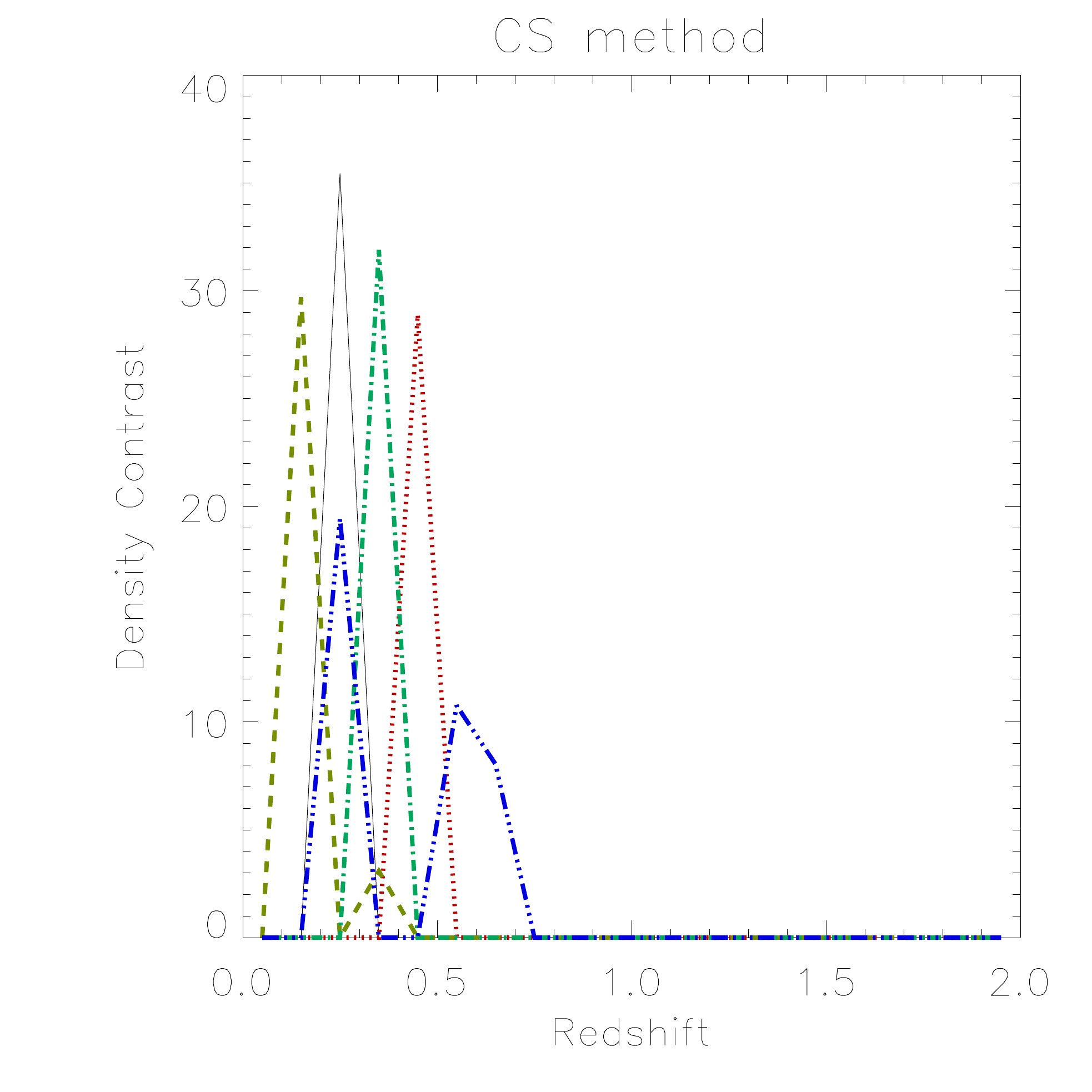}
  \includegraphics[width = 0.5\textwidth]{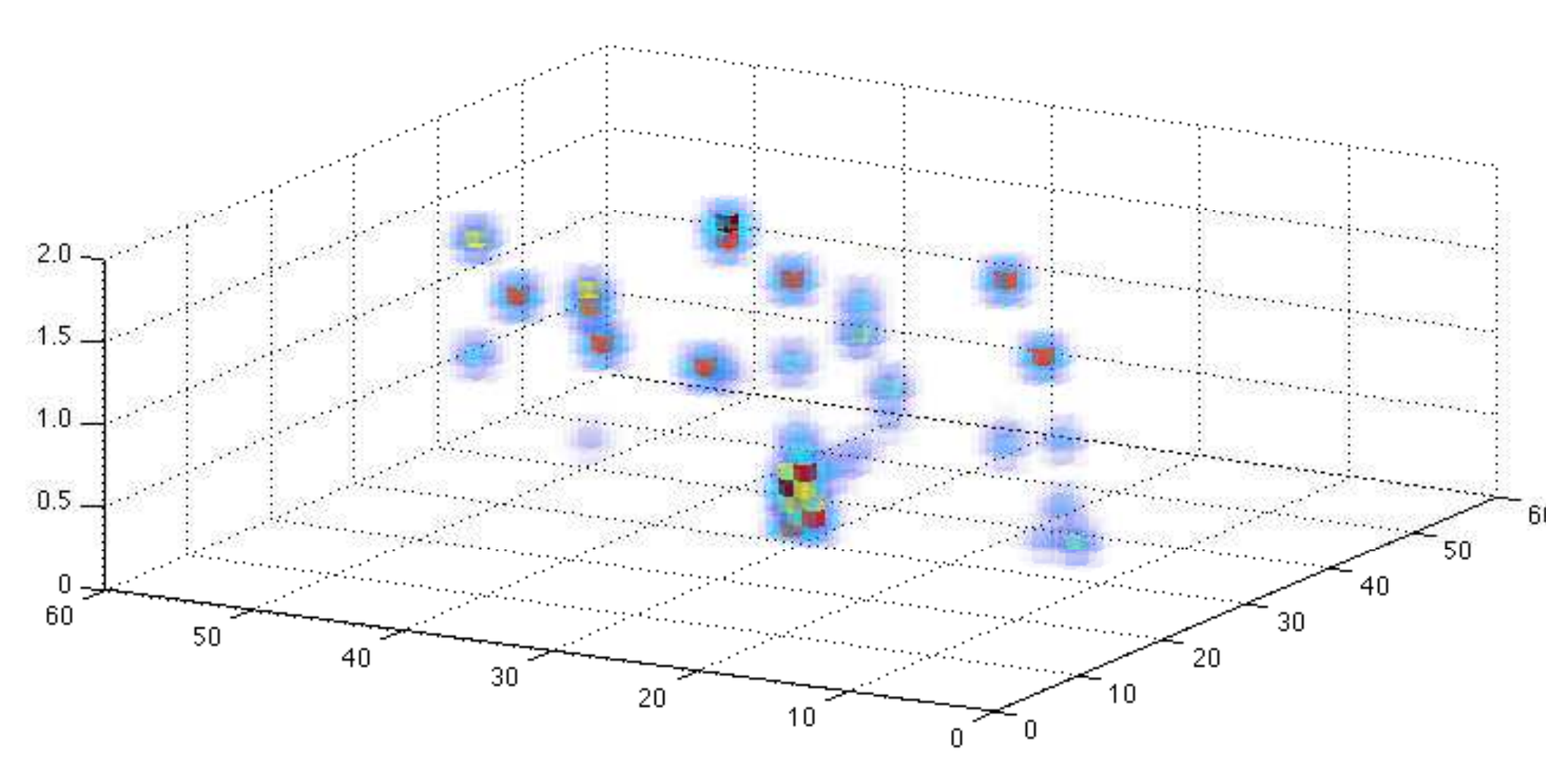}
\caption{Reconstructions of a cluster at reshift $z=0.25$ using noisy data with $n_g = 30$ galaxies per square arcminute. Shown above are the 2D projection of the reconstruction (top left), 1D line of sight plots for the four central lines of sight (top right) and smoothed 3D rendering of the reconstruction, as before.\label{fg:recn30}}
\end{figure}

The projections and 3D reconstruction obtained using this noisier data is shown in Figure \ref{fg:recn30}. The 2D projection shows a noisier reconstruction, with more false detections, and the central cluster appears less extended. This is to be expected, given the lower signal to noise in the data. The line of sight plots again show more noise, with the peaks of some lines of sight being shifted slightly from the true redshift. However, they are well localised around the true peak, and their amplitudes match quite well with the true underlying density.

The 3D rendering again shows a well-localised peak at the location of the cluster, albeit slightly extended along the line of sight. Note also that the false detections continue to appear as single pixel-scale detections, rather than as extended objects. As before, we believe that a fully three-dimensional implementation of our algorithm will reduce the number of such false detections by searching for coherent structures of larger angular extent than a single pixel.

While improvements can be made to the noise model used, it is clear that, with appropriate choice of parameters, effective noise control can be obtained with noisy data, and a very clean reconstruction obtained using our method. Here again, our method is seen to improve on the bias, smearing and amplitude damping problems seen in the linear methods presented above, and the results at this noise level are of a comparable or better standard than the results from linear methods at higher signal to noise, as presented above.

\subsection{Improving the Line of Sight Resolution}

Given the success of our method at reconstructing lines of sight at the same resolution as the input data, it is interesting to consider whether we are able to improve on the output resolution of our reconstructions. It is also worthwhile to test our ability to detect clusters at higher redshifts than that considered above, given that compressed sensing is specifically designed to tackle underdetermined inverse problems. Indeed, noise-free simulations suggest that a resolution improvement of up to a factor of 4 in the redshift direction may be possible with this method.

Therefore, we generate clusters as before, with our data binned into $N_{\rm sp} = 20$ redshift bins, but aim to reconstruct onto $N_{\rm lp} = 25$ redshift bins. We further consider clusters at redshifts of $z_{\rm cl} = 0.2,\ 0.6,\ \mbox{and}\ 1.0$. Given the changed reconstruction parameters, we modify our noise control parameters slightly and, for all the results which follow, take $\lambda = 7.5,\ \epsilon = 2.8$. 

\begin{figure}[h]
  \includegraphics[width=0.25\textwidth]{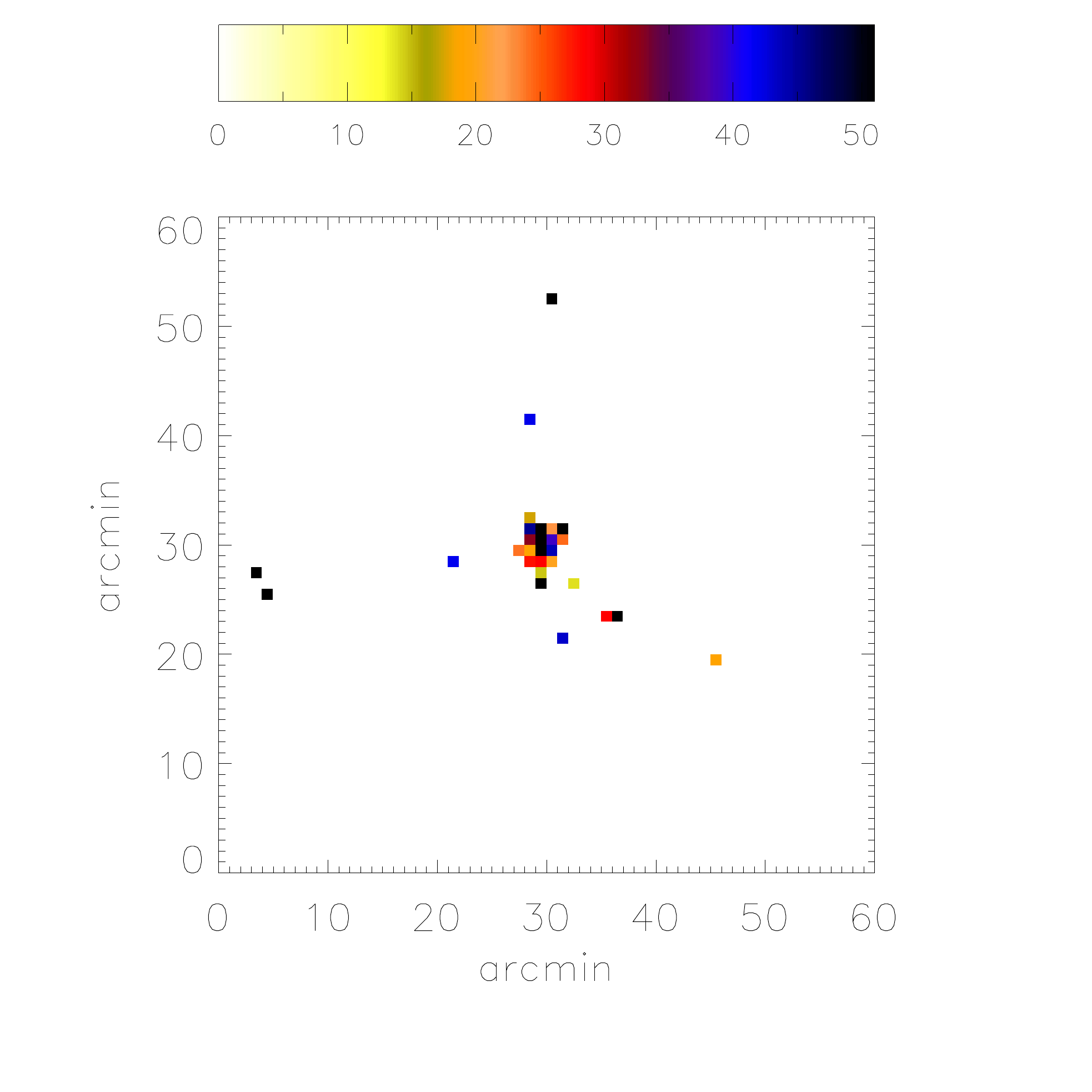}\includegraphics[width=0.25\textwidth]{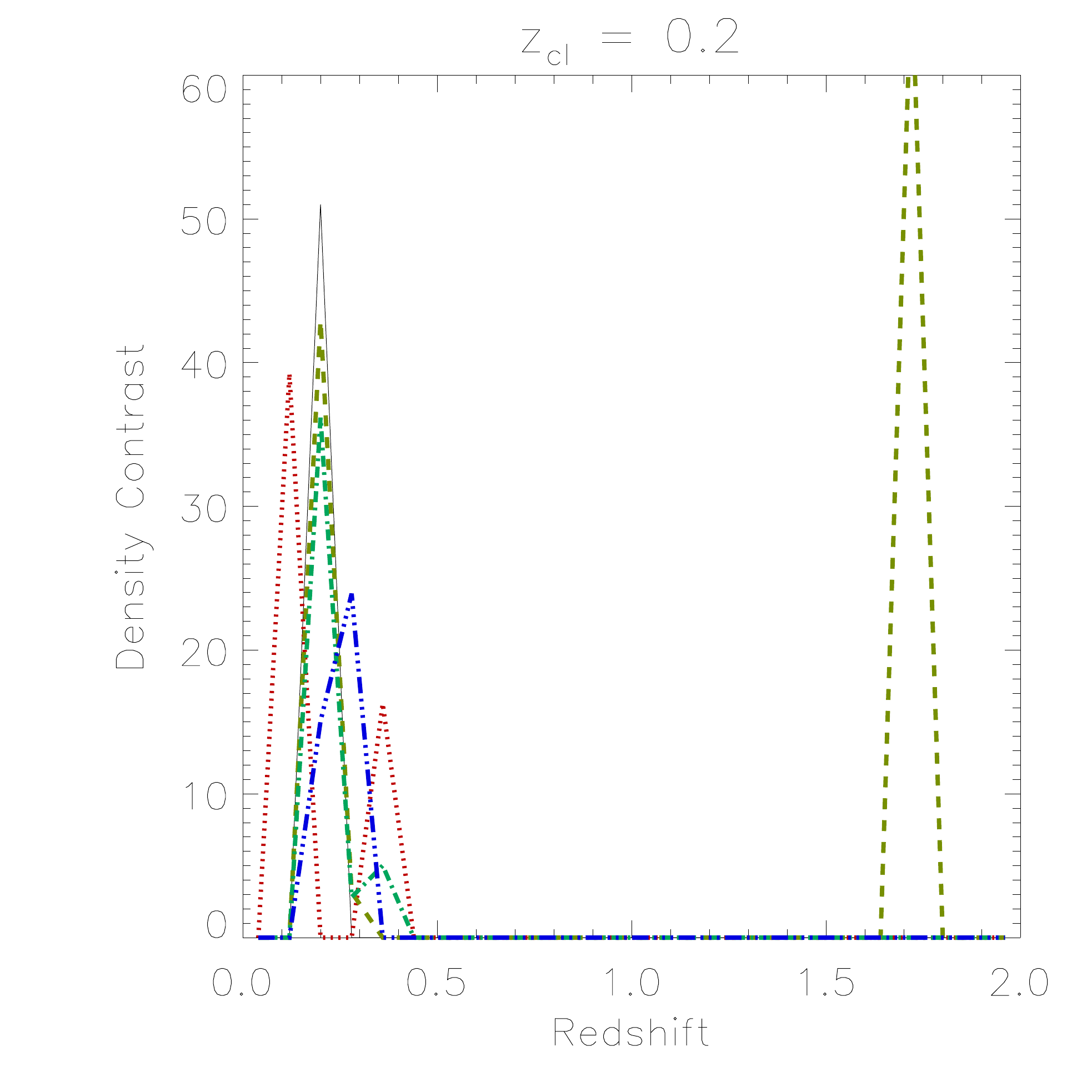}
  \includegraphics[width=0.25\textwidth]{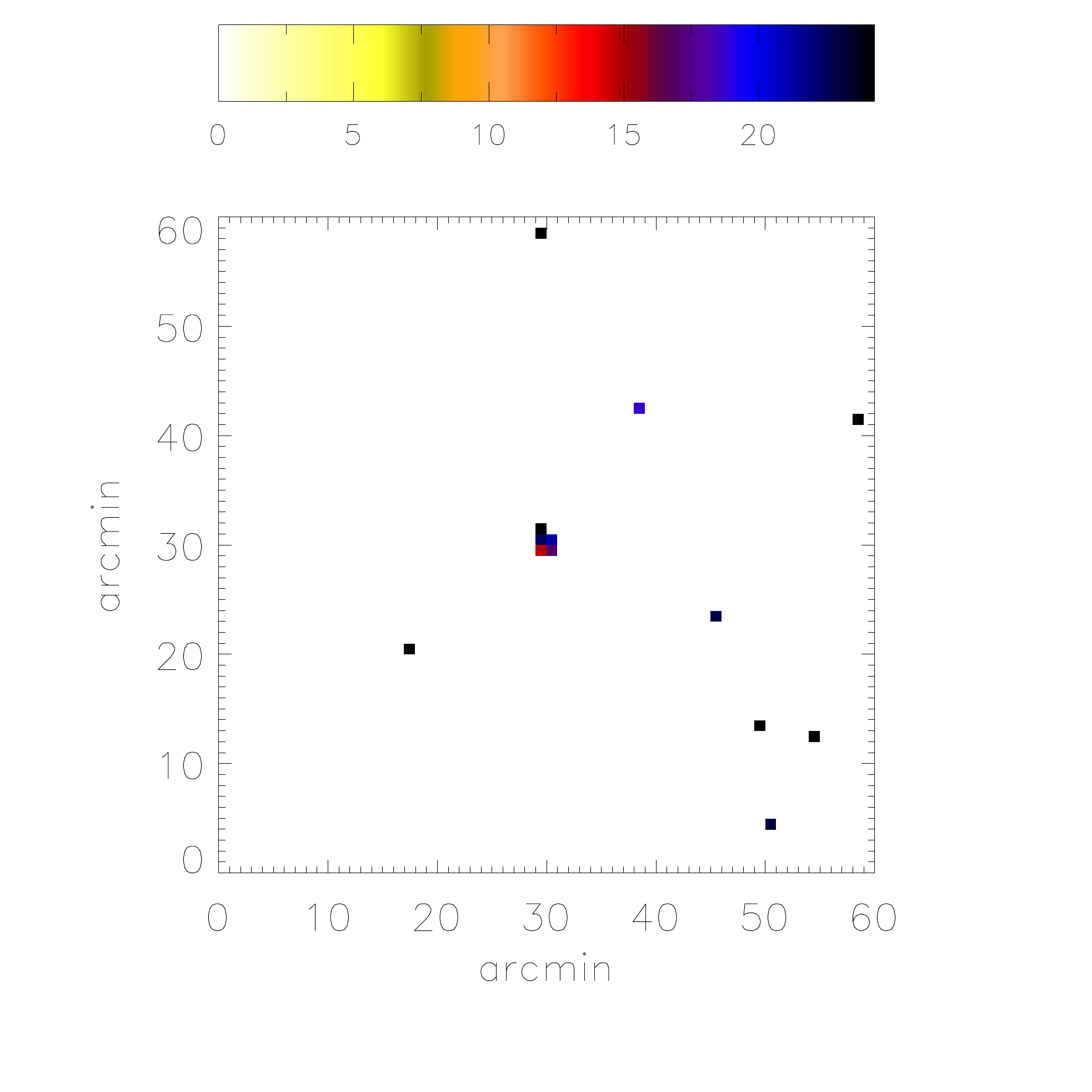}\includegraphics[width=0.25\textwidth]{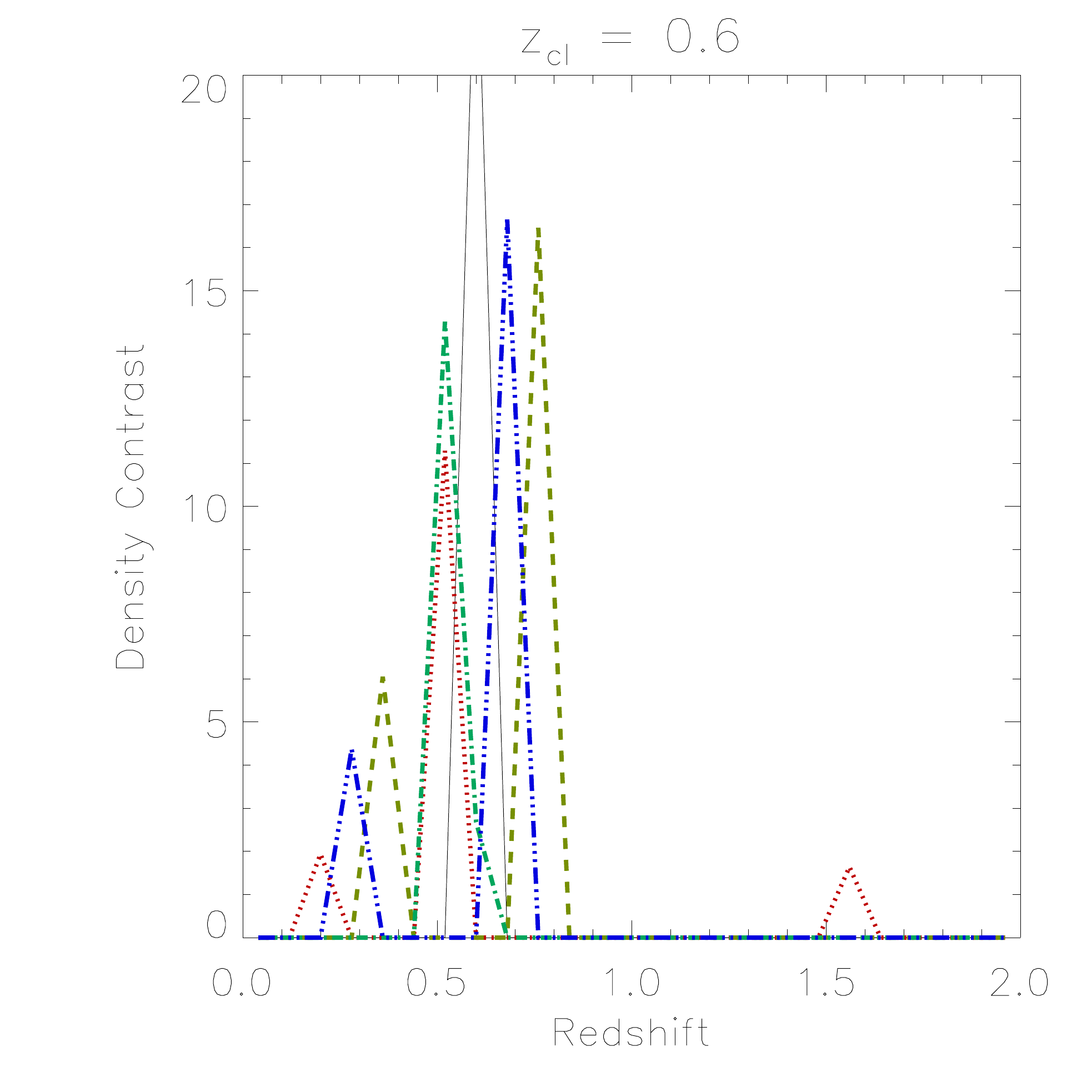}
  \includegraphics[width=0.25\textwidth]{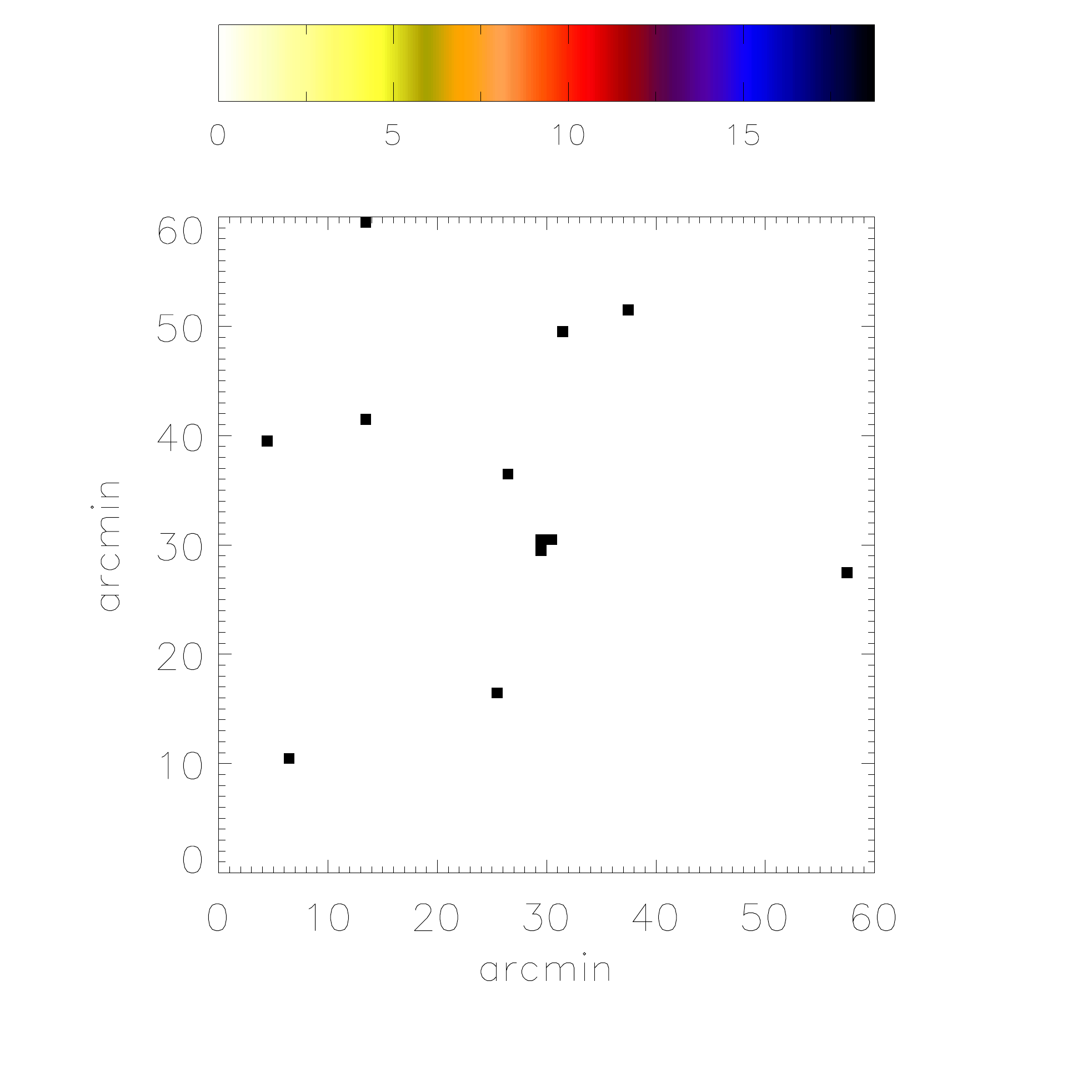}\includegraphics[width=0.25\textwidth]{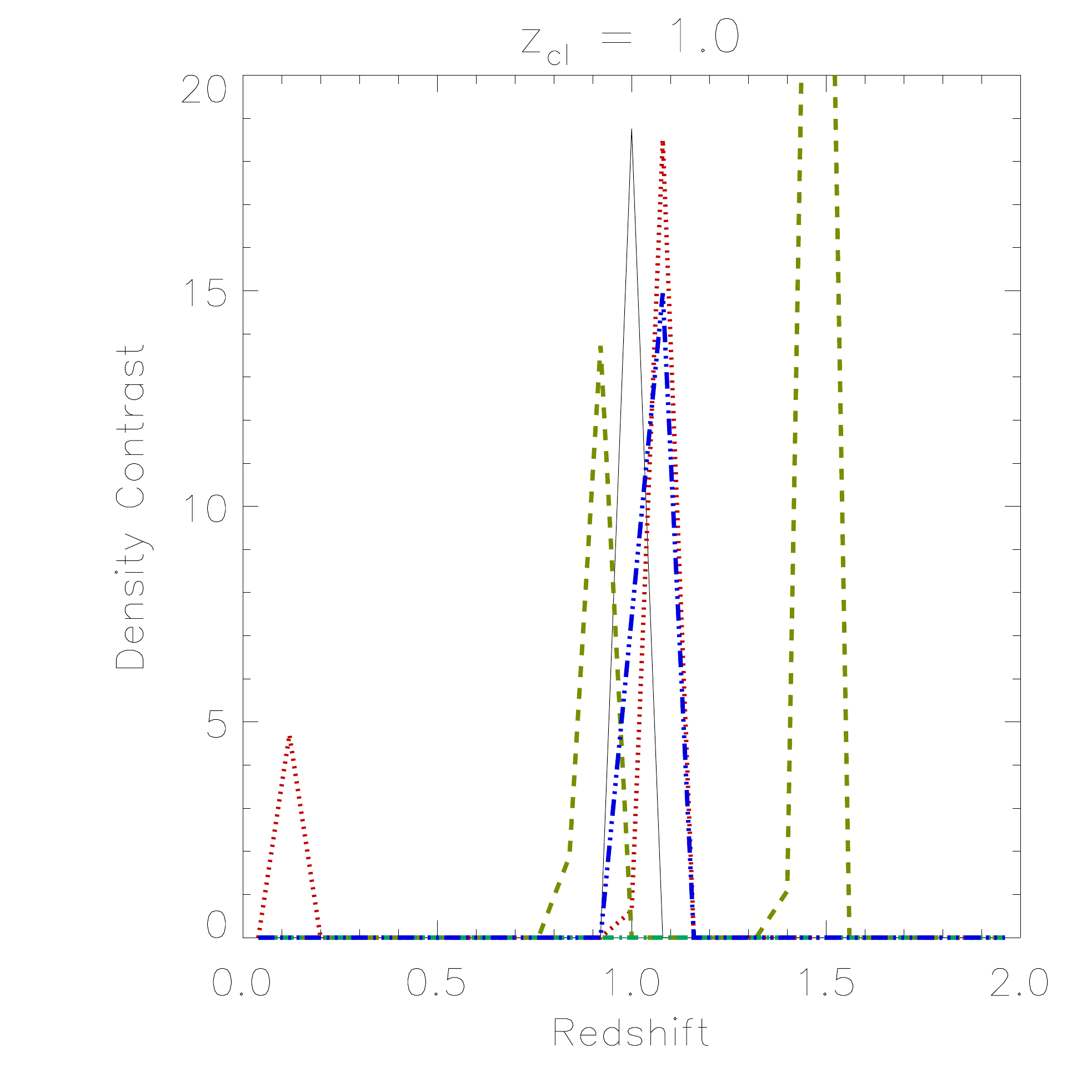}
  \caption{Reconstructions of single clusters located at a redshift of $z_{\rm cl}=0.2$ (top row), $z_{\rm cl} = 0.6$ (middle row) and $z_{\rm cl}=1.0$ (bottom row). As before, the left column shows the two dimensional integrated projection of the reconstruction, while the right panel shows the input density contrast along the line of sight (solid line) and the 1D reconstruction along each of the four central lines of sight (dashed lines). \label{fg:single_haloes1}}
\end{figure}

\begin{figure*}[hp]
  \centering
  \includegraphics[width=0.6\textwidth]{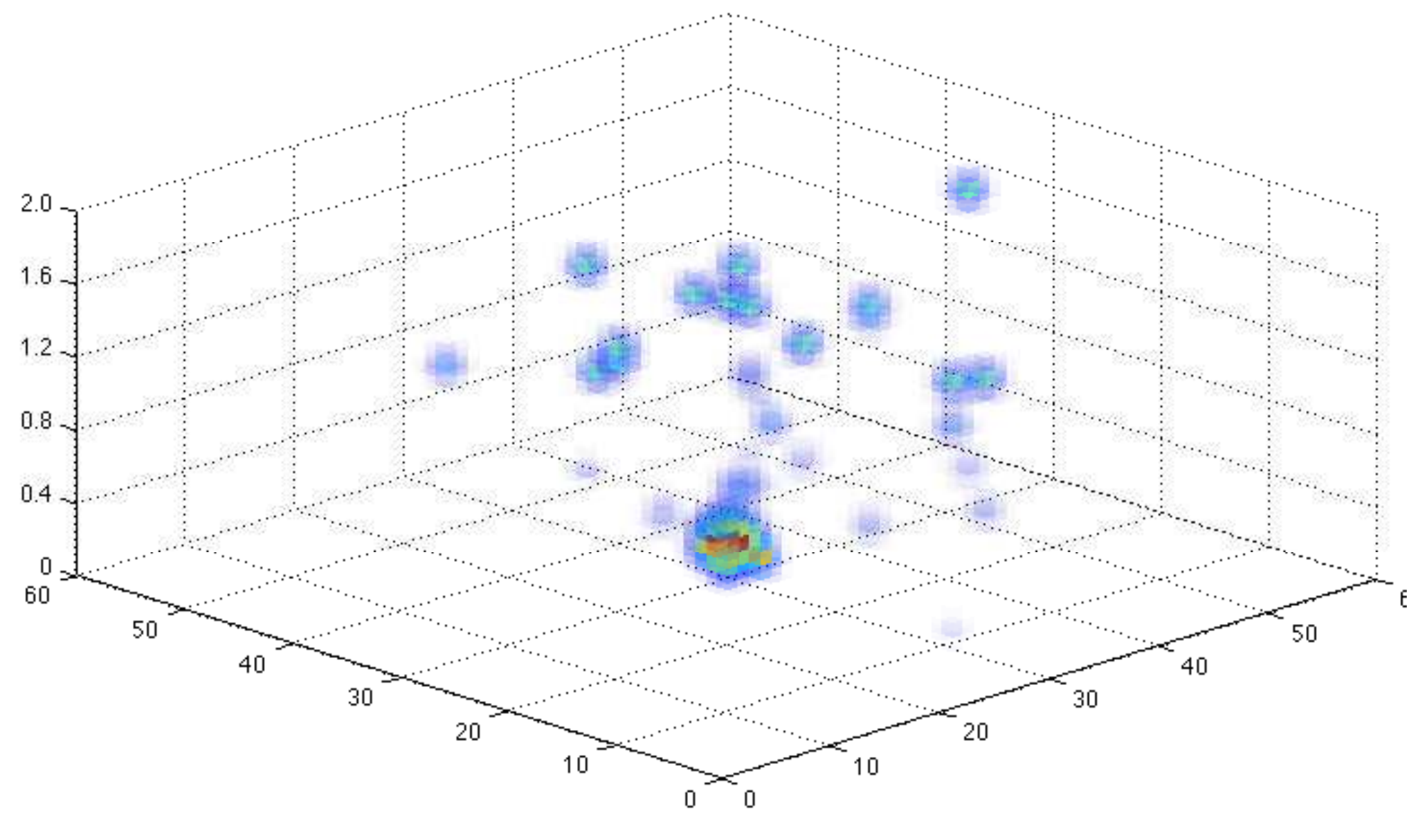}
  \includegraphics[width=0.6\textwidth]{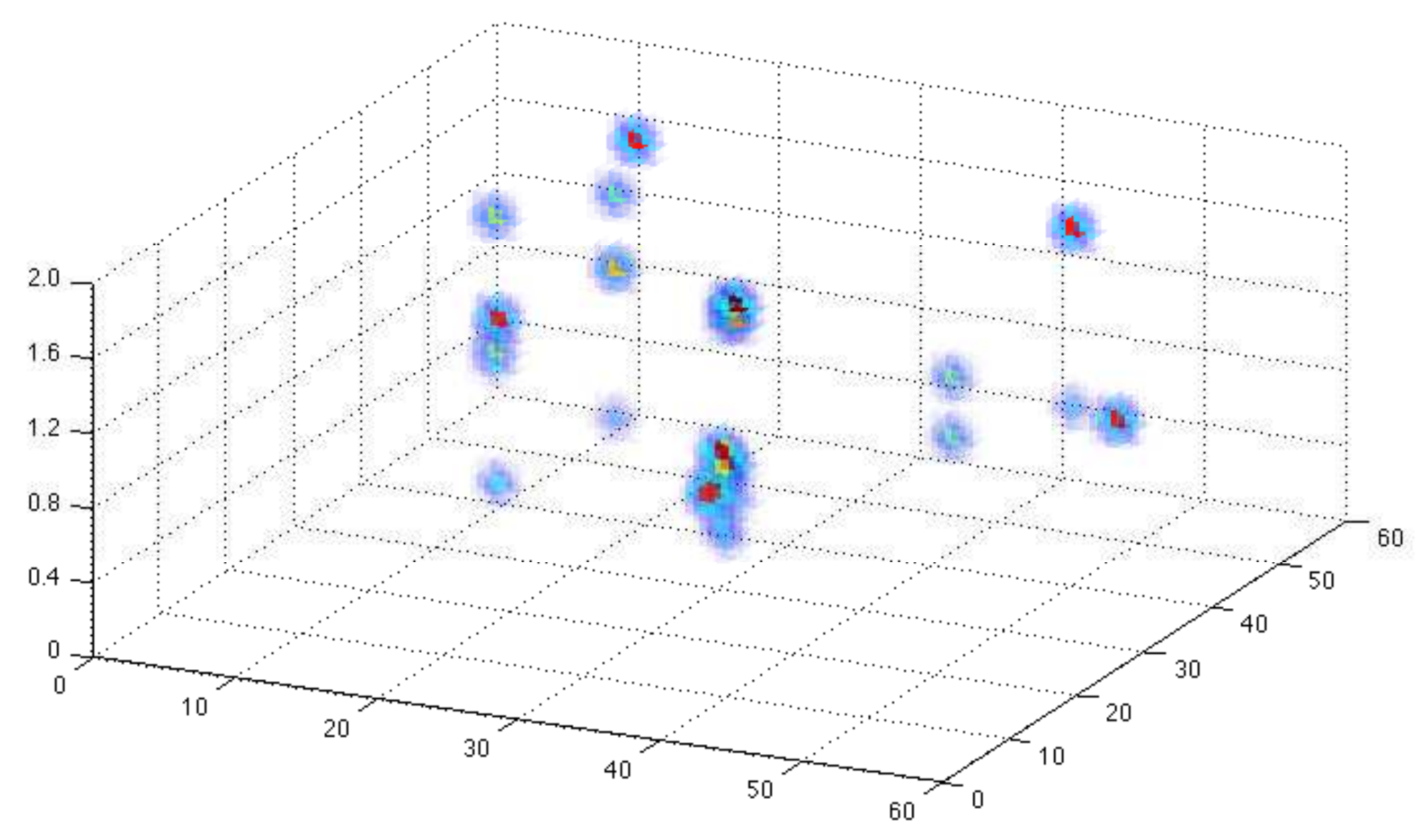}
  \includegraphics[width=0.6\textwidth]{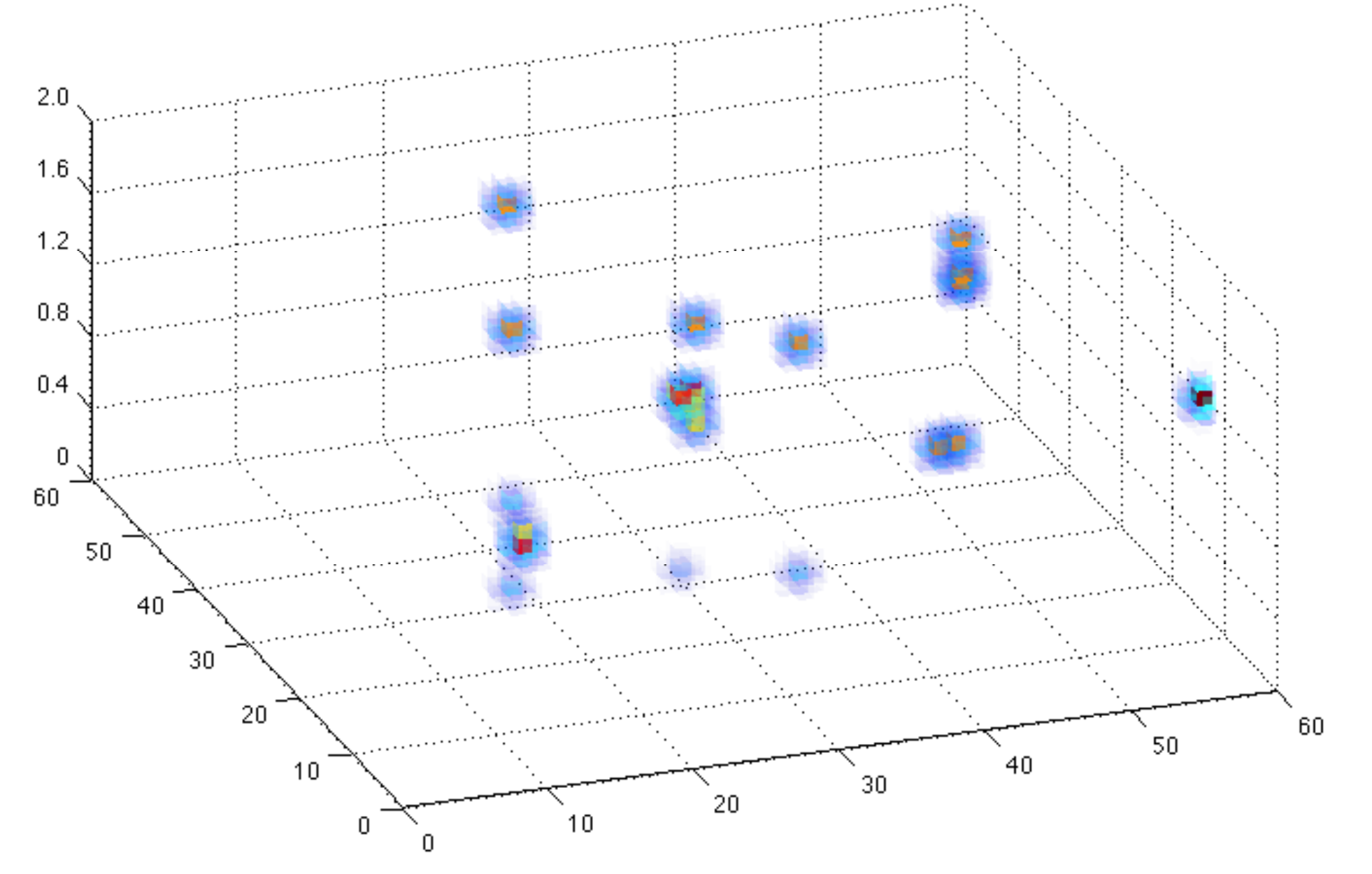}
  \caption{Reconstructions of single clusters located at a redshift of $z_{\rm cl}=0.2$ (top), $z_{\rm cl} = 0.6$ (middle) and $z_{\rm cl}=1.0$ (bottom). The reconstructions are thresholded at $\delta = 3$ and smoothed with a Gaussian of width $\sigma = 0.7\,$pix. \label{fg:single_haloes2}}
\end{figure*}

Figure \ref{fg:single_haloes1} shows the two-dimensional projection of the reconstruction and the 1D reconstruction of the four central lines of sight as before. Figure \ref{fg:single_haloes2} shows the reconstructions of these haloes using our method as a three-dimensional rendering. The three-dimensional rendering is, as before, thresholded at $\delta_{\rm rec} = 3$ and smoothed with a Gaussian of width $\sigma = 0.7\,$pix.

Several features are immediately apparent. Firstly, all three clusters are clearly identified by our method. This is particularly impressive in the case of the $z_{\rm cl}=1.0$ cluster, as linear methods have, thus far, been unable to reconstruct clusters at such a high redshift \citep[see, e.g.][]{sth09, vanderplasetal11}. We note that such a detection is dependent on the redshift distributions of sources, however, and the lack of detection in \cite{sth09} and \cite{vanderplasetal11} may be due, in part, to their choice of probability distribution. However, we also note that the background source density in our sample is highly diminished behind the $z_{\rm cl}=1.0$ cluster ($32.5\,$arcmin$^{-2}$).

Again, the three-dimensional renderings indicate that there is very little smearing of the reconstruction along the line of sight, in contrast with linear methods. This is further evidenced by the line-of-sight plots, in which the unsmoothed reconstructions show very localised structure. Furthermore, the reconstructions exhibit minimal redshift bias, and some lines of sight are seen to recover the amplitude of the density contrast without any notable damping.

However, again we see several prominent ``hot pixels" or false detections along noisy lines of sight. Such detections are more evident as the cluster moves to high redshift, and may be significantly larger than the expected density contrast of the cluster. This is expected, as the number density of sources behind the lens diminishes.

We note that these false detections often manifest at high redshift, arising out of the overfitting effect seen in Figure \ref{fg:epsilon}. We also note that the false detections are very well localised in both angular and redshift space, and do not form coherent large structures, making them easily identifiable as false detections; they tend to be localised to isolated pixels. Because of this lack of coherence, we expect that a fully three-dimensional implementation would suppress many of these false detections by aiming to detect coherent structure in three dimensions, and thereby seeking structures of larger extent than a single pixel.

\subsection{More Complex Line-of-Sight Structure}

\begin{figure*}[htp]
  \centering
  \includegraphics[width=0.7\textwidth]{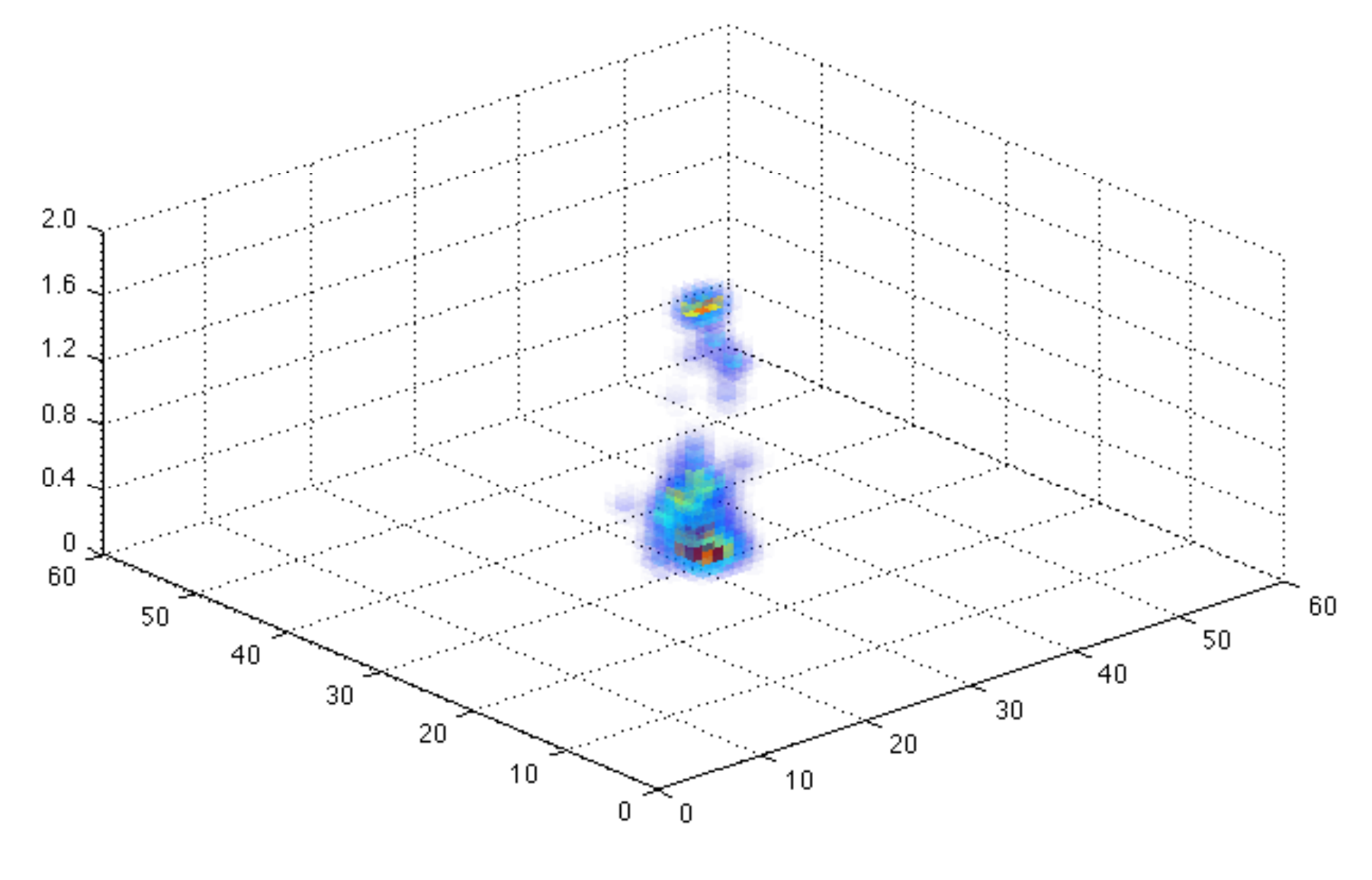}
  \includegraphics[width=0.75\textwidth]{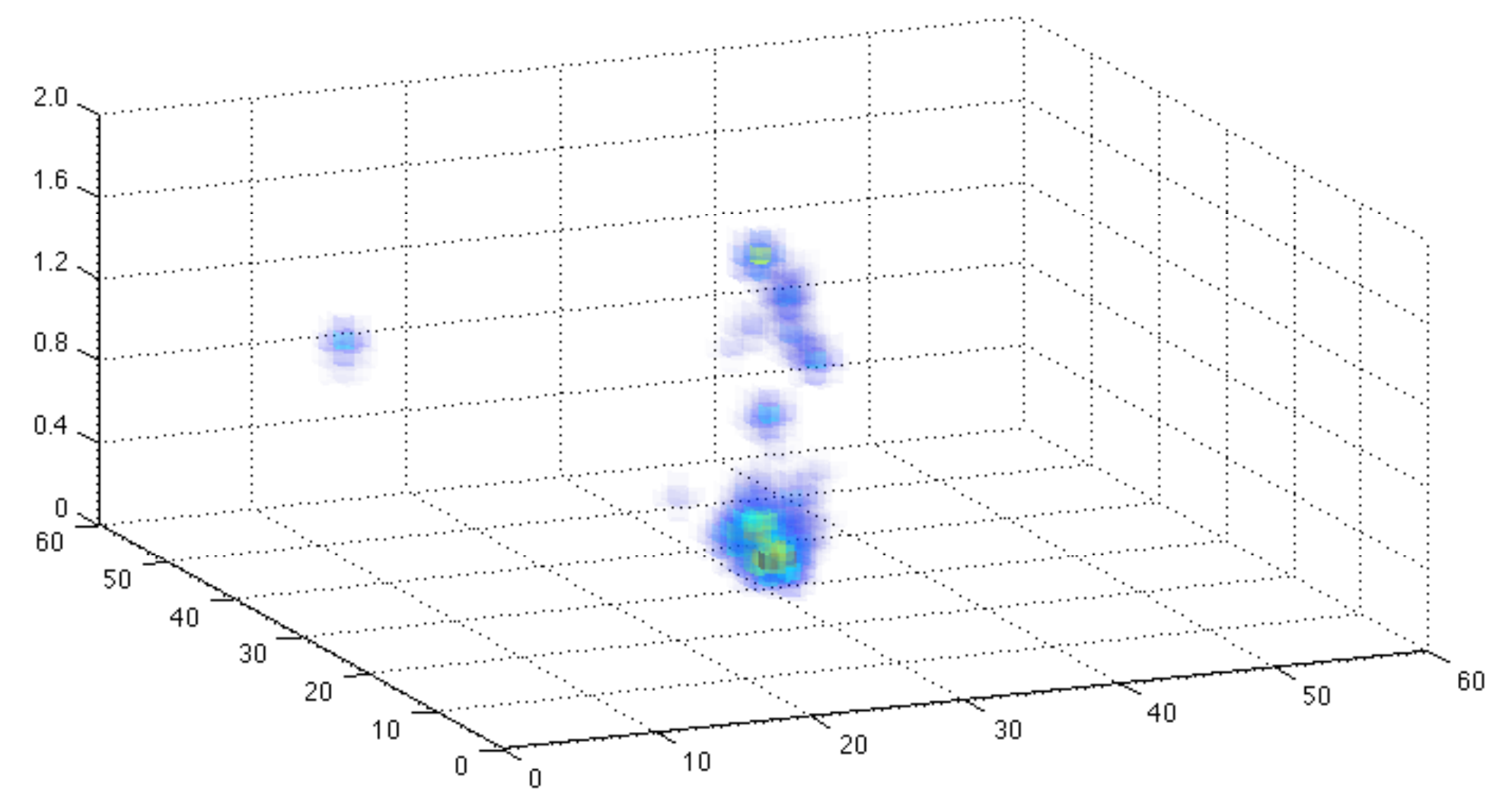}
  \includegraphics[width=0.75\textwidth]{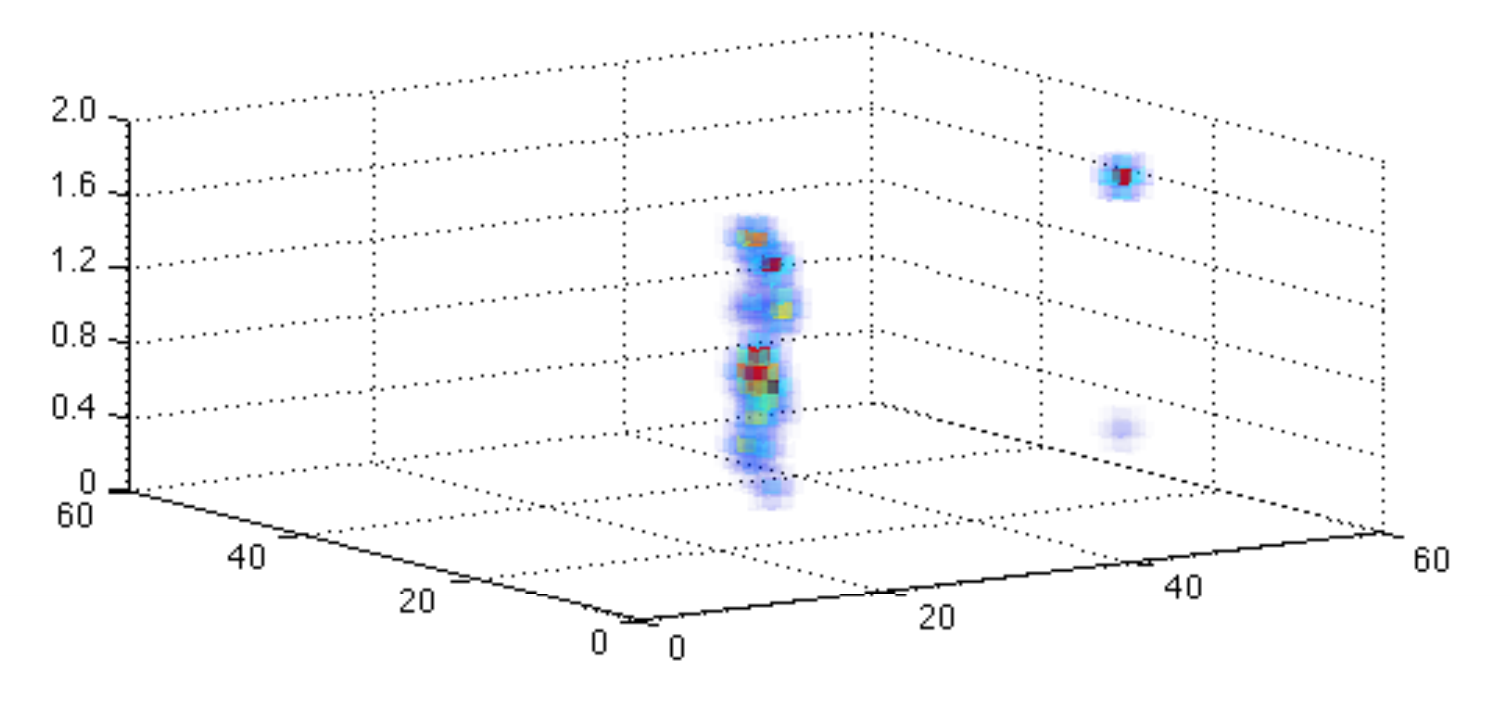}
  \caption{Reconstructions of two clusters along the line of sight, located at a redshift of $z_{\rm cl}=[0.2, 0.6]$ (top), $z_{\rm cl} = [0.2, 1.0]$ (middle) and $z_{\rm cl}=[0.6, 1.0]$ (bottom). The reconstructions are thresholded at $\delta = 3$ and smoothed with a Gaussian of width $\sigma = 0.7\,$pix. \label{fg:doubclust2}}
\end{figure*}

Given the ability of our method to localise structure in redshift space, it is interesting to consider whether the algorithm is able to disentangle the lensing signal from two clusters located along the same line of sight. We consider three different cluster pairings -- $z_{\rm cl} = [0.2, 0.6]$, $z_{\rm cl} = [0.2, 1.0]$, and $z_{\rm cl} = [0.6, 1.0]$ -- and the results obtained by our method are shown in three-dimensional rendering in Figure \ref{fg:doubclust2}. We find that the reconstructions of individual lines of sight in this case is significantly more noisy than before. 

The figure shows the $z_{\rm cl} = [0.2, 0.6]$ cluster pairing to be reconstructed as a single, coherent structure smeared out between the two redshifts. In the other two cases, two distinct clusters are observed, but their redshifts are slightly biased, and a moderate amount of smearing in the redshift direction is seen. This smearing arises from the fact that individual lines of sight detect the structures at different redshifts; the aggregate effect is an elongation along the line of sight.

While these results are by no means as clean as the results obtained for single clusters, it is promising to note that we are able to detect the presence of more complex line of sight structure, despite the reconstruction of individual lines of sight being fairly noisy. It seems clear that in this case, a fully three-dimensional treatment of the data which takes into account the correlations between neighbouring lines of sight and which seeks to reconstruct coherent structures would offer improvements in this area.

\section{Summary and Conclusions}
\label{sec:discuss}

Current approaches to 3D weak lensing involve linear inversion, where a pseudo-inverse operator is constructed incorporating prior constraints on the statistical distribution of the measurement noise and the underlying density. These methods are straightforward to construct and implement, make use of common tools, and are usually fairly fast. This makes them a convenient choice when approaching the 3D weak lensing problem. 

However, reconstructions obtained in this way suffer from line-of-sight smearing, bias in the detected redshift of structures, and a damping of the reconstruction amplitude relative to the input. It has further been noted \citep{vanderplasetal11} that the reconstructions obtained using these techniques may be fundamentally limited regarding the resolution attainable along the line of sight, due to smearing effects resulting from these linear methods. In addition, such methods are unable to treat an underdetermined inversion, and therefore are limited in their output resolution by the resolution of the input data which, in turn, is limited by the measurement noise.

We have presented a new approach to 3D weak lensing reconstructions by considering the weak lensing problem to be an instance of compressed sensing, where the underlying structure we aim to reconstruct is sparsely represented in an appropriate dictionary. Under such a framework, we are able to consider underdetermined transformations, thereby relaxing the constraints on the resolution of the reconstruction obtained using our method.

We have reduced the problem to that of one-dimensional reconstructions along the line of sight. Whilst this is clearly not optimal, as it throws away a lot of information, it allows us to simplify the problem and to employ a particularly simple basis through which we impose sparsity. We employ techniques recently developed in the area of convex optimisation to construct a robust reconstruction algorithm, and demonstrate that our method closely reproduces the position, radial extent and amplitude of simulated structures, with very little bias or smearing. This is a significant improvement over current linear methods. In addition, we have shown that our method produces clean results that demonstrate an improvement over linear methods, even when the signal to noise in our data is reduced by a factor of $\sim 2$ in line with that expected from current lensing surveys.

Furthermore, we demonstrate an ability to reconstruct clusters at higher redshifts than has been attainable using linear methods. Although our reconstructions exhibit false detections resulting from the noise, these noisy peaks do not form coherent structures, and are therefore well-localised and easily identifiable as noise peaks.

We have also tested the ability of our method to reconstruct multiple clusters along the line of sight. Whilst the reconstructions are noisy, and exhibit a stronger redshift bias, damping and smearing than that seen in the single cluster case, our method is seen to be sensitive enough to detect the presence of more than one structure along the line of sight. It is hoped that by improving our method, we will be able to more accurately reproduce these structures.

It is clear from our results that this method holds a lot of promise; even in the one dimensional implementation presented here, our results clearly show improvements to the bias, normalisation and smearing problems seen with linear methods. There is much room for improvement, however, and a fully three-dimensional implementation of the algorithm described here (which, as written, is entirely general) will be presented in future work. Such a treatment is expected to reduce the incidences of false detection in our reconstructions, the key to this improvement being the choice of an appropriate three-dimensional dictionary with which to sparsify the solution.

In addition, we note that the simulations presented here are idealised as compared to real data. Most notably, we do not take any account of photometric redshift errors, which will be present in any real dataset. This is an important source of error in lensing measurements, and in order for the method presented above to have any real value in reconstructing the matter distribution in a lensing survey, it must include treatment of these errors. This will be presented in a future work, included in the three-dimensional implementation of the algorithm presented here.

The quality of data available for weak lensing measurements continues to improve, and the methods by which we measure the weak lensing shear are becoming ever more sophisticated. So, too, must the methods we use to analyse the data in order to reconstruct the underlying density. While linear methods appear to be limited in resolution, and offer biased estimators, it is clear that a nonlinear approach such as ours does not, and -- in a fully three-dimensional implementation -- may therefore allow us to map the cosmic web in far greater detail than has previously been achieved.

\section{Acknowledgments}
We would like to thank Jalal Fadili for useful discussions, and Jake VanderPlas for supplying the code for the linear reconstructions and for helpful comments on an earlier version of this paper. We would also like to thank the anonymous referee, for their useful comments and suggestions. This work is supported by the European Research Council grant SparseAstro (ERC-228261).

\appendix

\section{Solving the optimization problem}
\label{sec:solv-optim-probl}

In this section, we explain the development of the reconstruction algorithm. First, we will present an overview of some key concepts and results from convex optimisation, before introducing the primal-dual scheme chosen to solve Equation \eqref{eq:minalt}, and finally discussing the convergence criterion for the algorithm. For a complete introduction to convex analysis, the reader is referred to \cite{Rockafellar70,LemarechalHiriart96,bv04}.

\subsection*{Notation and terminology}
\label{sec:notation}

Let $\mathcal{H}$ a real Hilbert space; in our case, a finite dimensional vector subspace of $\mathbb{R}^n$. We denote by
$\norm{.}$ the norm associated with the inner product in $\mathcal{H}$, $I$ is the identity operator on
$\mathcal{H}$, and $\norm{.}_p (p \geq 1)$ is the $\ell_p$ norm. A real-valued function $f$ is \textit{coercive} if $\lim_{\norm{\bx}  \to +\infty}f(\bx)=+\infty$, and is \textit{proper} if its domain is non-empty: 
\begin{equation*}
\dom f = \{ \bx\in\mathcal{H} \mid f(\bx) < +\infty\} \neq \emptyset\ .
\end{equation*} 
Lastly, $\Gamma_0(\mathcal{H})$ represents the class of all proper lower semicontinuous (lsc) convex functions from $\mathcal{H}$ to $(-\infty,+\infty]$.
  
\subsection{Convex optimisation}
\label{sec:convex-optimisation}

The theory of convex optimisation aims to solve problems of the form
\begin{equation}
  \label{eq:4}
  \hat{x} \in \argmin_{x \in \mathcal{H}} F(x)~,
\end{equation}
where $F : \mathcal{H} \to \mathbb{R}$ is a convex function. If $F$ represents a potential function, for example, then \eqref{eq:4}
will seek for the state of equilibrium, i.e. the state that minimises the potential.

Many algorithms exist for solving such problems, and an excellent introduction can be found in \citep{bv04}. Recently, methods have been developed which exploit the decomposition of $f$ into a sum of smaller convex functions. This can be very helpful if these functions show properties that are not preserved when the functions are summed. 

For example, if we assume that $F$ can be decomposed into a sum of $K$ functions all in $\Gamma_0(\mathcal{H})$, then solving \eqref{eq:4} is equivalent to solving
\begin{equation}
  \label{eq:sum}
  \argmin_{\bx \in \mathcal{H}} ~ \sum_{k=1}^K F_k(\bx)~.
\end{equation}
Such formulation is interesting when the $F_k$ functions show properties that are lost while considering the sum $F$, directly. In our case, we seek a formulation such that the functions we seek to minimise have proximal operators that are \textit{simple}, and have closed form. While the sum $F$ may not have this property, the functions $F_k$ can be chosen such that they satisfy this condition.

\subsection{Proximity operator}
\label{sec:proximity-operator}

We may now begin to describe the proximal splitting algorithm used in this work. At the heart of the splitting framework is the notion of the \textit{proximity operator}:
\begin{definition}[\citet{Moreau1962}]
  \label{def:1}
  Let $F \in \Gamma_{0}(\mathcal{H})$. Then, for every $\bx\in\mathcal{H}$, the function $\by \mapsto F(\by) + \norm{\bx-\by}^{2}_2/2$
  achieves its infimum at a unique point denoted by $\prox_{F} \bx$. The operator $\prox_{F} : \mathcal{H} \to \mathcal{H}$
  thus defined is the \textit{proximity operator} of $F$.
\end{definition}

Therefore, the proximity operator of the indicator function of a convex set $\mathcal{C}$:
\begin{equation*}
\imath_{\mathcal{C}} : \bx \mapsto \begin{cases} 0 & \text{if } \bx \in \mathcal{C}~, \\ \infty & \text{otherwise },\end{cases}\ 
\end{equation*} 
is merely its orthogonal projector. Some key properties are presented in the following lemma:
\begin{lemma}[\citet{cw05}]
  {~}\\ \vspace{-1em}
  \label{lem:decomp}
  \begin{itemize}
  \item \textit{Separability}: Let $F_k \in \Gamma_0(\mathcal{H}),\ k \in \{1,\cdots,K\}$ and let $G : (\bx_k)_{1\le k\le K} \mapsto \sum_k F_k(\bx_k)$. Then
    $\prox_{G} = (\prox_{F_k}) _{1 \le k \le K}$.
  \item \textit{Translation}: Let $F \in \Gamma_0(\mathcal{H})$ and $G \in \Gamma_0(\mathcal{H})$ such that $\forall x \in \mathcal{H}, G(\bx) =
    F(\bx - \by)$, $\by \in \mathcal{H}$. Then $\forall \bx \in \mathcal{H}, \ \prox_{G} \bx = \by + \prox_{F} (\bx - \by)$.
  \end{itemize}
\end{lemma}

\subsection{Primal-dual scheme}
\label{sec:dual-primal-scheme}

Consider the optimization problem:
\begin{equation}
  \label{eq:5}
  \hat{\bx} \in \argmin_{\bx\in\mathcal{H}} G\circ\mathbf{U} (\bx) + B(\bx)~,
\end{equation}
where $G$ and $B$ are two convex, proper and lower semi-continuous functions and $\mathbf{U}$ is a linear bounded operator. Equation \eqref{eq:5} can be see as a special case of Equation \eqref{eq:sum} where $K=2$ and one of the functions contains a linear bounded operator. Then the Algorithm~\ref{algo:primdual}, proposed in \cite{cp10}, will converge to the solution of \eqref{eq:5}, assuming that the proximity operators of $G$ and $B$ are easy to compute or known in closed form.

\begin{algorithm}[htb]
  \noindent{\bf{Parameters:}} The number of iterations
  $N_{\mathrm{iter}}$, proximal steps $\sigma > 0$ and $\tau > 0$. \\
  \noindent{\bf{Initialization:}}\\
  $\bx^0 = \bar{\bx}^0 = 0$
  $\boldsymbol{\xi}^0 = 0$. \\
  \noindent{\bf{Main iteration:}} \\
  \noindent{\bf{For}} $t=0$ {\bf{to}} $N_{\mathrm{iter}}-1$,
  \begin{itemize}
    \setlength{\topsep}{0pt}
    \setlength{\parskip}{0pt}
    \setlength{\itemsep}{0pt}
    \setlength{\partopsep}{0pt}
  \item \underline{Dual step}: $\boldsymbol{\xi}^{t+1} = (I - \sigma \prox_{G/\sigma})(\boldsymbol{\xi}^{t}/\sigma + \mathbf{U} \bar{\bx}^{t})$.
  \item \underline{Primal step}:
    $\bx^{t+1} = \prox_{\tau B} \left(\bx^t -\tau \mathbf{U}^*\boldsymbol{\xi}^{t+1} \right)$.%
  \item Update the coefficients estimate: $\bar{\bx}^{t+1} = 2 \bx^{t+1} - \bx^t$
 \end{itemize}
  \noindent{\bf{End main iteration}} \\
  \noindent{\bf{Output:}} Final solution $\bx^{\star} = \bx^{N_{\mathrm{iter}}}$.
  \caption{Primal-dual scheme for solving .}
  \label{algo:primdual}
\end{algorithm}
 
Adapting the arguments of \citep{cp10}, convergence of the sequence $(\bx^t)_{t\in \mathbb{N}}$ generated by Algorithm~\ref{algo:primdual} is ensured.
\begin{proposition}
  Suppose that $G$ and $B$ are two convex, proper and lower semi-continuous functions. Let $\zeta =  \norm{\boldsymbol{U}}^2$, choose $\tau > 0$ and $\sigma$ such that $\sigma\tau\zeta < 1$, and let  $(\bx^t)_{t\in\mathbb{R}}$ be that defined by Algorithm~\ref{algo:primdual}. Then, $(\bx^t)_{t\in\mathbb{N}}$ converges to a (non-strict) global minimiser of Equation \eqref{eq:5} .
\end{proposition}

\subsection{Application to solution of Equation \eqref{eq:minalt}}
\label{sec:appl-eqref}

First, we must rewrite the problem in Equation \eqref{eq:minalt} in an unconstrained form by replacing the constraints by the indicator functions of the corresponding constraint sets:
\begin{equation}
  \label{eq:unco}
  \min_{\boldsymbol{s}\in\mathbb{R}^n} \norm{\boldsymbol{\Phi}^*\boldsymbol{s}}_1 +
  \imath_{\ell_2(\varepsilon)} \left( \boldsymbol{\Sigma}^{-\tfrac{1}{2}} \boldsymbol{d} - \boldsymbol{\Sigma}^{-\tfrac{1}{2}} \mathbf{R}\boldsymbol{s} \right) +
  \imath_{\mathcal{C}}(\boldsymbol{s})~,
\end{equation}
where $\imath_{\ell_2(\varepsilon)}$ is the indicatrice function of the $\ell_2$-ball of radius $\varepsilon$ and
$\imath_{\mathcal{C}}$ the indicatrice function of a closed convex set $\mathcal{C}$.

Notice that \eqref{eq:unco} can be expressed in the form of \eqref{eq:5} with,
\begin{align}
  \label{eq:7}
  G(\bx,\by) &= \norm{\bx}_1 + \imath_{\ell_2(\varepsilon)}(\by - \boldsymbol{\Sigma}^{-\tfrac{1}{2}} \mathbf{d})~, \\
  B(\bx) &= \imath_{\mathcal{C}}(\bx)~, \\
  \mathbf{U} &= \begin{pmatrix} \boldsymbol{\Phi}^* \\ \boldsymbol{\Sigma}^{-\tfrac{1}{2}} \mathbf{Q} \end{pmatrix}~.
\end{align}
We now need only to apply Algorithm~\ref{algo:primdual} in order to compute the solution. This requires computation of the three proximity operators, which are given by the following proposition:
\begin{proposition}
  \label{prop:palg}
  Let $F \in \Gamma_0(\mathcal{H})$. Then,
  \begin{itemize}
  \item if $F : \bx \mapsto \norm{\bx}_1$, then its associated proximity operator, $\prox_{\lambda F}$, is the component-wise soft-thresholding operator with threshold $\lambda$ as defined by Equation \eqref{eq:softt};
  \item if $F : \bx \mapsto \imath_{\mathcal{C}}(\bx) = \begin{cases} 0 & \text{if } \bx \in \mathcal{C}~, \\
        \infty & \text{else ,} \end{cases}$ \\then its associated proximity operator, $\prox_{F}$, is the
    Euclidian projector to the convex set $\mathcal{C}$, $\mathrm{Prj}_{\mathcal{C}}$;
    \item if $F : \bx \mapsto \imath_{\ell_2(\varepsilon)}(\bx) = \begin{cases} 0 & \text{if } \norm{\bx}_2
          \le \varepsilon~, \\ \infty & \text{else ,} \end{cases}$ \\then its associated proximity operator,
      $\prox_{F}$, is the projector onto the $\ell_2$-ball with radius $\varepsilon$ defined as:
      \begin{equation*}
        \prox_{F} \bx = \begin{cases} \bx & \text{if } \norm{x}_2 \le \varepsilon~, \\ 
        \frac{\varepsilon}{\norm{\bx}_2} \bx & \text{else~.} \end{cases}
      \end{equation*}
  \end{itemize}
\end{proposition}

The application of Proposition~\ref{prop:palg} and Lemma~\ref{lem:decomp} to the optimization problem in Equation \eqref{eq:unco}, using the the primal-dual scheme in Algorithm \ref{algo:primdual}, yields the reconstruction algorithm expressed in Algorithm \ref{alg:inversion}.

\section{Practical Considerations}
\label{sec:implementation}
Application of the method described in Appendix \ref{sec:solv-optim-probl} to the specific case of 3D lensing leads us to Algorithm~\ref{alg:inversion}, where one can note the projection over the two constraints described above (data fidelity and minimum value of the solution), and the operator ${\cal S}t_{\lambda}$, which imposes a soft threshold at a level of $\lambda/\omega$ as:
\begin{gather}
  \label{eq:softt}
  \mathcal{S}t_{\lambda}\ :\ \alpha \mapsto (g_{\lambda}(\alpha_i))_{1\le i \le L}, \\
  g_{\lambda}\ : \ \eta \mapsto \begin{cases} (\abs{\eta} -\lambda)\mathrm{sign}(\eta) & \abs{\eta} > \lambda \\ 0 & \mbox{otherwise}\end{cases}\ . \nonumber
\end{gather}

\begin{algorithm}[h]
  \small
  \noindent{\bf{Parameters:}} Choose $\omega, \tau > 0$ such that $\omega\tau \Theta^2 \le 1$ where $\Theta \equiv \norm{\boldsymbol{\Sigma}^{-\tfrac{1}{2}}\mathbf{R}}_2 + \norm{\boldsymbol\Phi^*}_2$,
  where $\norm{\mathbf{R}}_2$ is the spectral norm of the operator $\mathbf{R}$. \\
  \noindent{\bf{Initialization:}}\\
  $\boldsymbol{y}_1^0 = \boldsymbol{y}_2^0 = \mathbf{0},\ \boldsymbol{\hat{s}}^0 = \mathbf{0},\ \boldsymbol{x}^0 = \mathbf{0}$~.\\
  $N = $ number of elements in $\boldsymbol{d}$~. \\
  $\boldsymbol{\hat{d}}=  \boldsymbol{\Sigma}^{-{\tfrac{1}{2}}}\boldsymbol{d}$~.\\
  Choose $\lambda > 0$~. \\
  \noindent{\bf{Main iteration:}} \\
  \noindent{\bf{For}} $n=0$ {\bf{to}} $N_{\mathrm{iter}}-2$,
 
  \begin{itemize}
    \setlength{\topsep}{0pt}
    \setlength{\parskip}{0pt}
    \setlength{\itemsep}{0pt}
    \setlength{\partopsep}{0pt}
  \item \underline{Initialise auxiliary variables}: \\ \vspace{5pt}
    $\boldsymbol{t}_1^n = \boldsymbol{y}^n_1 + \omega \boldsymbol{\Phi}^*\hat{\boldsymbol{s}}^n.$ \\
    $\boldsymbol{t}_2^n = \boldsymbol{y}^n_2 + \omega \boldsymbol{\Sigma}^{-\tfrac{1}{2}}\mathbf{R}\hat{\boldsymbol{s}}^n.$\vspace{5pt}
  \item \underline{Sparsity-promoting penalty}: \\ \vspace{5pt}
    $\boldsymbol{y}_1^{n+1} = \boldsymbol{t}_1 - \omega\mathcal{S}t_{\lambda/\omega}\left(\boldsymbol{t}_1^n/\omega\right)$.\vspace{5pt}
  \item \underline{Data fidelity term}: \\ \vspace{5pt}
    $\boldsymbol{y}_2^{n+1} = \boldsymbol{t}_2^n -\omega \left(\boldsymbol{\hat{d}} + \boldsymbol{\zeta}\right)$ \\
    $\boldsymbol{\zeta} = \begin{cases} \boldsymbol{t}_2^n/\omega - \boldsymbol{\hat{d}}/\omega & \text{if } \norm{\boldsymbol{t}_2^n/\omega - \boldsymbol{\hat{d}}/\omega}_2 < \epsilon\sqrt{N}~, \\ \\ 
      \frac{\epsilon\sqrt{N}(\boldsymbol{t}_2^n/\omega - \boldsymbol{\hat{d}}/\omega)}{ \norm{\boldsymbol{t}_2^n/\omega - \boldsymbol{\hat{d}}/\omega}_2} & \mbox{otherwise.} \end{cases}$ 
    \vspace{5pt}
  \item \underline{Projection on the convex set $\mathcal{C}$}: \\\ \\
    $\boldsymbol{x}^{n+1} = \mathrm{Prj}_{\mathcal{C}} \left[ \boldsymbol{x}^n - \tau\left(\boldsymbol{\Phi}\boldsymbol{y}_1^{n+1} + \mathbf{R}^\dagger \boldsymbol{\Sigma}^{\tfrac{1}{2}} \boldsymbol{y}_2^{n+1} \right) \right]$.\vspace{5pt}
  \item Update current estimate: \\\ \\
    $\boldsymbol{\hat{s}}^{n+1} = 2\boldsymbol{x}^{n+1} - \boldsymbol{x}^n$.
  \end{itemize}
  \noindent{\bf{End main iteration}} \\
  \vspace{5pt}
  \caption{Nonlinear iterative algorithm for solving Equation \eqref{eq:minalt}.}
  \label{alg:inversion}
\end{algorithm}

\subsection{Convergence criteria}
\label{sec:convergence-criteria}

The main difficulty with the primal-dual scheme described above -- indeed, with any iterative algorithm -- is to define an appropriate convergence criterion. In this case, the difference between
two successive iterates $\bx^{t}$ and $\bx^{t+1}$ is not bounded. However, the partial primal-dual gap $\mathfrak{G}_{pd}$, defined by:
\begin{equation}
  \label{eq:6}
  \begin{split}
    \mathfrak{G}_{pd} (\bx,\by) = & \max_{\by'}  \scalp{\by'}{\mathbf{U}\bx} - G^*(\by') + B(\bx) -\\ & \min_{\bx'} \scalp{\by}{\mathbf{U}\bx'} - G^*(\by) + B(\bx')~,
  \end{split}
\end{equation}
(when solving~\eqref{eq:5}) is bounded. Here, $G^*$ is the convex conjugate of convex function $G$:
\begin{equation}
  \label{eq:8}
  G^* : \bx \mapsto \max_{\by} \scalp{\bx}{\by} - G(\by)~.
\end{equation}

Let us define two variables from the sequences $(\bx^t)_t$ and $(\boldsymbol{\xi}^t)_t$ produced by Algorithm~\ref{algo:primdual}: 
\begin{equation*}
\bx^N = \tfrac{1}{N} \sum_{t=0}^{N-1} \bx^t\mbox{ and }\boldsymbol{\xi}^N = \tfrac{1}{N} \sum_{t=0}^{N-1} \boldsymbol{\xi}^t\ ,
\end{equation*} 
which are the \textit{accumulation variables} at iteration $N-1$. \citet{cp10} have shown that the sequence defined by $\left(\mathfrak{G}_{pd} (\bx^N,\boldsymbol{\xi}^N)\right)_{N \in \mathbb{N}}$ is bounded, and decreases at a rate of $\mathcal{O}(1/t)$, where $t$ is the iteration number.

In order to use \eqref{eq:6} in our context, the two indicatrice functions inside \eqref{eq:unco} are not considered, as they play little role. Therefore, in our case, we may rewrite $\mathfrak{G}_{pd} (\bx,\by)$ as:
\begin{align}
  \label{eq:9}
  \mathfrak{G}_{pd} (\bx,\by) & \approx \max_{\by'}  \scalp{\by'}{\mathbf{U}x} - G^*(\by') - G^*(\by)~, \\
  &\approx G(\mathbf{U} \mathbf{\bx}) - G^*(\by)~, \\
  &\approx \lambda \norm{\boldsymbol{\Phi}^* \mathbf{\bx}}_1~.
\end{align}
We determine the algorithm to have converged when
\begin{equation}
\Delta \mathfrak{G}_{pd}^N \equiv \frac{\mathfrak{G}_{pd} (\bx^N,\boldsymbol{\xi}^N) - \mathfrak{G}_{pd} (\bx^{N-1},\boldsymbol{\xi}^{N-1})}{\mathfrak{G}_{pd} (\bx^{N-1},\boldsymbol{\xi}^{N-1})} < \alpha ,
\end{equation}
where the appropriate $\alpha$ can be determined from simulations, and is dependent on a tradeoff between the desired level of accuracy of the reconstructed data and the time taken to complete the reconstruction. Choosing $\alpha$ to be large will result in an estimate of the solution that may be some distance away from the solution that would be obtained if absolute convergence were reached, but which is obtained in a small number of iterations.

\begin{figure}[htbp]
\includegraphics[width = 0.45\textwidth]{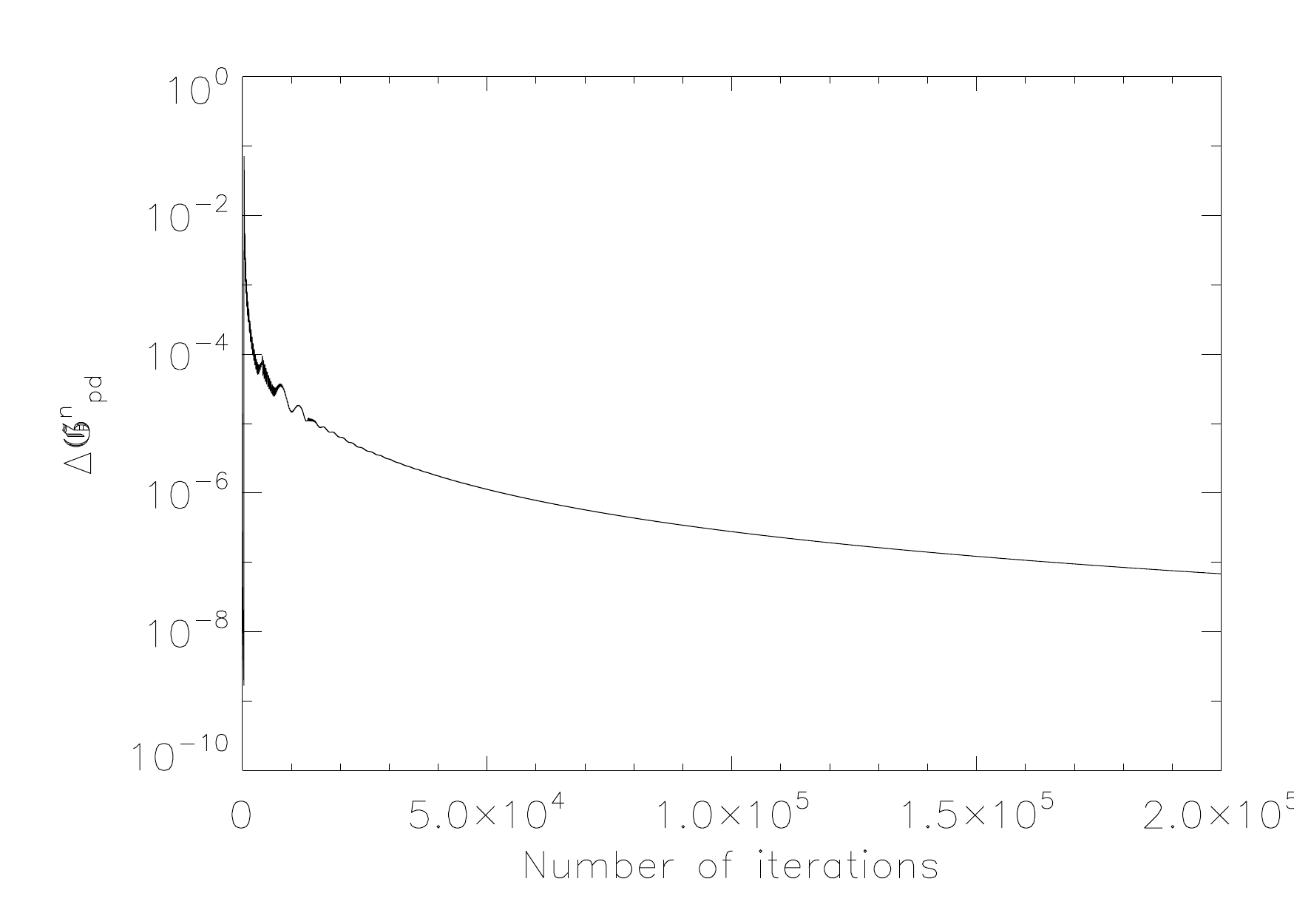}
\caption{The function $\Delta\mathfrak{G}_{pd}^N$, as defined in the text, as a function of iteration number.}
\label{fg:conv}
\end{figure}

Figure~\ref{fg:conv} shows a characteristic example of the function $\Delta\mathfrak{G}_{pd}^N$ as a function of iteration number $N$ for the simulations described in \S~\ref{sec:results}. This function is clearly a smooth, largely steadily decreasing function of iteration number, and thus an appropriate choice for defining convergence. Note that the curve shows some oscillations with iteration number. Such oscillations arise on lines of sight where the noise results in the algorithm having difficulty fitting the data within the constraints. With appropriate choice of parameters, these oscillations are relatively small and the curve eventually becomes smooth.

We find from experimentation that $\alpha = 10^{-6}$ yields a solution that is sufficiently accurate for our purposes, and is sufficient to largely remove the oscillations, thus we set this to be our threshold for all reconstructions presented above. In addition, we require that each line of sight undergoes at least 1500 iterations, to avoid misidentification of convergence due to early oscillations of $\Delta\mathfrak{G}_{pd}^N$, specifically a strong dip seen in this curve at the start of iteration.

It is possible, along certain lines of sight, for the estimate of the solution to remain constant at zero due to the soft thresholding while $\Delta\mathfrak{G}_{pd}^N$ still varies. To account for this, if the current cumulative estimate $x^N$ does not vary for 200 iterations, we assume that convergence has been reached for that line of sight.

We now consider several practical issues involved in the implementation of this algorithm.

\subsection{Choice of Step Sizes $\omega,\ \tau$}

$\omega$ and $\tau$ control the step size in the evolution of the algorithm, and are required to be positive and to satisfy the inequality
\begin{equation}
\omega\tau\Theta^2 \le 1
\label{eq:ineq}
\end{equation}
where $\Theta \equiv \norm{\boldsymbol{\Sigma}^{-\tfrac{1}{2}}\mathbf{Q}}_2+\norm{\boldsymbol{\Phi}^*}_2$ is the sum of the $\ell_2$ norms of the operators used in the algorithm. This sum is dominated by the second term, as the elements of the $\mathbf{Q}$ lensing efficiency matrix are small. We have chosen a $\delta$-function dictionary, therefore the application of the transformation $\boldsymbol{\Phi}^*$ represents a multiplication by the identity matrix. 

Thus we have $\Theta \sim \norm{\boldsymbol{\Phi}^*}_2\sim 1$, which implies that $\omega\tau \sim 1$ is appropriate. For all the results that follow, we choose $\omega = \tau = 1$. Smaller values of these parameters may be used with little effect on the resulting solution. However, the smaller the values chosen, the smaller the steps taken in each iteration of the algorithm. This results in a slower convergence than seen with larger values of $\omega$ and $\tau$. Choosing larger values than implied by Equation \eqref{eq:ineq} is not advised, as convergence of the algorithm is not guaranteed in this case; the algorithm may give rise to a strongly oscillating estimator, and the resulting solution may not be the one that best fits the data given the prior constraints. 



\bibliographystyle{aa}

\bibliography{refs}

\begin{thebibliography}{41}
\expandafter\ifx\csname natexlab\endcsname\relax\def\natexlab#1{#1}\fi

\bibitem[{Aitken(1934)}]{aitken34}
Aitken. 1934, Proc. R. Soc. Edinb., 55, 42

\bibitem[{{Albrecht} {et~al.}(2006){Albrecht}, {Bernstein}, {Cahn}, {Freedman},
  {Hewitt}, {Hu}, {Huth}, {Kamionkowski}, {Kolb}, {Knox}, {Mather}, {Staggs},
  \& {Suntzeff}}]{albrechtetal06}
{Albrecht}, A., {Bernstein}, G., {Cahn}, R., {et~al.} 2006, ArXiv
  astro-ph/0609591

\bibitem[{{Bacon} \& {Taylor}(2003)}]{bt03}
{Bacon}, D.~J. \& {Taylor}, A.~N. 2003, MNRAS, 344, 1307

\bibitem[{{Barbey} {et~al.}(2011){Barbey}, {Sauvage}, {Starck}, {Ottensamer},
  \& {Chanial}}]{Barbey11}
{Barbey}, N., {Sauvage}, M., {Starck}, J.-L., {Ottensamer}, R., \& {Chanial},
  P. 2011, \aap, 527, A102+

\bibitem[{{Bobin} {et~al.}(2008){Bobin}, {Starck}, \& {Ottensamer}}]{bobin2008}
{Bobin}, J., {Starck}, J.-L., \& {Ottensamer}, R. 2008, IEEE Journal of
  Selected Topics in Signal Processing, 2, 718

\bibitem[{Boyd \& Vandenberghe(2004)}]{bv04}
Boyd, S. \& Vandenberghe, L. 2004, Convex Optimization (Cambridge University
  Press)

\bibitem[{Cand{\`e}s \& Tao(2006)}]{CandesTao04}
Cand{\`e}s, E. \& Tao, T. 2006, {IEEE} Transactions on Information Theory, 52,
  5406--5425

\bibitem[{Candes \& Plan(2010)}]{candes2010}
Candes, E.~J. \& Plan, Y. 2010, A Probabilistic and RIPless Theory of
  Compressed Sensing, submitted

\bibitem[{{Castro} {et~al.}(2005){Castro}, {Heavens}, \& {Kitching}}]{chk05}
{Castro}, P.~G., {Heavens}, A.~F., \& {Kitching}, T.~D. 2005, Phys. Rev. D, 72,
  023516

\bibitem[{Chambolle \& Pock(2011)}]{cp10}
Chambolle, A. \& Pock, T. 2011, Journal of Mathematical Imaging and Vision, 40,
  120

\bibitem[{Combettes \& Wajs(2005)}]{cw05}
Combettes, P.~L. \& Wajs, V.~R. 2005, SIAM Multiscale Model. Simul., 4, 1168

\bibitem[{Donoho(2006)}]{donoho:cs}
Donoho, D. 2006, IEEE Transactions on Information Theory, 52, 1289--1306

\bibitem[{Fadili \& Starck(2009)}]{fs09}
Fadili, M. \& Starck, J.-L. 2009, in Proceedings of the International
  Conference on Image Processing, ICIP 2009, 7-10 November 2009, Cairo, Egypt
  (IEEE), 1461--1464

\bibitem[{{Hoekstra} \& {Jain}(2008)}]{hj08}
{Hoekstra}, H. \& {Jain}, B. 2008, Annual Review of Nuclear and Particle
  Science, 58, 99

\bibitem[{{Hu} \& {Keeton}(2002)}]{hk02}
{Hu}, W. \& {Keeton}, C.~R. 2002, Phys. Rev. D, 66, 063506

\bibitem[{{Ivezic} {et~al.}(2008){Ivezic}, {Tyson}, {Allsman}, {Andrew},
  {Angel}, \& {for the LSST Collaboration}}]{lsst1}
{Ivezic}, Z., {Tyson}, J.~A., {Allsman}, R., {et~al.} 2008, ArXiv: 0805.2366

\bibitem[{{Kitching} {et~al.}(2011){Kitching}, {Heavens}, \&
  {Miller}}]{kitchingetal11}
{Kitching}, T.~D., {Heavens}, A.~F., \& {Miller}, L. 2011, MNRAS, 426

\bibitem[{{Larson} {et~al.}(2011){Larson}, {Dunkley}, {Hinshaw}, {Komatsu},
  {Nolta}, {Bennett}, {Gold}, {Halpern}, {Hill}, {Jarosik}, {Kogut}, {Limon},
  {Meyer}, {Odegard}, {Page}, {Smith}, {Spergel}, {Tucker}, {Weiland},
  {Wollack}, \& {Wright}}]{larsonetal11}
{Larson}, D., {Dunkley}, J., {Hinshaw}, G., {et~al.} 2011, ApJS, 192, 16

\bibitem[{Lemar\'echal \& Hiriart-Urruty(1996)}]{LemarechalHiriart96}
Lemar\'echal, C. \& Hiriart-Urruty, J.-B. 1996, Convex Analysis and
  Minimization Algorithms I, 2nd edn. (Springer)

\bibitem[{{Levy} \& {Brustein}(2009)}]{lb09}
{Levy}, D. \& {Brustein}, R. 2009, JCAP, 6, 26

\bibitem[{{Li} {et~al.}(2011){Li}, {Cornwell}, \& {de Hoog}}]{Cornwell2011}
{Li}, F., {Cornwell}, T.~J., \& {de Hoog}, F. 2011, \aap, 528, A31+

\bibitem[{{LSST Science Collaboration} {et~al.}(2009){LSST Science
  Collaboration}, {Abell}, {Allison}, {Anderson}, {Andrew}, {Angel}, {Armus},
  {Arnett}, {Asztalos}, {Axelrod}, \& et~al.}]{lsst2}
{LSST Science Collaboration}, {Abell}, P.~A., {Allison}, J., {et~al.} 2009,
  ArXiv: 0912.0201

\bibitem[{{Ma} {et~al.}(2006){Ma}, {Hu}, \& {Huterer}}]{maetal06}
{Ma}, Z., {Hu}, W., \& {Huterer}, D. 2006, ApJ, 636, 21

\bibitem[{{Massey} {et~al.}(2007{\natexlab{a}}){Massey}, {Rhodes}, {Ellis},
  {Scoville}, {Leauthaud}, {Finoguenov}, {Capak}, {Bacon}, {Aussel}, {Kneib},
  {Koekemoer}, {McCracken}, {Mobasher}, {Pires}, {Refregier}, {Sasaki},
  {Starck}, {Taniguchi}, {Taylor}, \& {Taylor}}]{masseyetal07b}
{Massey}, R., {Rhodes}, J., {Ellis}, R., {et~al.} 2007{\natexlab{a}}, Nature,
  445, 286

\bibitem[{{Massey} {et~al.}(2007{\natexlab{b}}){Massey}, {Rhodes}, {Leauthaud},
  {Capak}, {Ellis}, {Koekemoer}, {R{\'e}fr{\'e}gier}, {Scoville}, {Taylor},
  {Albert}, {Berg{\'e}}, {Heymans}, {Johnston}, {Kneib}, {Mellier}, {Mobasher},
  {Semboloni}, {Shopbell}, {Tasca}, \& {Van Waerbeke}}]{masseyetal07a}
{Massey}, R., {Rhodes}, J., {Leauthaud}, A., {et~al.} 2007{\natexlab{b}}, ApJS,
  172, 239

\bibitem[{Moreau(1962)}]{Moreau1962}
Moreau, J.-J. 1962, CRAS S\'er. A Math., 255, 2897

\bibitem[{{Munshi} {et~al.}(2008){Munshi}, {Valageas}, {van Waerbeke}, \&
  {Heavens}}]{munshietal08}
{Munshi}, D., {Valageas}, P., {van Waerbeke}, L., \& {Heavens}, A. 2008, Phys.
  Rep., 462, 67

\bibitem[{{Peacock} {et~al.}(2006){Peacock}, {Schneider}, {Efstathiou},
  {Ellis}, {Leibundgut}, {Lilly}, \& {Mellier}}]{peacocketal06}
{Peacock}, J.~A., {Schneider}, P., {Efstathiou}, G., {et~al.} 2006, {ESA-ESO
  Working Group on ''Fundamental Cosmology''}, Tech. rep., Ed. Universirty

\bibitem[{{Refregier} {et~al.}(2010){Refregier}, {Amara}, {Kitching}, {Rassat},
  {Scaramella}, {Weller}, \& {Euclid Imaging Consortium}}]{euclid}
{Refregier}, A., {Amara}, A., {Kitching}, T.~D., {et~al.} 2010, ArXiv:
  1001.0061

\bibitem[{Rockafellar(1970)}]{Rockafellar70}
Rockafellar, R. 1970, Convex analysis (Princeton University Press)

\bibitem[{{Schneider}(2006)}]{schneider06}
{Schneider}, P. 2006, {Weak Gravitational Lensing} (Springer), 269--+

\bibitem[{{Simon} {et~al.}(2011){Simon}, {Heymans}, {Schrabback}, {Taylor},
  {Gray}, {van Waerbeke}, {Wolf}, {Bacon}, {Barden}, {B{\"o}hm},
  {H{\"a}u{\ss}ler}, {Jahnke}, {Jogee}, {van Kampen}, {Meisenheimer}, \&
  {Peng}}]{simonetal11}
{Simon}, P., {Heymans}, C., {Schrabback}, T., {et~al.} 2011, MNRAS, 1789

\bibitem[{{Simon} {et~al.}(2009){Simon}, {Taylor}, \& {Hartlap}}]{sth09}
{Simon}, P., {Taylor}, A.~N., \& {Hartlap}, J. 2009, MNRAS, 399, 48

\bibitem[{Starck {et~al.}(2010)Starck, Murtagh, \& Fadili}]{smf10}
Starck, J.-L., Murtagh, F., \& Fadili, J.~M. 2010, {Sparse Image and Signal
  Processing} (Cambridge University Press)

\bibitem[{{Taylor} {et~al.}(2004){Taylor}, {Bacon}, {Gray}, {Wolf},
  {Meisenheimer}, {Dye}, {Borch}, {Kleinheinrich}, {Kovacs}, \&
  {Wisotzki}}]{tayloretal04}
{Taylor}, A.~N., {Bacon}, D.~J., {Gray}, M.~E., {et~al.} 2004, MNRAS, 353, 1176

\bibitem[{{Taylor} {et~al.}(2007){Taylor}, {Kitching}, {Bacon}, \&
  {Heavens}}]{tayloretal07}
{Taylor}, A.~N., {Kitching}, T.~D., {Bacon}, D.~J., \& {Heavens}, A.~F. 2007,
  MNRAS, 374, 1377

\bibitem[{{Tegmark}(1997)}]{tegmark97}
{Tegmark}, M. 1997, ApJL, 480, L87+

\bibitem[{{Tegmark} {et~al.}(1997){Tegmark}, {de Oliveira-Costa}, {Devlin},
  {Netterfields}, {Page}, \& {Wollack}}]{tegmarketal97}
{Tegmark}, M., {de Oliveira-Costa}, A., {Devlin}, M.~J., {et~al.} 1997, ApJL,
  474, L77+

\bibitem[{{Van Waerbeke} \& {Mellier}(2003)}]{vwm03}
{Van Waerbeke}, L. \& {Mellier}, Y. 2003, ArXiv astro-ph/0305089

\bibitem[{{VanderPlas} {et~al.}(2011){VanderPlas}, {Connolly}, {Jain}, \&
  {Jarvis}}]{vanderplasetal11}
{VanderPlas}, J.~T., {Connolly}, A.~J., {Jain}, B., \& {Jarvis}, M. 2011, ApJ,
  727, 118

\bibitem[{{Wiaux} {et~al.}(2009){Wiaux}, {Jacques}, {Puy}, {Scaife}, \&
  {Vandergheynst}}]{wiaux2009}
{Wiaux}, Y., {Jacques}, L., {Puy}, G., {Scaife}, A.~M.~M., \& {Vandergheynst},
  P. 2009, \mnras, 395, 1733

\end{thebibliography}


\end{document}